\documentclass[12pt]{article}

\usepackage[utf8]{inputenc} 
\usepackage{newunicodechar} 
\usepackage{amssymb,amsmath,amsfonts,graphicx,color}
\graphicspath{{./4_Output/}} 
\usepackage{eurosym,geometry,ulem,setspace,comment}
\usepackage[colorlinks=true,citecolor=blue,linkcolor=blue,urlcolor=blue]{hyperref} 
\usepackage[symbol]{footmisc}
\usepackage{lscape, tikz, afterpage} 
\usepackage{pbox}
\usepackage{float,enumitem}
\usepackage{subcaption,bbm}
\usepackage[space]{grffile}
\usepackage{import}
\usepackage{multicol, setspace} 

\usepackage{chngcntr} 
\usepackage{pdfpages} 

\usepackage{natbib}
\usepackage[utf8]{inputenc} 
\usepackage{newunicodechar} 
\usepackage[flushleft]{threeparttable}
\usepackage[labelsep=period]{caption} 
\usepackage{multirow,booktabs, tabularx}
\usepackage{dcolumn} 
  \newcolumntype{d}[1]{D{.}{.}{#1}}

\definecolor{paleblue}{rgb}{0.439,0.654,0.863}
\definecolor{midpurple}{rgb}{0.616,0.451,0.686}

\usepackage{pgfplots}
\usepgfplotslibrary{groupplots}
\pgfplotsset{compat = newest}
\usetikzlibrary{positioning, arrows.meta}
\usepgfplotslibrary{fillbetween}

\usepackage{longtable,booktabs,threeparttablex}

\title{Medical Bill Shock and Imperfect Moral Hazard\footnote{
The authors are grateful to the editor, Sarah Miller, and three helpful referees for their constructive comments and guidance on this paper. In addition, we gratefully acknowledge Zarek Brot-Goldberg, Laura Derksen, Tal Gross, Tim Layton, Ryan McDevitt, Petra Rasmussen, Paul Shafer, Michael Stepner, and Jonathan Zhang for very helpful comments on this project, as well as participants in the ASHEcon 2023 Conference, the AcademyHealth 2023 Annual Research Meeting, and the APPAM 2022 Research Conference and seminar participants at Brigham Young University, Cornell University, Duke University, and the University of Toronto.}}

\author{David M. Anderson\thanks{Duke University, Department of Population Health Sciences, 215 Morris St., Durham, North Carolina, 27701. \textcolor{blue}{\underline{\href{mailto:dma34@duke.edu}{dma34@duke.edu}}}.} \and Alex Hoagland\thanks{Corresponding author. University of Toronto, Institute of Health Policy, Management, and Evaluation, 155 College St, Toronto, Ontario, Canada. \textcolor{blue}{\underline{\href{mailto:alexander.hoagland@utoronto.ca}{alexander.hoagland@utoronto.ca}}}. Website: \textcolor{blue}{\underline{\href{https://alex-hoagland.github.io}{alex-hoagland.github.io}}}.} \and Ed Zhu\thanks{Boston University, Department of Economics, 270 Bay State Road, Boston, MA 02215. \textcolor{blue}{\underline{\href{mailto:edzhu1@bu.edu}{edzhu1@bu.edu}}}.}}

\begin{document}

\bgroup
\let\footnoterule\relax

\begin{singlespace}
\maketitle

\begin{abstract}
\noindent Consumers are sensitive to medical prices when consuming care, but delays in price information may distort moral hazard. We study how medical bills affect household spillover spending following utilization, leveraging variation in insurer claim processing times. Households increase spending by 22\% after a scheduled service, but then reduce spending by 11\% after the bill arrives. Observed bill effects are consistent with resolving price uncertainty; bill effects are strongest when pricing information is particularly salient. A model of demand for healthcare with delayed pricing information suggests households misperceive pricing signals prior to bills, and that correcting these perceptions reduce average (median) spending by 16\% (7\%) annually. 
\vspace{0.0in}\\
\vspace{0.05in}\vspace{.00in} 
\vspace{0in}\\
\bigskip
\end{abstract}
\end{singlespace}
\thispagestyle{empty}

\clearpage
\egroup
\setcounter{page}{1}
\renewcommand{\thefootnote}{\arabic{footnote}}

\section{Introduction}\label{sec:intro}
Health insurance plays a vital role in protecting consumers against the risk of volatile, unpredictable health and financial shocks. However, incomplete information plagues health insurance markets, ultimately leading both public institutions (e.g., governments) and private organizations (e.g., insurers) to provide sub-optimal coverage \citep{einav_moral_2018,dave_health_2009}. 

A primary information friction in healthcare markets is ``\textit{ex-post} moral hazard,” the extent to which consumption is price sensitive.\footnote{Referring to elastic healthcare demand as ``moral hazard" is a now widely-used abuse of notation, beginning with \cite{arrow_uncertainty_1963}. Specifically, moral hazard refers to how, conditional on health, consumption adapts to care prices \citep{pauly_moral_2008,cutler_anatomy_2000}. Previous work has underscored its role in decision-making \citep{kowalski_censored_2016-1,duarte_price_2012,dunn_health_2016}.} Moral hazard ultimately justifies exposing consumers to out-of-pocket (OOP) cost-sharing for health services \citep{chandra_patient_2010,goldman_integrated_2007}, for example, through increasing enrollment in high-deductible health plans (HDHPs) to limit high consumption of potentially low-value health services \citep{geyman_cost-sharing_2012}. Price pressures may harm households who either delay or forego necessary medical care\footnote{Consumers exposed to higher rates of cost-sharing are more likely to report delaying medical care, a finding exacerbated among low-income households \citep{kullgren_health_2010} or those with high-cost chronic conditions \citep{fu_out--pocket_2021,gaffney_high-deductible_2020}.} or reduce consumption of high-value health services such as preventive care.\footnote{While value-based insurance designs---where certain high-value services are carved out of cost-sharing obligations---have become more prevalent \citep{chernew_value-based_2007-1}, confusion about insurance contracts may still affect take-up \citep{hoagland_out--pocket_2021,shafer_trends_2021}.} 

While consumers are responsive to prices when making healthcare consumption decisions, there is ongoing uncertainty about the extent to which consumers actually have access to accurate and timely information about the marginal costs of care \citep{lieber_does_2017}. Recent work has focused on consumer knowledge of the \textit{ex-ante} OOP price of a service, such as how consumers search across multiple medical providers offering the same service at different prices \citep{brown_empirical_2017}. In this paper, we highlight an overlooked feature of medical demand under price uncertainty with significant implications for models of moral hazard: lack of timely \textit{ex-post} pricing information. 

Consumers are rarely given accurate information about service prices at the point of consumption, much less information about their own expected OOP contribution.\footnote{Notably, health services are characterized by a total amount billed by physicians (a ``sticker" price); a negotiated total price approved by the patient's insurer; and the relative fraction of that negotiated price that is the patient's OOP responsibility. Importantly, price transparency for medical pricing must take into account these various prices, including the relative lack of information contained in sticker prices.} Service prices vary across healthcare organizations, payers, and service bundles \citep{fronsdal_variation_2020,patel_financial_2023}, even for very common services \citep{gruber_financing_2022,cooper_price_2019}.\footnote{Appendix Figure \ref{figax:price-hist} illustrates some of the variations in prices for common services in our sample.} Furthermore, \textit{ex-post} haggling over reimbursement between insurers and providers may prolong consumers' price uncertainty. Patient OOP costs vary over contracts and time as a function of prior consumption at the household level; hence, residual uncertainty about realized spending affects marginal prices for care consumed even before bills arrive, and consumers must form expectations about their already realized expenses in the interim. 


We isolate the causal impact of receiving a medical bill on household spillovers in healthcare consumption, by studying how households with employer-sponsored insurance (ESI) in the US make collective spending decisions after one household member incurs a significant health expenditure.\footnote{Specifically, we assess spillover household responses following the use of a health service classified as ``shoppable" by the Centers for Medicare \& Medicaid Services (CMS) \citep{cms_medicare_2019}; see Section \ref{sec:data} for details. We exclude the household member who received the service in order to estimate spillover responses among the unaffected household members and identify the causal effect of a bill's arrival in changing these responses. Our results are robust to using a broader definition of index health events, including unplanned injuries and surgeries.} We use exogenous variation in the time consumers wait for their bills---driven by variation at the insurer-clinician level in claim submission and processing times---to identify the effect a bill on spending, compared to household spending prior to the event and spending during the household's ``interim period" between a service and a bill. We formalize this three-way comparison between post-bill spending and both pre-event and interim spending using a modification of a triple differences regression. 

We find strong evidence that a bill's arrival changes household behavior, implying that patients and households face real uncertainty about their bills. In the interim period between the service and its bill, household members increase their total health spending by about 21.8\% (roughly \$27 per person per week). However, once the bill arrives, consumption drops significantly by 10.9\%, half of the post-service increase. We provide evidence that these bill effects are concentrated among bills that provide meaningful pricing information to households, including information about the expected OOP costs for care consumed between the service date and the bill's arrival. In particular, households who learn that they have not yet met a deductible (meaning their marginal cost-sharing is 100\% for services not carved out of coverage) exhibit larger estimated effects than those whose index service spending meets the deductible threshold.

We show evidence suggesting bill effects are driven primarily by the pricing information they provide, rather than by other channels. We observe the largest responses to a bill when it is most informative of prices, such as for households just shy of meeting their deductible. Second, we demonstrate that households appear to learn about their marginal prices over time. Other potential mechanisms, including liquidity constraints or supplier-induced responses, do not sufficiently explain observed spending changes. Finally, bills shift \textit{where} households seek these services (e.g., from a hospital to an outpatient clinic). 

These features of observed bill effects---which suggests households face real uncertainty about marginal prices prior to a bill's arrival---may lead to deviations in medical decision-making from typical model predictions under full information. We therefore develop and estimate a model of ``imperfect moral hazard" in which households face delays in information about their OOP spending to date. We use the exogenous variation in our data to identify the implied distribution of spending signals and household learning over time. The model estimates suggest---in keeping with our reduced-form results---that households receive noisy spending signals prior to a bill's arrival. On average, these signals are larger than the true bill, meaning that households are likely to incorrectly assume that they have met their deductible while waiting for a bill. Delayed resolution of price uncertainty results in 85\% of households spending more on care than they would under real-time claims adjudication. We estimate that the average (median) affected household spends \$364 (\$94) more per household member per plan year.\footnote{Although this leads to greater total spending, this is not to say that increased consumption is normatively of lower value to the household. For example, this increase in consumption could be an increase in services that are beneficial to the household but were previously delayed or foregone due to liquidity constraints.} We also find strong evidence of consumer learning, with price uncertainty affecting households significantly more at the beginning of a plan year. 

We present the first model of healthcare demand under delayed information and highlight its implications for consumption and welfare; hence, our work makes several important contributions. First, we add to an ongoing literature on dynamic responses to cost-sharing, including the strategic delay of services such as dental care \citep{cabral_claim_2017-1} and models of ``forward-looking" moral hazard \citep{aron-dine_moral_2015,baicker_behavioral_2015}. In contrast to previous work modeling uncertainty in expected end-of-year prices \citep{einav_response_2015}, we highlight the role of pricing uncertainty in the short-run demand for care. Recent work has found patients will defer care when anticipating future price changes \citep{hettinger_intertemporal_2022,johansson_reductions_2023}. Although there is strong evidence for the role of dynamic moral hazard in healthcare \citep{klein_response_2022,diaz-campo_dynamic_2022}, our results highlight that information about \textit{ex-post} prices changes real-time decisions about both when and where to receive care, even for services which cannot be strategically delayed.

Our findings also fit into a larger discussion of the usefulness of price transparency policies in mitigating large levels of healthcare consumption in the United States \citep{muir_clarifying_2012,zhang_impact_2020}. In contrast to previous work---which highlighted how the availability of price information may change the strategic decisions of patients shopping for a service \citep{gondi_early_2021-1, reed_care-seeking_2005}---we highlight a new mechanism through which price transparency may affect \textit{future} care decisions, even across household members. Our findings suggest that shortening periods of price uncertainty may affect decision-making for the entire household. Policies that decrease the information gap between consumption and accurate price information---such as real-time claims adjudication for physical health claims, similar to prescription drug claims adjudication \citep{hartzema_utilizing_2011}---would reduce fluctuations in consumption decisions, improving household welfare to the extent that unexpected price shocks are eliminated. 


Finally, we contribute to a broad and growing literature related to bureaucracy in medicine \citep{brot-goldberg_rationing_2023,shi_monitoring_2022,league_administrative_2022}. In addition to large expenses related to administrative burden in the healthcare system, our work highlights that administrative frictions directly affect medical consumption decisions, both for the affected enrollee and others in a household. Idiosyncratic differences in clinician practice management and physician-insurer interactions, including hospital or physician delays in submitting claims, insurer delays or errors in following provider group contracts, or disagreements between insurers and providers, may exacerbate these frictions. Faced with these uncertainties, households are left to form unreliable expectations about the costs of their care while making demand decisions. In this sense, our work is closely related to recent work that found providers are less willing to treat patients with greater degrees of billing uncertainty, such as publicly-insured patients \citep{dunn_costs_2020}. We complement this work by highlighting these effects on the demand side, where patients and their families have less choice over administrative frictions, but still face real costs for delayed information.


Health care is not the only setting where marginal price uncertainty affects consumption decisions. ``Bill shock" is common in other industries including household utilities, cell phone services, and even college education financing \citep{grubb_cellular_2015}. Our work furthers models of demand under marginal price uncertainty by providing a tractable estimation of consumer beliefs and learning. Our model is related to others where individuals learn about uncertain prices of goods (including financial assets and agricultural goods) \citep{ngangoue_learning_2021,boyd_microeconomics_2020}; however, our model does not rely on consumer inattentiveness to past consumption, but underscores the role of delayed information arising from complex contracts involving multiple parties.\footnote{Our work is also related to literature on learning models with delays in belief updating \citep{karlsson_ostrich_2009,peng_learning_2005}. However, in these models, delays typically arise endogenously as consumers either choose to delay learning or have limited information processing abilities. In contrast, our model exploits exogenous variation in the delayed \textit{arrival} of information outside the consumer's control, but which still affects the marginal utility and costs associated with choices retroactively.} Studying these complex contracts has the added advantage that we avoid concerns about endogenous price setting at the supplier level, given that bill shock arises as a disconnect between insurers and physicians rather than from a single supplier such as a cell phone provider \citep{grubb_consumer_2015}.\footnote{This is in contrast to endogenous price setting in the context of \textit{ex-ante} prices for specific medical services, as discussed in \cite{brown_empirical_2017}.} Finally, studying bill shock in ESI is particularly salient given that it comprises roughly 6\% of US GDP. 

We discuss the setting of shoppable services and the data in Section \ref{sec:data}. We then present our methods and identifying assumptions in Section \ref{sec:methods}, followed by our empirical results in Section \ref{sec:rf-evidence}. We incorporate these findings into a model of imperfect moral hazard in Section \ref{sec:model}. Finally, Section \ref{sec:conclusion} highlights the relevance of these findings for the optimal design of insurance contracts. 

\section{Setting and Data}\label{sec:data}
\subsection{Data}

We use data on household healthcare utilization from the Merative (formerly IBM Watson Health) Marketscan \textit{Commercial Claims and Encounters} Database, which contains detailed inpatient, outpatient, and pharmaceutical claims for a sample of households enrolled in ESI through large U.S. firms. Each observation includes diagnostic, procedural, and payment information, including the date of service and the corresponding date on which the insurer paid their portion of the claim. The data also includes household, firm, and insurance plan identifiers and characteristics.\footnote{We use the empirical set of enrollees in a plan to estimate insurance plan characteristics, following previous literature \citep{hoagland_ounce_2022,marone_should_2022}. See \cite{hoagland_ounce_2022} Appendix A for a detailed description of this methodology and an evaluation of the quality of these inferences.}

We limit our analytical sample to enrollees in one of eight large firms between 2006 and 2018.\footnote{Firms were randomly selected from a larger sample of firms with plan identifiers available; plans include both high-deductible health plans (HDHPs) and other plan types (HMOs and zero-deductible plans), and have a start date of January 1 for all observed years. Note that insurance plan identifiers are only available through 2013, as discussed below.} Our final sample includes 386,240 households with two or more members, full eligibility, and continuous enrollment across their window of observation. Throughout, spending data has been normalized to 2022 USD using the Consumer Price Index for All Urban Consumers series.

\begin{table}[htb]
\centering
\begin{threeparttable}
\begin{tabular}{l|cc}
\toprule
& \multicolumn{1}{c}{Full Sample} & \multicolumn{1}{c}{Plan-Identified Sample} \\
\midrule
\textbf{Panel A:} Demographics \\
Age (individual) &31.67 (0.000) &31.15 (0.000) \\
\% female (individual) & 0.51 (0.000) & 0.51 (0.000) \\
Risk score & 0.29 (0.000) & 0.29 (0.000) \\
Family size & 3.08 (0.000) & 3.10 (0.000) \\
\midrule
\textbf{Panel B:} Medical Utilization \\
Total medical spending (individual) & \$4,764 [\$975] (0.002) & \$4,406 [\$887] (0.002) \\
\% of individuals with no spending & 0.17 (0.000) & 0.20 (0.000) \\
OOP medical spending (individual) & \$650 [\$198] (0.000) & \$562 [\$167] (0.000) \\
Household deductible $|$ deductible $>0$ & --- & \$1,040.24 (0.001) \\
\% Households with zero deductible & --- & 0.26 (0.000) \\
Household coinsurance rate & --- & 0.29 (0.000) \\
\% individuals with shoppable services & 0.06 (0.000) & 0.06 (0.000) \\
Total cost, shoppable service & \$5,572 [\$3,721] (0.011) & \$5,645 [\$3,814] (0.015) \\
OOP, shoppable service & \$691 [\$388] (0.002) & \$574 [\$290] (0.002) \\
\midrule
Years & 2006--2018 & 2006--2013 \\
$N_\text{families}$ & 368,237 &367,445 \\
$N_\text{individuals}$ & 1,357,392 & 1,311,554 \\
\bottomrule
\end{tabular}
\begin{tablenotes}
\small
\item \textit{Notes}: Enrollees include employees and their covered dependents. Risk scores are calculated using the CMS-HCC 2014 community model. Household plan characteristics are calculated as discussed in Section \ref{sec:data}. Spending values are reported in 2022 USD. Standard errors are reported in parentheses; medians (when reported) are in brackets.
\end{tablenotes}
\caption{\label{tab:sumstats-hh} Household Summary Statistics}
\end{threeparttable}
\end{table}

Table \ref{tab:sumstats-hh} presents summary statistics for the full sample as well as the sample subset with insurance plan identifiers.\footnote{The Marketscan data includes insurance plan identifiers for a subset of plans for which information can easily be abstracted from summary plan description booklets made available by the insurer \citep{hansen_truven_2017}. Typically, these plans represent the largest firms in the data, and include detailed information on the number and different types of plan offering made available to employees (including at what level), as well as ``financial provisions, health service benefits, managed care features and health coverage types." In the current data, this practice ended after 2013.} Households tend to be young and relatively low-risk, with an average age of 31.7 years and between 3 and 4 household members. Insurance coverage is more generous than average, although the conditional average deductible is over \$1,000, and household members who select into shoppable services typically spend close to a full year's OOP costs on that service alone. Note that the sub-sample with plan identifiers does not appear substantially different from the full sample, an important fact given that we use the plan-identified sample in our structural approach (Section \ref{sec:model}). Households in the plan-identified sample incur slightly lower OOP costs than the full sample; however, this is likely indicative of decreasing insurance coverage generosity over time, given that the latest 5 years of data are excluded in this sub-sample.

\subsection{CMS Shoppable Services}
To evaluate how household utilization responds to the resolution of pricing uncertainty, we study household services that are both planned and expected to incur a significant---but unknown---OOP cost. These two criteria are useful to ensure that we are studying a setting where household strategic decision-making is most likely to occur as well as one where bills provide useful information to households. We therefore identified 30 CMS ``shoppable services," common services that patients can schedule in advance and for which there exists substantial variation in charges across providers \citep{cms_medicare_2019,white_reference_2014}.\footnote{We identified these services in the claims data using Current Procedural Technology (CPT) codes for outpatient and inpatient services and Diagnostic Related Groups (DRGs) for inpatient hospitalizations. The complete list of services is available in Appendix Table \ref{tab:procs}. CMS shoppable services also include commonly-used hospital evaluation and management (E\&M) codes; we did not include these in our sample due to the substantially lower average cost of these services compared to other categories.} Our services broadly include pathology (biopsies), radiology (electrocardiograms), and surgical services (spinal fusions). Overall, CMS shoppable services constituted 16\% of overall OOP spending for individuals on ESI in 2017 \citep{bloschichak_cms-specified_2020}, making them an important area of study for both patients and policymakers.\footnote{Effective January 1, 2021, hospitals must publish standard charges for these services online, including negotiated rates. This does not affect our analytical sample (which goes through 2018). Prior to implementing this rule, there has been little empirical evidence found that patients engage in price shopping for these procedures ahead of time \citep{mehrotra_americans_2017}. }

We selected shoppable services as index events to prioritize several attributes of household decision-making, particularly in the context of planned service consumption. First, we focus on settings where bills provide meaningful price information to consumers; our results are not informative about services where co-payments, which are typically collected by the clinical provider at the point of service, constitute all cost-sharing. Co-payments as the sole source of cost-sharing are common for low-acuity acute primary care visits. Second, we select planned, scheduled services as index events as households may form expectations---even before the service is realized---that they will meet a deductible, giving time for scheduled increases in household medical demand. This plausibly strategic behavior motivates a focus on shoppable services over other health shocks such as hospitalizations or injuries, as it most closely mirrors the ideal experiment for our research question (i.e., how is household demand affected by price uncertainty?). Finally, our service selection is not based on the relative quality or value of a service (such as urgent or non-urgent hospitalizations); we evaluate how affected households select into high- or low-value services post-exposure in Section \ref{sec:rf-evidence} \citep{fadlon_family_2021,card_does_2009}. 
Taken together, using shoppable services as index events best allows us to estimate how pricing information changes household demand for healthcare by focusing on services where bills are most likely to meaningfully drive decision-making. 

However, our results are robust to more general inclusion criteria for the index events. For example, we show in Appendix Table \ref{axtab:expanded-services} that our main effects appear even in settings when \textit{ex-ante} strategic service delays may not be as apparent, such as following an unplanned hospitalization. We consider separately how household members respond to an unplanned hospitalization of another family member for injuries, poisonings, or appendectomies.\footnote{See Appendix Table \ref{axtab:expanded-services} for details on services included in this robustness check.} Even in this broad set of events---which now encompasses over 200,000 index events and 126 million individual-week observations---we continue to find robust evidence that household members reduce their spending by comparable amounts when a bill arrives; we discuss this more in Section 4.1.

We construct the baseline sample by identifying all shoppable services consumed among the households remaining in our analytic sample after requiring two or more members, full eligibility, and continuous enrollment. In the primary specifications, we limit index services to the first shoppable service consumed within a household plan-year---this removes 2.79\% of the 814,795 identified shoppable services. However, this restriction is relaxed in our learning specification (Subsection \ref{subsubsec:learning}) and the structural model of imperfect moral hazard (Section \ref{sec:model}). Throughout, we use both never-treated and not-yet-treated household-years as control groups (see Section \ref{sec:methods} for in-depth discussion).


\subsection{Bill Dates \& Waiting Times}
One limitation of our data is that we do not view the exact date consumers received their first bill for their index event; instead, we observe the date the insurance plan adjudicated the claim and initiated payment to the clinical provider. An Explanation of Benefits (EOB) is generated at this point, making it the earliest possible date when a patient will receive definitive information about their deductible and coinsurance obligations. Clinicians seldom collect cost-sharing that is not co-payments before the adjudication of the claim by the insurer \citep{ippolito_how_2023,mohama_more_2021}. Hence, patients as consumers are unlikely to have perfect pricing information before this date.\footnote{Over our study period, a greater number of healthcare facilities and insurers began to offer price transparency tools to consumers. Takeup of these tools by consumers remains extremely low over the course of our sample \citep{zhang_impact_2020}; in addition, these tools do not remove all pricing uncertainty (especially for OOP prices), but rather provide patients with credible intervals of expected OOP costs.}

We, therefore, use the earliest EOB generation date for a bundle of claims as a proxy for the arrival of new pricing information for consumers. Some consumers may receive same-day electronic notifications of EOB generation, while others will have a delay imposed by the postal service. In addition, we do not observe when patients \textit{consume} the new information that either an electronic or paper EOB provides. The effects of any measurement error here are expected only to attenuate our findings.\footnote{Note that in our context, measurement error only goes in one direction, as EOBs are not generated by payers prior to plan payment decisions \citep{denning_kansas_2014,davis-jacobsen_real-time_2008}. Hence, there are no instances in which consumers will learn precise information about non-copayment OOP obligations prior to the plan payment date. Note that individual line items with delayed adjudication do not change the information arrival date in our context.} Since our proxy measures the earliest possible date at which households have exposure to accurate pricing information, noise in our context always leads to a misclassification of the post-bill indicator to be 1 when it should be 0, rather than the other way around. Hence, the resulting coefficient on the post-bill indicator will be a weighted average of true post-bill effects and contamination from the interim period for any misclassified treatment dates; therefore, as long as the effects of a bill's arrival are of opposite sign than the effects of service (for example, if spillover household consumption increases following a service but then declines after the bill arrives), any contamination bias will attenuate the correction parameter towards zero. Similar logic applies to the use of any price transparency tools, as any use of these tools will attenuate the value of new and accurate information on household decision-making. 

\begin{figure}[htbp]
    \caption{Variation in Wait Times Between Service Date and Bills' Arrival}
    \label{fig:waittimes}
    \centering

    \includegraphics[width=4in]{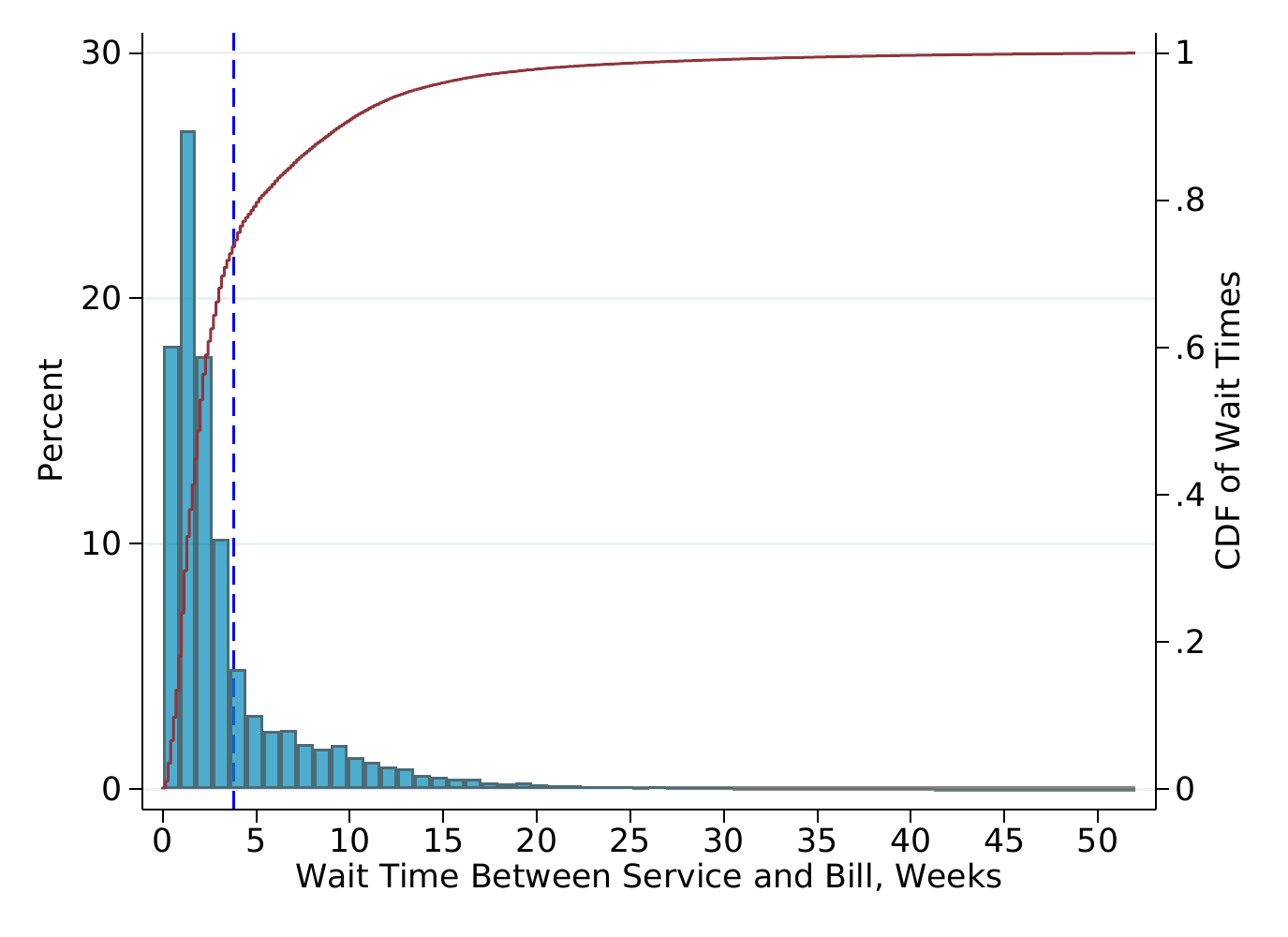}
    \vspace{0.2cm}
    \begin{minipage}{0.95\textwidth} 
	{\footnotesize \textit{Notes}: Figure depicts distribution of wait times between a service date and the date the insurer paid their portion of the bill, measured in weeks. Only services included as shoppable health events in our analytical sample are shown here. Vertical dashed blue line indicates the average duration of the waiting period (3.9 weeks). Red curve and secondary $y$-axis indicate the cumulative fraction of bills with a waiting time less than or equal to $x$.
	\par
	}
	\end{minipage}
\end{figure}

Figure \ref{fig:waittimes} presents the distribution of wait times (in weeks) between a shoppable service and the date the plan paid their portion of the bill. There is substantial variation in this wait time, with roughly 60\% of bills being paid by insurers within the first four weeks, and the rest taking longer than a month for payment to be settled. 

We claim that the length of the waiting period between a service and a bill is exogenous to households, allowing us to identify the causal impact of receiving information on spillover utilization. Appendix Figure \ref{figax:waittimes-months} illustrates substantial variation in this length both within and across years. Waiting times tend to be higher at the beginning of a calendar year and the first month of each quarter, when insurers have billing changes and new policies to incorporate into their processing algorithms.\footnote{Waiting times are also affected by more general health policies, such as the national transition to the International Classification of Diseases, 10th Revision, Clinical Modification (ICD-10-CM), in October 2015. This transition increased billing complexity by roughly five times and, subsequently, the rate of administrative frictions in processing billing information \citep{caskey_transition_2014}. Even major health disruptions, such as the COVID-19 pandemic, can overwhelm payer processing of claims, occasionally even drastically increasing wait times for bills \citep{snowbeck_risk_2022}. } Waiting times are also affected by other time-varying features of the healthcare system that are exogenous to the household, including the rate at which physicians submit claims to insurers for reimbursement. The exact variation in bill waiting times is therefore the result of interactions between an insurer---typically chosen at the employer level in our context, rather than the household level---and specific physicians or hospitals. Even if households attempted to choose general practice providers based on the relative efficiency of billing with their specific insurer, this is unlikely to be a driver in household choice of physicians and hospitals from whom they receive the shoppable services in our data (e.g., the surgeon who performs a mastectomy). Hence, the variation in the length of time a household waits for their bills is both unpredictable and exogenous at the consumer level; we discuss this more in Section \ref{sec:methods}. 

\subsection{Descriptive Evidence for Bill Shock}\label{susbsec:desc-evidence}
Finally, we document the extent to which bills in our sample vary, with particular emphasis on how bills might convey new pricing information that patients were unable to access prior to its arrival. This is important as residual variation in OOP payments or total service costs---after incorporating information about providers, service types, and dates of services---gives a sense of how a bill's arrival may generate ``bill shock" altering consumer choices. 

\begin{figure}[htbp]
    \caption{Quantifying Bill Shock: Difference between Realized and Predicted Service Prices}
    \label{fig:billshock-desc}
    \centering

	\subfloat[Total Costs]{
	    \includegraphics[width=3.0in]{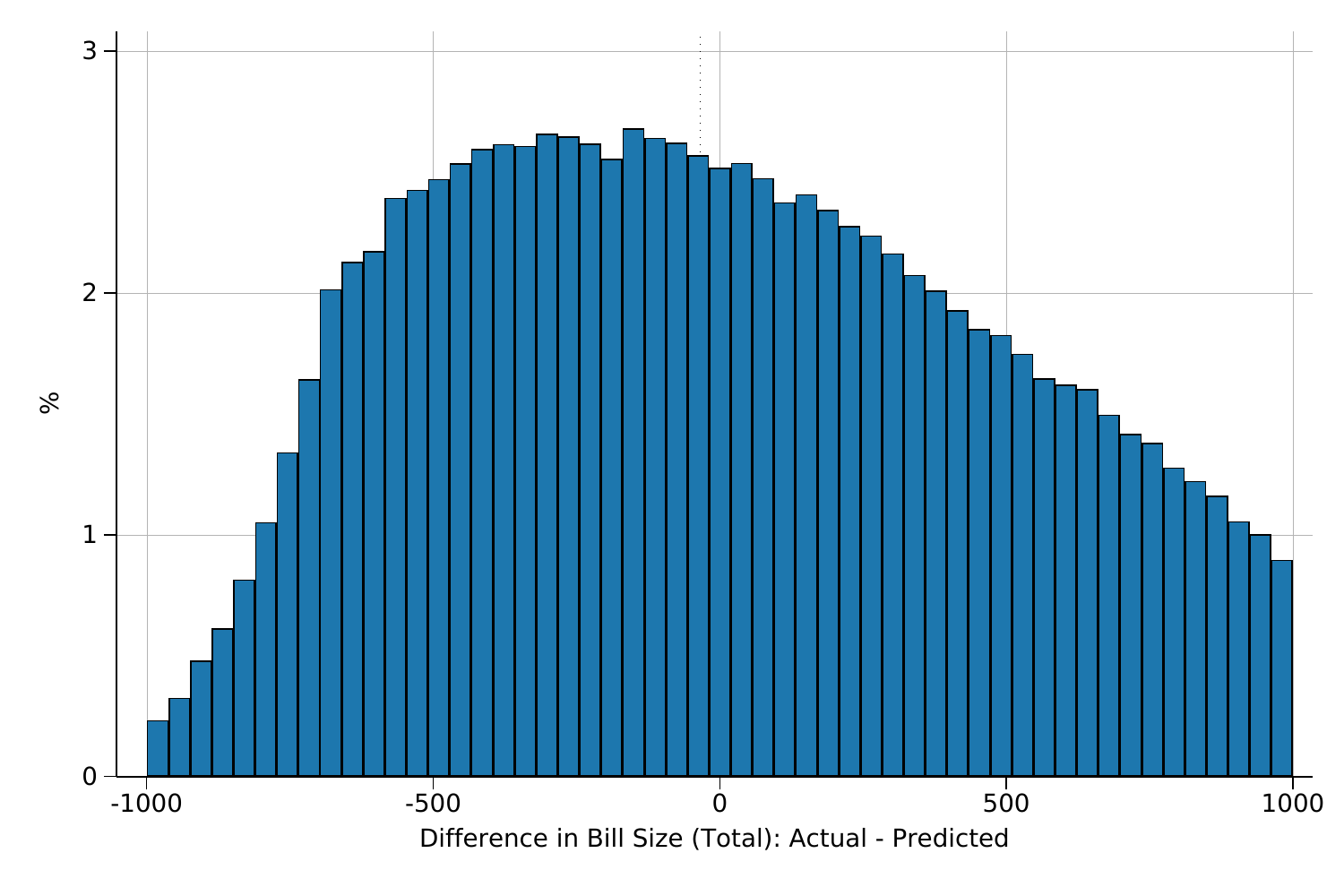}
	 }
	\subfloat[OOP Costs]{
	    \includegraphics[width=3.0in]{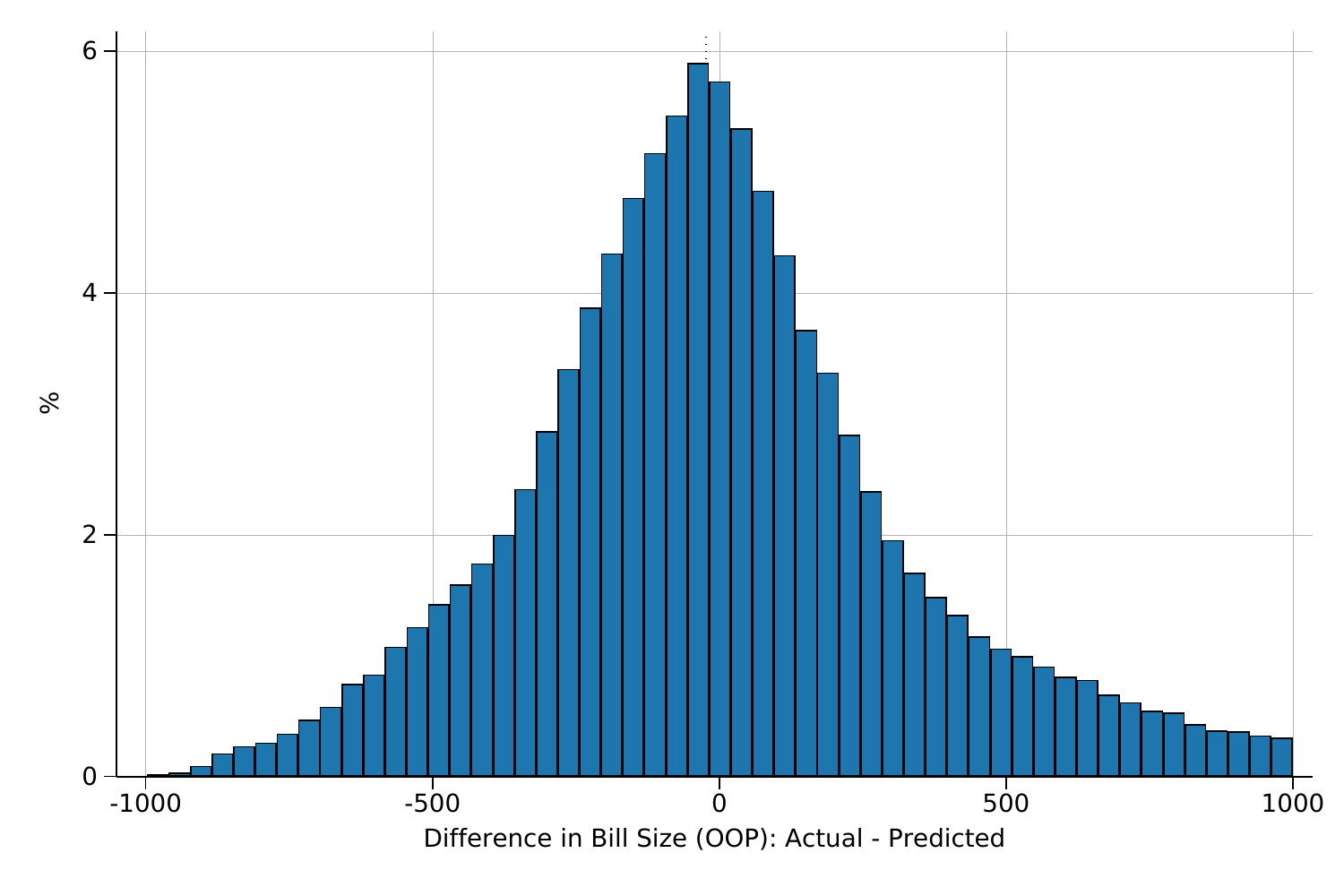}
	 }
    \vspace{0.2cm}
    
    \begin{minipage}{0.95\textwidth} 
	{\footnotesize \textit{Notes}: Figures each show differences between realized prices and predicted prices across all shoppable services in analytic data. Predicted prices use regression on year, week of year, service type, and provider fixed effects. Panel (a) shows differences in total billed costs, while panel (b) shows differences in OOP payments. Figures are truncated with absolute differences below \$1,000 (all measured in 2022 USD). 
	\par
	}
	\end{minipage}
\end{figure}

Appendix Figure \ref{figax:price-hist} illustrates the variation in prices (both total billed and OOP) for a particular service in our sample (routine vaginal delivery). To investigate this further, we regress prices on provider, service type, year, and relative week of year fixed effects and compare realized to predicted prices. Figure \ref{fig:billshock-desc} reports the results, showing broad variation in predicted prices both for insurers and patients. Note that predictions for total bill size are slightly right-skewed---reflecting the highly skewed nature of overall prices---while the distribution of predicted OOP costs exhibits far less skew. Importantly, conditional on these fixed effects, the average (median) absolute residual is \$443.07 (\$416.03) for total costs, and \$262.50 (\$197.20) for patient OOP costs. Appendix Figure \ref{axfig:billshock-desc} performs a similar analysis on wait times to highlight that the bill's arrival time also displays large variation even conditioning on these fixed effects; we observe that the average (median) absolute residual in the time spent between the service date and the paid date is 3.4 (2.7) weeks.

This evidence motivates the ideal comparisons underlying our proposed estimation. To test for bill effects on household medical decision-making, an ideal experiment would compare households who received the same service on the same date from the same provider. These households would implicitly have the same \textit{ex-ante} price distribution; however, one household would randomly receive their bill (their \textit{ex-post} price realization) faster than the other. Direct comparison of household behaviors during this time when only one household has received their bill would identify estimated bill effects.

\section{Methods}\label{sec:methods}
An important feature of delayed pricing information is that household responses to shoppable services necessarily take place in two stages: first, households must respond to the event based on \textit{expectations} about costs relative to a deductible; second, those responses evolve only after new information---in the form of a bill or EOB---resolves uncertainty. We leverage these two distinct periods in a triple-differences regression framework to estimate spillover responses \textit{separately} for the periods before and after a bill's arrival.\footnote{We refer to this as a ``triple difference" regression because of the multiple margins of treatment arising from the bill's delay post-service. That is, a truer expression of Equation \ref{eq:reg} as a DDD regression is $\mathbb{E}[\text{spend}_{ity}] = \exp\left\{\beta_1\mathbbm{1}(\text{service} \times \text{post})_{ity}+\beta_2\mathbbm{1}(\text{service} \times \text{post} \times \text{bill})_{ity} + FEs\right\}$, with the bill arrival dividing the post-treatment period into two.} 

We estimate the causal impact of a bill on total \textit{spillover} spending for all household members excluding the focal consumer of the shoppable service.\footnote{There is strong evidence that individual health events generate spillovers affecting spillover utilization, including for this data \citep{hoagland_ounce_2022,fadlon_family_2019}. There may be situations where information is not shared fully across a household (for example, young adults in the household still covered on a plan but no longer living at home). This would tend to bias our estimated results towards zero, a problem discussed in other work \citep{kowalski_censored_2016}.} We estimate how a bill affects total spillover spending (measured per week per household-member) in household $i$ at week $t$ of year $y$ as given by Equation \ref{eq:reg}:
\begin{equation}
    \mathbb{E}[\text{spend}_{ity}] = \exp\left\{\beta_1\mathbbm{1}(\text{post\_service}_{ity})+\beta_2\mathbbm{1}(\text{post\_bill}_{ity})+ \alpha_\mathcal{I}+\tau_t + \delta_y + \mu_\text{S} + \xi_\text{MD} \right\},
    \label{eq:reg} 
\end{equation}

\noindent where the two main regressors are indicators for the consumption of a shoppable service and, subsequently, receipt of a bill. We consider robustness to controlling for various time-invariant fixed effects, including for households ($\alpha_\mathcal{I}$), years ($\tau_t$), relative week of the year ($\delta_y$, to account for within-year seasonality in spending), index service type ($\mu_\text{S})$ and provider fixed effects ($\xi_\text{MD}$, for the providers performing the focal service). 

Since our results are consistent across these specifications, the identifying variation driving our results comes from \textit{idiosyncratic variation} in billing delays within an organization-service level at a given time. That is, we assume that bill delays are exogenous conditional on household, time (year and week of year), event type, and provider fixed effects (consistent with the descriptive evidence discussed in Section \ref{susbsec:desc-evidence}). This assumption maps intuitively to the thought experiment described above, comparing households consuming the same service from the same provider at the same point in time, but with different wait times for pricing realizations. As households and providers have no control in their bill wait times at this level, this exogeneity assumption is plausible; we discuss potential identification threats below. 

We use Poisson regression to estimate multiplicative effects on spending. This allows us to deal with the skewed nature of our (non-negative) spending data while appropriately including weeks with zero spending.\footnote{Prior to consuming any shoppable services, an average of 31\% of households consume any health services in a given week, spanning inpatient and outpatient care and pharmaceutical purchases. Hence, at least \textit{ex-ante}, we do not expect that our regressions suffer from too many zeros, which might make it difficult to detect treatment effects given small variations in bill timing \citep{wooldridge_distribution-free_1999}. Our results are also robust to zero-inflated Poisson methods.} Our estimator will be consistent as long as the conditional mean of the dependent variable is correctly specified, as is the case in ordinary least squares (OLS) regression; additionally, Poisson regression allows us to avoid the inconsistency of regression coefficients induced by heteroskedasticity in a log-linear transformed model \citep{santos_silva_log_2006} and concerns from nonlinear transformations of the dependent variable \citep{mullahy_why_2022}.\footnote{Regressions use ``ppmlhdfe" to handle high-dimensional fixed-effects \citep{correia_fast_2020}. }

Our parameter of interest, $\beta_2$, is identified from billing delays; however, one may be concerned our estimator also reflects time-varying responses in utilization relative to the initial medical event. Given that the shoppable service may affect both the focal individual and others in the household to the extent that it represents a health shock, one might be concerned that $\beta_2$ reflects a general downward spending trend for the family in event time as the shock fades. We address this in two ways: first, we directly estimate dynamic treatment effects in Section \ref{subsec:dynamic}, accommodating time-to-event dummies in a typical ``event-study" design; second, we show throughout that our results are robust to the inclusion of linear time trend controls before and after the index event (Appendix Tables \ref{axtab:ddd-table-ext} and \ref{axtab:timetrends}). Our estimates are unchanged across specifications, possibly because shoppable events do not generally reflect health ``shocks" affecting a household in the way unplanned or more expensive shocks do \citep{fadlon_family_2019,hoagland_ounce_2022}; hence, there is no evidence of a downward trend in spending between the event and the bill's arrival.

\subsection{Testing for Possible Selection into Bill Delays}
A critical identifying assumption of our parameter of interest ($\beta_\text{post\_bill}$ in Equation \ref{eq:reg}) is that a bill's arrival is exogenous at the household level. Previous work has highlighted the potential endogeneity inherent when attempting to estimate demand elasticities to major health events, especially for planned consumption \citep{duarte_price_2012}. We do not estimate demand elasticities in these models (that is, $\beta_\text{post\_service}$ is not a demand elasticity); instead, we are only measuring how the total volume of household responses---including both strategic and non-strategic responses---change following a bill. Therefore, our estimation has no potential endogeneity concerns, as long as bill arrival times are exogenous to the household. 

A first-order concern for this exogeneity is that bill wait times may be associated with underlying patient risk, which potentially introduces selection bias. For example, if a claim for a more medically-complex patient takes longer for insurers to process (even within procedure types), waiting times may be systematically longer for the most at-risk patients in our sample. This could artificially inflate our regression coefficients if risk is correlated across households and riskier households spend more on average. 

However, we find no evidence that billing delays are associated with patient risk. We can test these claims directly by comparing the price an individual paid for a service and the time they waited for their bill. We regressed spending measures on time bins for bill delays of 5 days each (that is, a wait time of 0--4 days, 5--9 days, and so on), adjusting for year, week of year, event type, and provider fixed effects (Appendix Figure \ref{axfig:billbalance}). We would expect to observe significant relationships in the coefficients for these bins if there were correlations between more medically complex patients and billing delays. We do not observe any such relationships for either total cost or patient OOP.\footnote{Results are robust to measuring bill size in levels or relative to an individual's average pre-event spending (percentage changes). Results are also robust to excluding physician fixed-effects, addressing the potential concern that perhaps physicians, rather than patients, drive bill delay in strategic ways.} 

We also present comparisons of the trends in billing delays across procedure types in Appendix Figure \ref{axfig:corr-wait-cost}. Across spending measures, we consistently find a lack of correlation between the cost of a procedure and wait times. This is true even for services with the largest variation in expected costs, such as vaginal deliveries and arthroscopy.\footnote{Appendix Figure \ref{axfig:corr-wait-cost} shows shoppable services with total costs under \$1,000 may take about 3 days longer to be processed; these constitute less than 4\% of services in our data. This effect runs in the opposite direction of the initial concern: if riskier/more expensive patients receive bills faster, we would not expect to find large swings in spending attributable only to higher-spending households waiting longer for their bills. Finally, throughout the paper, our results are robust to controlling for procedure-specific time trends, to accommodate any concerns about variation in bill wait times across shoppable service types.} Taken together with the regression results above, these findings confirm that patient health status appears uncorrelated with the amount of time for a claim to be adjudicated. 

Finally, a remaining concern is that intra-household correlations in health status may bias our results \citep{meyler2007health,cawley2011economics}. Given that the shoppable service claim could be plausibly correlated with the health needs of other household members, households with more costly focal services may also tend to have greater healthcare spending than other households. Hence, if these households also wait longer for their bills (for example, because waiting times are endogenous to health), we may overestimate a bill’s effect on spending due to simply the prolonged waiting period for high-spending households. However, this concern arises only if there are remaining endogeneity issues between bill delays and size, as discussed above; otherwise, level differences in healthcare spending needs across households are adjusted for in Equation \ref{eq:reg} using standard parallel trends assumptions and household fixed effects. We would only overestimate treatment effects in this context if affected household members’ health were to worsen idiosyncratically between the (planned) shoppable service and the bill, an unlikely outcome given the short time intervals analyzed. 

\section{Empirical Results}\label{sec:rf-evidence}
\subsection{Effect of Bills on Spending}
\label{subsec:correct}


We investigate how households exhibit differential spending responses to an index service before and after a bill's arrival, estimating Equation \ref{eq:reg}. Table \ref{tab:ddd-table} presents the regression results: we find robust evidence that although spillover spending increases after a focal member receives care, a medical bill's arrival causally affects these responses. Without including information on bill arrival timing, the overall spending increase is roughly 40.2\% of average weekly per-person household spending (about \$48 per person-week). Such spending increases are consistent with prior literature, but may be driven by a number of factors including correlated health within households \citep{meyler2007health}, strategic delays in seeking care until after a high-cost event \citep{cabral_claim_2017,kowalski_censored_2016}, or responses to new information about health systems or health risk \citep{hoagland_ounce_2022}.\footnote{Note that this spending is normalized to the per-person level, and results are robust to excluding newborns (following childbirth) from the sample.}

\begin{table}[htb]
\centering
\begin{threeparttable}
\begin{tabular}{l|cc|cccc}
\toprule
& \multicolumn{2}{c}{Main Models} & \multicolumn{4}{c}{Alternative Specifications} \\
\midrule
Post Service &  0.402*** & 0.218*** &  0.597*** &  0.472*** &  0.486*** &  0.464*** \\
& (0.0032) & (0.0032) & (0.0032) & (0.0032) & (0.0033) & (0.0032) \\
Post Bill & & -0.109*** & -0.080*** & -0.096*** & -0.076*** & -0.077*** \\
& & (0.0030) & (0.0030) & (0.0030) & (0.0031) & (0.0030) \\
\midrule
$\overline{\text{spend}_{it}}$ & \$120.49 & \$120.49 & \$120.49 & \$120.49 & \$120.49 & \$120.49 \\
Household FEs & X & X & X & X & X & X \\
Year FEs & X & X &  & X & X & X \\
Week of Year FEs & X & X &  & & X & X \\
Provider FEs & X & X & & & & X\\
Event Type FEs & X & X &  & & & \\
Observations & 61,860,735 &   61,860,735 &  61,860,735 &   61,860,735 & 61,860,735 & 61,860,735 \\
\bottomrule
\end{tabular}

\begin{tablenotes}
    \small
    \item \textit{Notes}: Table presents triple-difference Poisson regression estimates for a bill's effect on households' spillover health spending. Focal consumers are excluded from the outcome. Regression coefficients illustrate the expected change in log household spending (measured per person-week) associated with the service date and bill arrival (both measured as dummy variables). Throughout, standard errors are clustered at the household level. For robustness, see Appendix Tables \ref{axtab:ddd-table-ext} and \ref{axtab:timetrends}.
    
    $^{*} p < 0.05, ^{**} p < 0.01, ^{***} p < 0.001$ 
    \end{tablenotes}
    \caption{\label{tab:ddd-table} Estimated Impact of Bill Arrival on Household Health Spending}
\end{threeparttable}
\end{table}

Surprisingly, our results suggest bills meaningfully affect spillover spending. Before pricing information, we estimate spillover spending increases by 21.8\% post-service in our preferred specification; however, a bill causes reductions in this increase by 10.9\% from baseline, roughly half of the increase. This decline is consistently estimated across specifications and robust to multiple approaches.\footnote{Alternative approaches include (a) incorporating procedure-specific fixed effects and time trends, allowing for potential differences in household behavior following different types of shoppable services (Appendix Table \ref{axtab:ddd-table-ext}); (b) excluding household fixed effects; (c) excluding the small fraction of households who switch plans soon after a shoppable service (i.e., whose spending is censored at the end of the plan year); and (d) controlling for linear time trends before and after the service or dynamic treatment effects for the shoppable service independent of bill arrival time (Appendix Table \ref{axtab:timetrends}).} Bill effects amount to spending reductions of \$13 per person per week for the average household, or about \$317 in per-person annual spending. 

We also test to ensure that our results are robust to potential measurement error in the date bills actually arrive, given our use of plan payment dates as a proxy. If plan payment dates are unrelated to household information, we may be simply splitting the post-service period into two random periods and attaching significance to spurious differences between them. To test this possibility, we conducted placebo tests, estimating Equation \ref{eq:reg} on artificial data that randomly assigned new wait times based on the empirical distribution of bills.\footnote{For each shoppable service, we fixed the service date and artificially varied the bill arrival date as the service date plus a random draw of a wait time, drawn from the empirical distribution of waits (Figure \ref{fig:waittimes}).} The results of 1,000 placebo regressions are reported in Appendix Figure \ref{fig:placebo}; placebo coefficients are centered close to zero and generally indistinguishable from a null effect. Taken with the results from Table \ref{tab:ddd-table}, this suggests that it is unlikely that our results are spurious correlations from a semi-random splitting of the post-service period.

Finally, we show that these results are not dependent on the choice of shoppable services as the index event, but that bill effects are present in a broader set of events including unplanned hospitalizations for injuries and appendix surgeries (Appendix Table \ref{axtab:expanded-services}). Even in an expanded set of events---which now encompasses over 200,000 index events and 126 million individual-week observations---we continue to find robust evidence that household members reduce their spending when a bill arrives. Estimated effects are 7.4\% in the unplanned hospitalization sample, compared to 10.9\% in the main specification, and are robust to pooling across both samples or further separating injuries and appendix surgeries. Interestingly, we note that the post-service coefficient approximately doubles in value when considering planned index events; this supports the hypothesis that households engage in strategic behavior, co-planning their healthcare utilization conditional on the realization of some health needs such as shoppable services. However, this strategic behavior does not appear to drive the responses to a bill's arrival. 

These findings are consistent with a model where consumers overestimate their progress towards crossing a cost-sharing threshold, such as a deductible. Once a medical bill provides definitive information about contributions, however, individuals correct their spending in response to information that they are not as close to these discrete changes in marginal costs as they initially anticipated. 

\subsection{Dynamic Treatment Effects}\label{subsec:dynamic}
We next assess the dynamic treatment effects induced by a bill's arrival using a two-way fixed-effects (TWFE) approach. The static estimation (Equation \ref{eq:reg}) inherently estimates many pairwise comparisons between households who have received a bill and households still waiting for one; to the extent that these pairwise comparisons are different at different points in relative time due to heterogeneous treatment effects, our static estimation procedure may lead to biased regression coefficients \citep{goodman2022bacondecomp}. For example, if there are larger bill effects in the short-run that dissipate quickly, our static comparisons may over- or under-report the true effects depending on the distribution of these comparisons in our sample. To address these concerns, we estimate a dynamic version of Equation \ref{eq:reg}.

Such an approach is complicated by the fact that there are two events in the triple-differences specification with potentially differential timing: the consumption of the index event and the bill's arrival. As we are interested only in the effects of the latter, we estimate dynamic effects on a matched sample of households consuming the same service at the same time and organization. Once this matched sample is constructed, we estimate the differential impact of the first bill to arrive on household spillover spending, compared to households who had similar index service experiences but who have not yet received their bill. 

We use a nearest-neighbor matching process at the household level, matching on index service type, provider, year and event date. The final matched sample consists of 35,734 matched sets, with an average (median) of 6 (4) households in each set. That is, 214,404 of the 368,240 (58\%) households are represented in the matched sample. Once matched, we compare the effect of the first bill's arrival within a group on spillover household spending using a TWFE version of Equation \ref{eq:reg}: 
\begin{equation}
    \mathbb{E}[\text{spend}_{ity}] = \exp\left\{\sum_{k=-T}^T \gamma_k \mathbbm{1}\left\{t-E_{it}=k\right\} + \alpha_\mathcal{I}+\tau_t + \delta_y + \mu_\text{Service} + \xi_\text{MD} \right\},
    \label{eq:twfe}
\end{equation}
where $E_{it}$ indicates a set of time-to-first-bill dummies within the matched groups, with the week before the first bill's arrival as the omitted reference group.\footnote{Note that by virtue of the matching process, these results are virtually consistent across the set of fixed-effects included in the regressions.} 

\begin{figure}[htb]
    \caption{Dynamic Treatment Effects of Bill Arrivals on Household Spending}
    \label{fig:event-study}
    \centering

	    \includegraphics[width=.85\textwidth]{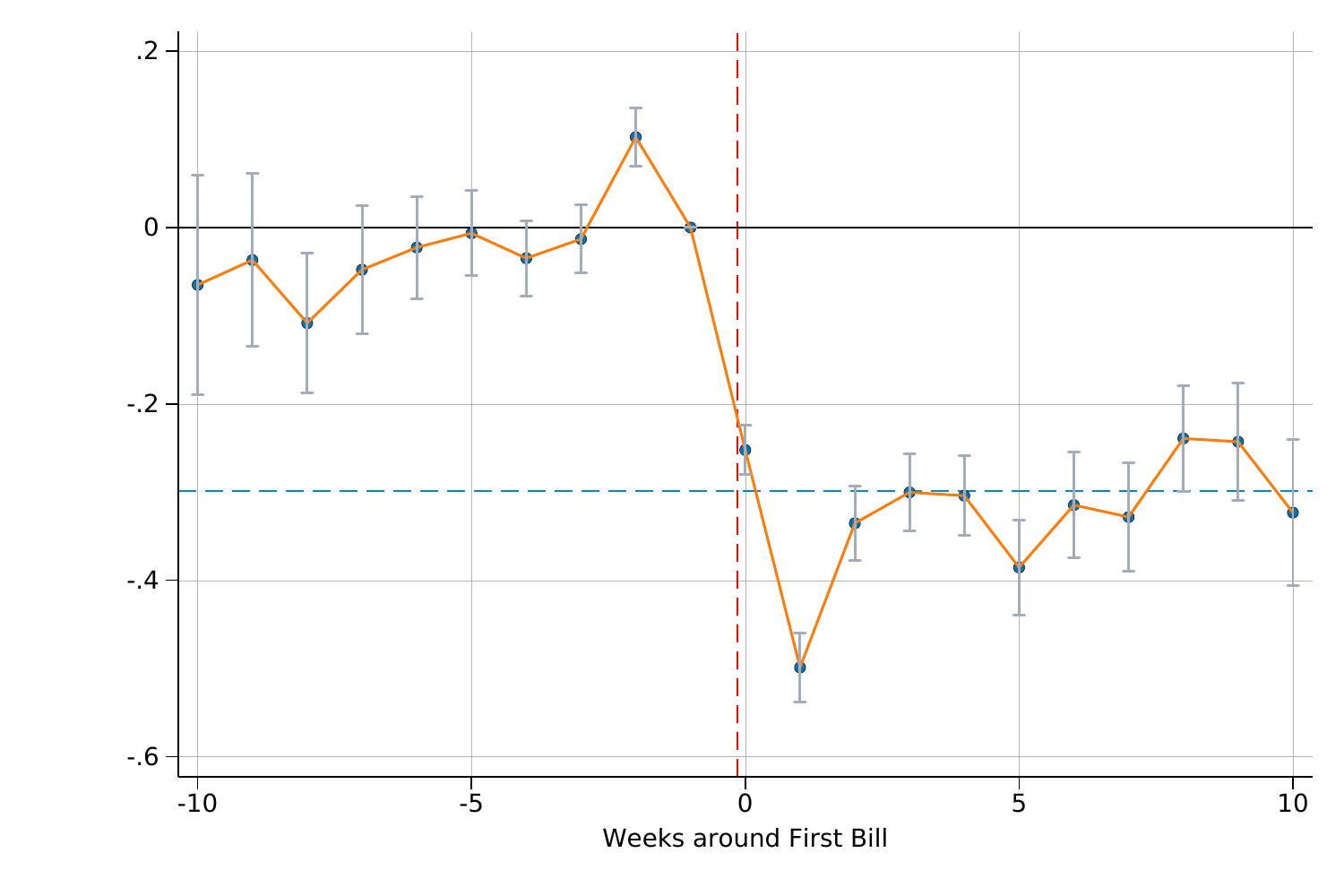}
	 
    \vspace{0.2cm}
    
    \begin{minipage}{0.95\textwidth} 
	{\footnotesize \textit{Notes}: Figure shows estimated coefficients and 95\% confidence intervals for the TWFE regression estimated in Equation \ref{eq:twfe} on the matched sample, as discussed in Section \ref{subsec:dynamic}. Standard errors are clustered at the household level. 
	\par
	}
	\end{minipage}
\end{figure}

Figure \ref{fig:event-study} presents the results. The pre-trend period---defined as the time between the index service consumption date and the first bill's arrival---reflects very little change in household spending across groups.\footnote{There is some evidence of differential spending patterns two weeks prior to a bill's arrival, with early-treated households increasing their spending slightly before a bill arrives; however, note that this moves in the opposite direction from a pre-trend that would bias our result, and hence is likely attributed to noise in the reduced sample.} Beginning in the week the bill arrives, affected household members decrease their spending by over 20\%, with average decreases of approximately 32\% over the following weeks. This is consistent with the 32.8\% reduction in spending between the post-service and post-bill periods estimated in Equation 1 (Table \ref{axtab:timetrends}).\footnote{Note that Equation \ref{eq:twfe} mechanically includes linear time trends around the time of the event; hence, we compare the coefficients to the version of Equation \ref{eq:reg} that included these linear time trends.} In particular, the strongest effects are observed in the first week immediately following the bill, where spending is estimated to decline by roughly 50\% relative to the post-service period. However, note that the average untreated household in this matched data waits an additional 3 weeks after the first bill's arrival for their own bill; hence, there is substantial attrition across the time periods limiting the ability to extrapolate long-term dynamic treatment effects. 

Given the variation in the timing of index service consumption, as well as the possibility for heterogeneous treatment effects, we consider the robustness of our regression results from Equation \ref{eq:twfe} in the context of the novel staggered treatment design estimators \citep{goodman2022bacondecomp}. In particular, we follow the approach of \cite{wooldridge2021two}, the only updated TWFE specification to be shown to be robust to nonlinear modeling such as Poisson regression. This approach incorporates a fully flexible regression capturing full heterogeneity in treatment effects by including household/year/week-of-year interactions in the main model. By doing so, the ATT can be consistently estimated, including with dynamic weighted averages in the style of Equation \ref{eq:twfe}. We estimate this approach and compare our results in Appendix Figure \ref{axfig:event-study-wooldridge}; we do not find significant differences across our estimates. In particular, the Mundlak estimator reveals a similar pattern of spending reductions following a bill's arrival; this estimator, if anything, finds stronger spending reductions than a typical TWFE framework, estimating an average treatment effect closer to a 60\% reduction in spending.\footnote{Note that estimating the Mundlak estimator in this large dataset with high-dimensional fixed-effects Poisson regression is quite computationally intensive; hence, results here are presented only for on a 50\% sample of the (matched) data, which has already been significantly reduced in size from the full analytic dataset based on the matching process. Hence, the discrepancy between the regressions likely arises from (a) differences in comparisons of heterogeneous treatment effects where a typical TWFE regression places more weight on less likely comparisons (such as households who wait more than 6 weeks for a bill) and (b) sampling differences across the two estimators.}

Given the differences in sample construction and estimation, it is important to use caution when interpreting the estimated dynamic treatment effects in context of the broader results of both our reduced-form evidence and structural approach. In particular, this estimation does not include any pre-service spending as a control group (including years in which households did not consume an index service). Similarly, interpretation of dynamic treatment effects is limited by the relatively short wait times between services and their bills, as well as relatively short differences in wait times across households consuming similar services. Despite these caveats, however, the matching and event study approach corroborates our earlier findings that the arrival of a bill meaningfully changes household health consumption in the short run, with particularly strong effects immediately after a bill's arrival. 

\subsection{Pricing Information as a Mechanism for Bill Effects}\label{subsec:pricing}
Our results show robust evidence that households change their spending patterns following a bill's arrival, observable in the raw data as well as static and dynamic regression specifications. We next turn to a discussion of the potential mechanisms generating these effects. 

Bills may affect the healthcare utilization of households for multiple reasons. Perhaps most saliently, households may learn about (OOP) prices for services; pricing information may affect spending decisions either through updated beliefs about household cost-sharing or due to distortions from liquidity constraints. However, other features of a bill---unrelated to prices---may also contribute to observed results. A bill may provide information about whether contracts (do not) cover specific procedures or include certain providers in their networks, or reveal discrepancies between a patient's understanding of a service and a provider's billing (for example, up-coding practices). This information may alter future healthcare spending if it erodes trust in the healthcare system \citep{webb_hooper_understanding_2019}. Finally, bill effects may be driven by a focal \textit{provider} rather than the household, who may respond to insurance denials or low reimbursement rates by affecting the provision of future services. 

We consider each of these potential mechanisms in turn. First, we show that households are most responsive to bills that provide meaningful pricing information, particularly when viewed through the lens of a household's deductible. Second, we show that households exhibit less bill shock after repeated health events, suggesting households learn from bills and update their beliefs accordingly. Finally, we show that our results are not driven by alternative concerns including liquidity constraints and provider-driven effects. 

Even considering all the information contained in a bill, however, may not be sufficient to resolve all pricing uncertainty at the household level. Residual uncertainty may persist to the extent that patients are not given negotiated prices for services, do not know which of their services are subject to their deductible, or may be under-informed about how deductibles or cost-sharing more generally works in their contexts. In each of these cases, household decision-making may be affected by persistent uncertainty even after the bill arrives. 

\subsubsection{Heterogeneity Across Deductible Contributions}
We first assessed how bill effects differed across household spending histories, relying on the intuition that these households may find bills differentially informative.

\begin{figure}[htb]
    \caption{Heterogeneous Bill Effects Across Household Deductible Status at Time of Service}
    \label{fig:dedhet}
    \centering

	    \includegraphics[width=.85\textwidth]{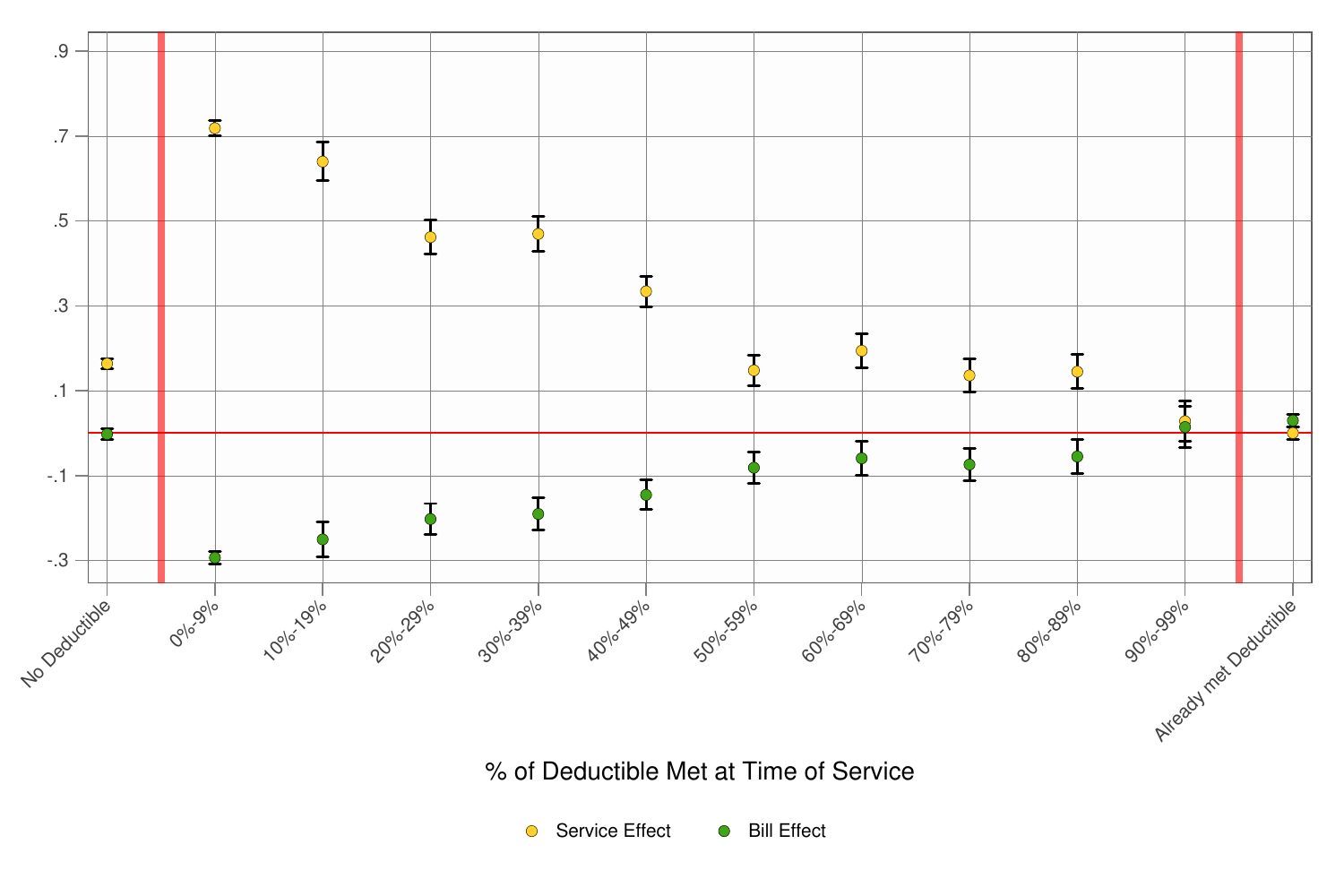}
	 
	 
    \vspace{0.2cm}
    
    \begin{minipage}{0.95\textwidth} 
	{\footnotesize \textit{Notes}: Figure shows estimated coefficients and 95\% confidence intervals for  $\mathbbm{1}\{\text{Post\_Service}_{it}\}$ and $\mathbbm{1}\{\text{Post\_Bill}_{it}\}$ in Equation \ref{eq:reg} by decile of household deductible spending prior to the event. Standard errors are clustered at the household level. Regressions adjust for total size of bill and linear time trends by group. Deductibles are imputed based on previous literature \citep{hoagland_ounce_2022,zhang_does_2018}. 
	\par
	}
	\end{minipage}
\end{figure}

\begin{figure}[htb]
    \caption{Heterogeneous Bill Effects By Household Deductibles and Service Cost}
    \label{fig:dedhet-twoway}
    \centering

    \includegraphics[width=5in]{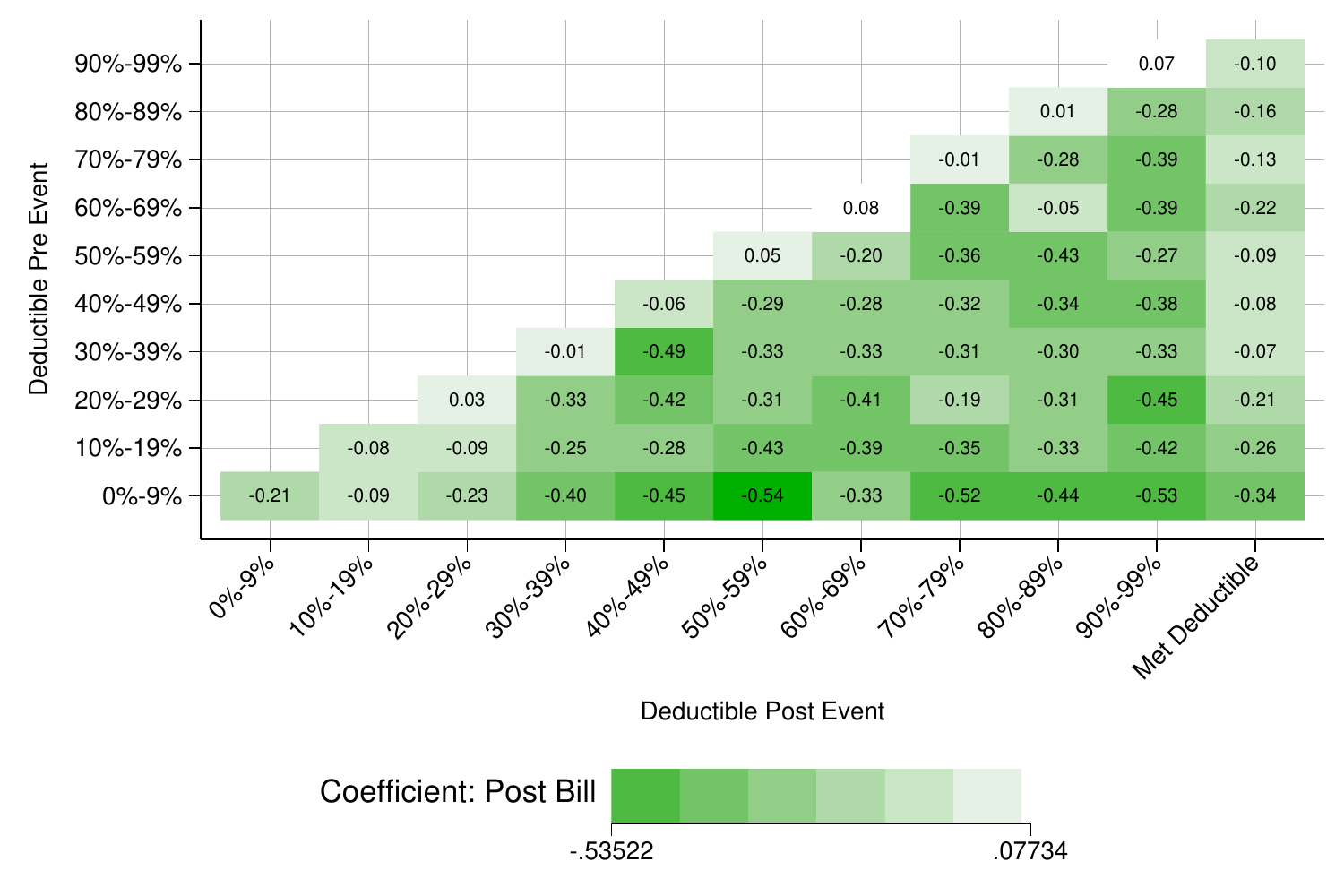}
    \vspace{0.2cm}
    \begin{minipage}{0.95\textwidth} 
	{\footnotesize \textit{Notes}: Figure depicts coefficients for $\mathbbm{1}\{\text{Post\_Bill}_{it}\}$ in Equation \ref{eq:reg} across deciles of household deductible spending prior to \textit{and} following events. Sample is restricted to those with non-zero, unmet deductibles. Each row indicates a different decile of deductible spending prior to the event, while each column indicates deciles following the event. Regressions include linear time trends by group as additional control for overall bill size. Standard errors are clustered at the household level.
	\par
	}
	\end{minipage}
\end{figure}

Figure \ref{fig:dedhet} presents results stratified by decile of household deductible spending prior to the event.\footnote{Appendix Figure \ref{axfig:ded-dollars} shows the equivalent versions of Figures \ref{fig:dedhet} and \ref{fig:dedhet-twoway} measured in levels (dollars) of deductible remaining rather than percentages.} The spending responses for both the interim period (yellow) and the post-bill correction (green) are shown. Both responses are largest for households who have spent little towards their deductible before the event: post-service spending increases are estimated to be over 70\% for households with less than 10\% of their deductible met, and become statistically insignificant for households close to their deductible. Correspondingly, post-bill corrections are estimated to be as high as 30\% (nearly 50\% of the post-spending increase) for households with less than 10\% of their deductible met; these effects similarly converge to precise zeroes for households close to meeting their deductibles. Finally, households whose marginal cost is not bill dependent---including those in zero-deductible plans as well as those with no remaining deductible---exhibit no responses to bills.\footnote{Even among those enrolled in zero-deductible plans, there is an average post-service spending increase of roughly 15\%. This could be due to the fact that household health is correlated across members \citep{meyler2007health,cawley2011economics}, so that household members may jointly increase health spending even without taking into account changes in marginal prices. However, these results also suggest that while households unaffected by changes in prices may increase spending post-service, these changes are not reflected in post-bill effects, as this coefficient precisely overlaps zero. This makes intuitive sense given the quasi-random timing of bills; as post-service increases---even due to correlated household health---are randomly distributed over post-service time, they are not affected by the quasi-random timing of the bill's arrival in settings where that bill did not provide meaningful pricing information.}


These results suggest that households find bills especially relevant when it contributes meaningful information about future expected cost-sharing; these effects are magnified for households who have yet to contribute much to their deductible, who may be relatively under-informed of their spending relative to a deductible.

While variation in pre-event spending provides useful information, further insight can be gained by leveraging a second dimension of variation: the relative OOP cost of the shoppable service itself. Exploring these two dimensions simultaneously leverages the fact that bills may be uninformative for specific plan types (e.g., those with zero deductible, as shown), but may also be more or less informative depending on where households end up on a cost-sharing curve. Price information is most valuable to households when it communicates whether they have crossed the threshold of their deductible; hence we identify the salience of pricing information---separate from other forms of learning---by comparing household responses to high- and low-cost events, considering both pre-event deductible contributions \textit{and} the resulting change in deductible spending after the scheduled health consumption. 

Figure \ref{fig:dedhet-twoway} presents the results. We restrict our attention to households with a non-zero, unmet deductible at the time of service, and separately estimate Equation \ref{eq:reg} across cells of households who have similar deductible spending both before and immediately following the shoppable service. The figure depicts a two-way heat map of estimated bill responses across cells.\footnote{For each regression, the control group is households who did not consume a shoppable service over the course of the year. See Appendix Figure \ref{axfig:dedhet-twoway-svc} for the corresponding figure of service effects.} Consistent with Figure \ref{fig:dedhet}, we find that households starting at lower levels of their deductible exhibit greater sensitivity to their bill; in the figure, this is seen by comparing bill responses as one moves down a given column. We directly observe that the OOP cost of a service changes responses; moving to the right for a given row, we observe coefficients much closer to zero for households that do not move across deciles of spending, but that weakly increase as the service becomes more expensive. 

Most notably, comparing household responses across the discontinuous threshold of meeting the deductible is particularly informative about mechanisms. Across all levels of pre-event spending, estimated coefficients are at least 55\% higher when a bill left households just short of meeting their deductible, rather than services that pushed households into a lower marginal-cost region of their contract. The average effect of receiving a bill for the shoppable service declines by 39.5\% from $-0.42$ to $-0.17$. The effects on either side of the deductible threshold are statistically distinct and statistically indistinguishable from 0 on the right side of the cutoff (Appendix Table \ref{axtab:metded}).

Finally, we conduct a falsification test exploring the impact of a bill's arrival for households who have already met their OOP max prior to the index event. The intuition for this test is that not only should the index service be free to the household, but households who have met their OOP max before the event should not expect the bill's arrival to convey any new information about their spending patterns or outstanding payments. Hence, we do not expect that a bill's arrival---or, for that matter, service consumption---should meaningfully alter the spending of other household members. This is exactly what we find; the results are reported in Appendix Table \ref{axtab:oop-max}. In alternative specifications, we observe \textit{positive} bill effects for this subsample; however, once we condition on the fixed effects used in our primary specification (reflecting the identifying assumptions required for quasi-random bill arrival timing), we do not observe any significant changes in household consumption after a bill arrives, if the OOP max has already been met. 

Overall, these findings illustrate that households are much more responsive to a bill when it contains important information about future marginal costs, consistent with price sensitivity driving household responses. Accounting for heterogeneity in household responsiveness to bills---including both spending histories and the cost of services---suggests that households are most responsive to a bill when its pricing information is highly relevant. 

\subsubsection{Evidence of Household Learning from Bills}\label{subsubsec:learning}
Although these results suggest households respond to more informative bills, it may not be that households learn from bills over time. Hence, we assess belief updating in two ways: first, by showing that households respond differently to bills for an initial index event compared to subsequent ones; and second, by considering how households exhibit forward-looking behavior in bill responses.

\begin{table}[htb]
\centering
\begin{threeparttable}
\begin{tabular}{l|c|c}
\toprule
& \multicolumn{1}{c}{Full Specification}  &  \multicolumn{1}{c}{Repeated Services} \\
\midrule
Post Service &  0.218***  & 0.268***   \\
& (0.0032) &  (0.0058)  \\
Post Bill & -0.109*** & -0.083*** \\
& (0.0030) &  (0.0056) \\
\midrule
2$^\text{nd}$ Service * Post Service & & -0.176*** \\
&  &(0.0070)  \\ 
2$^\text{nd}$ Service * Post Bill & & 0.118***  \\
&  &(0.0071) \\ 
\midrule
$\overline{\text{spend}_{it}}$ & \$120.49 &\$135.78 \\
Observations & 61,860,735 & 29,344,385 \\
\bottomrule
\end{tabular}
\begin{tablenotes}
    \small
    \item \textit{Notes}: Column 1 presents results from Table \ref{tab:ddd-table}. Column 2 restricts treated group to households with $\geq 2$ shoppable services in different years. Throughout, standard errors were clustered at the household level, and regressions included family, year, relative week of year, and provider (of shoppable service) fixed effects.
    
    $^{*} p < 0.05, ^{**} p < 0.01, ^{***} p < 0.001$ 
    \end{tablenotes}
    \caption{\label{tab:ddd-table-learn} Reduced-Form Evidence of Learning}
\end{threeparttable}
\end{table}

First, we consider how bills impact households differently for an initial shoppable service than for subsequent consumption, defined as index services that the household consumes in later years.\footnote{We limit attention to households who experience at least two shoppable events across different years in order to avoid capturing forward-looking behavior in this exercise.} Table \ref{tab:ddd-table-learn} presents the results. We observe spending corrections from a bill's arrival only for \textit{initial} shoppable services, with negligible effects for subsequent exposures. On the other hand, we continue to observe significant---although muted---post-service responses for households even for later services.\footnote{One might be concerned those with repeated events are higher-spending and higher-risk on average. We do not observe this to be true on average (Table \ref{tab:ddd-table-learn}); in addition, these results hold even considering childbirths as the only shoppable service. In this case, risk is less likely to be a predictor of multiple births, yet we observe bill effects are still reduced for subsequent deliveries compared to initial ones.} 

Second, we consider how bill responses vary across a calendar year; this, importantly, tests whether households are forward-looking in their response to bills. Households may react to bills' pricing information either as it conveys information about spot prices or as it updates household beliefs about end-of-year prices \citep{aron-dine_moral_2015}. Examining how households respond to services based on the time they were consumed sheds light on how households weigh these different prices. Figure \ref{fig:calendar-year} presents results from stratifying our sample by the week a shoppable service was consumed.\footnote{Results are comparable if instead, we stratify by the week a bill arrived.} We observe a strong gradient: households respond much more to services consumed in the first quarter of the year than in the last quarter. This is likely mechanical, given that as a year progresses, more households will have crossed their deductible thresholds, weakening or eliminating any incentives to respond to cost-sharing for shoppable services towards the end of the year. However, it does suggest households are more responsive to ``deductible-crossing events," and have incentives to attempt to strategically delay some care until after significant health events. 

On the other hand, the bill effects presented in panel (b) of Appendix Figure \ref{fig:calendar-year} are constant across the calendar year, except for the final 7 weeks of a year, in which the effects dissipate.\footnote{Most likely, this is due to the reduced information content of a bill arriving at the end of the year, given that plans and marginal prices reset at the beginning of January. This may also be influenced by the increased difficulty of scheduling services towards the end of a plan year \citep{shukla_retrospective_2021}.} This suggests that although households' strategic behavior evolves over a year, bill effects correcting misperceptions in that behavior are constant across the year. These findings are interesting, given that they show bill responses are not attenuated over a year by changes in forward-looking household behavior.\footnote{Importantly, each of the firms with plan identifying information in our sample has a plan start date of January 1; hence, we are unable to perform the relatively simple test of re-running our analysis on a subsample of affected households whose deductible contributions are not weakly increasing over the calendar year. This is an interesting exercise for future work in this area.} 

\subsubsection{Considering Alternative Mechanisms}
The evidence presented above suggests households respond to the pricing information contained in a bill \textit{and} update their beliefs about how spending affects future cost-sharing; however, additional mechanisms may affect our estimates of the bill effect. In particular, a remaining concern is that household spending decisions may be particularly affected by liquidity constraints \citep{gross_liquidity_2022}. Households facing liquidity constraints may respond to a bill by curtailing their medical spending not because they corrected beliefs about their position on a cost-sharing curve, but simply because information about prices---particularly if those prices were under-estimated---may affect household purchasing power. 

Although our data does not provide us with rich information on household incomes or consumption, we follow the approach of previous literature to investigate whether variations in the timing of expected household income shocks may change bill effects \citep{gross_liquidity_2022}. We examine heterogeneity across the week of month of a bill's arrival, relying on the intuition that if income effects drive our results, bills arriving just before the end of the month should have stronger observed effects than those arriving near the beginning. We do not find evidence that this is the case (Appendix Table \ref{axtab:liquidity}). Household medical spending increases in the middle of a month, with spending in the second week of the month approximately 50\% higher than spending at the end of a month; this is consistent with income affecting the timing of household health purchases. However, the corresponding bill effects are statistically indistinguishable from one another, suggesting that households reduce their spending by a consistent amount across a month. This evidence is largely suggestive, however, and should be carefully considered in future research with billing and income data. 

Finally, observed bill effects may be driven by a single provider, rather than by changes to household demand. However, we observe that our results are robust to excluding all spillover claims obtained from the provider of the index service (e.g., a general practice physician or a hospital system). Appendix Table \ref{axtab:focal-provider} reports strong bill effects even after removing these claims, suggesting that bill effects are driven by households rather than physicians.\footnote{Note that this analysis has several limitations. First, we are not able to identify all providers of shoppable services, and there is little documentation for why Marketscan data has some provider IDs missing and not others. Second, pharmaceutical spending cannot be reliably assigned to a physician identifier. Finally, control group years without an index service are omitted from this calculation.}
 
\subsection{What Services Are Affected?}

\subsubsection{Do Bills Only Affect Strategic Delaying of Services?}
One important question when considering our results is whether bill effects merely represent delays in the use of care or a more fundamental change in the quantity and type of medical care households seek. For example, a bill informing households they have yet to meet a deductible may generate a new wave of strategic delays, where services are further postponed until cost-sharing thresholds are crossed. However, we find evidence that households alter \textit{where} they seek care even for services that cannot be strategically delayed.

\begin{table}[htb]
\centering
\begin{threeparttable}
\begin{tabular}{l|cc|cc}
\toprule
& \multicolumn{2}{c}{\textbf{Regression Coefficients}} & \multicolumn{2}{c}{\textbf{Pre-Treatment Averages}} \\
& \multicolumn{1}{c}{Post Service} & \multicolumn{1}{c}{Post Bill} &\multicolumn{1}{c}{\% $\geq 0 $} & \multicolumn{1}{c}{Conditional Mean} \\
\midrule
\multicolumn{2}{l}{\textbf{All Spending: Injuries and Infections}} \\
Total Bill Effect & 0.151*** & -0.014  & 3.6\% & \$251 \\
& (0.0128) & (0.0131) \\ 
\hspace{.2cm} Physician Office & -0.008 & 0.040*** & 3.2\% & \$127  \\
& (0.0087) & (0.0090) & \\
\hspace{.2cm} Hospital Campus  & 0.258*** & -0.040*  & 0.5\% & \$996 \\
\hspace{.25cm} (incl. outpatient) & (0.0199) & (0.0201) \\ 
\bottomrule
\end{tabular}
\begin{tablenotes}
    \small
    \item \textit{Notes}: Table presents estimated coefficients evaluating bill effects on non-strategic health spending (Appendix Table \ref{tab:infections}). ``Physician office" includes outpatient non-hospital offices and urgent care centers; ``hospital campus" includes outpatient hospital services, inpatient admissions, and emergency departments. Column (3) indicates the fraction of pre-treatment weeks with positive spending; column (4) presents pre-treatment weekly averages, conditional on positive spending. $^{*} p < 0.05, ^{**} p < 0.01, ^{***} p < 0.001$ 
    \end{tablenotes}
    \caption{\label{tab:infections-table} Bill Effects on Care for Injuries and Infections}
\end{threeparttable}
\end{table}

We define a set of such services spanning injuries---including broken bones, dislocations, and sprains---and common respiratory and gastrointestinal infections (defined in Appendix Table \ref{tab:infections}). Using these services as an outcome (rather than an index event, as in Appendix Table \ref{axtab:expanded-services}), we assess bill effects (Equation \ref{eq:reg}) and stratify by the place of service were care was administered. Table \ref{tab:infections-table} presents the results, highlighting that post-service effects are entirely contained in hospital-based care for injuries (a 26\% increase), while bill effects constitute a transition back to physician office-based care. This exercise is merely descriptive, as there are multiple hypotheses to consider. One possible rationalization of these results, however, is that households may perceive reduced financial barriers post-service to seeking care at a hospital until bills provide additional information.

\subsubsection{Heterogeneous Effects by Type of Care} \label{subsubsec:het}

Finally, we explore how bills affect consumption across broad categories of medical services, including hospital care, outpatient services, and pharmaceutical spending.\footnote{Appendix Table \ref{tab:outpatient-services} includes detailed descriptions of the construction of each of these variables.} This decomposition allows us to examine whether household responses to a bill vary with any measure of perceived or real quality of care. Particularly, we examine how bills affect future utilization of typically high-value health services such as preventive screenings or behavioral health services as well as typically low-value care such as unnecessary pre-operative screenings or imaging services. 

Households consistently respond to bills by reducing consumption across a variety of services (Appendix Table \ref{tab:services}). Spillover use of inpatient services---including preventable hospitalizations---falls by roughly 17.0\% post-bill; however, this effect is not significant ($p=0.053$) and provides only suggestive evidence of effects on low-value inpatient services \citep{agency_for_healthcare_research_and_quality_guide_2007}.\footnote{Whether this increase is an over-utilization of unnecessary care or simply increased access to relevant hospital services---particularly considering the ``layperson standard" for hospital care---is an open question which warrants future research \citep{siegfried_adult_2019}.} However, outpatient services---such as general practitioner and specialist visits, labs, and preventive care---exhibit stronger bill responses.\footnote{Some outpatient services, such as chiropractic care, are affected neither by the consumption of a shoppable service nor its accompanying bill; others, such as behavioral health services, exhibit the opposite pattern from the overall bill effects. This is presumably because these services have more inelastic demand and lower rates of cost-sharing generally.} We do not observe bill effects among household demand for prescription drugs, perhaps because of already high pre-event levels of pharmaceutical consumption.

Surprisingly, we do not observe that households change their utilization of low-value care following a bill's arrival. These services, which include services such as imaging for lower-back pain, misuse of prescription medications to manage migraines and bacterial infections, or unnecessary pre-operative screenings, are determined by the recommendations  of the Choosing Wisely campaign \citep{colla_choosing_2015}. We find that households increase their use of low-value care by 6.6\% following a major service, and then further by another 2.8\% once the bill arrives. This may be a result of a ``cascade of care" effect associated with increased consumption of general medical care, which in turn prompts downstream increases in physician ordering of low-value services \citep{ganguli_assessment_2020}. Physicians typically retain control over when low-value services are performed, in order to reduce their own uncertainty, liability, or ``just to be safe" \citep{colla_choosing_2017}.

\section{Model}\label{sec:model}
Based on the empirical findings from our reduced-form analysis, we propose a simple model of ``imperfect" moral hazard. In this model, consumers with imperfect beliefs about prices and spending are price responsive in demanding care; however, accurate pricing information may lag consumption by weeks or months while still affecting the spot prices of care in the interim. Given these delays, consumers form expectations about their realized OOP spending and the implied marginal cost of care in each period. The modeling exercise presented here is therefore a useful one when considering how models of moral hazard in healthcare may be augmented to incorporate the effects of delayed pricing information. 

We focus our attention on noisy signals of household spending to construct a model of elastic demand for healthcare services under pricing uncertainty. That is, we assume that health shocks are perfectly observed in each period \citep{einav_selection_2013}, leaving only patient OOP costs unknown. This uncertainty can be decomposed into two parts: first, consumers have uncertainty about the OOP spending they have incurred to date, due to quasi-random delays in receiving pricing information through bills; and second, consumers have residual uncertainty about price variation and plan coverage information that persists over time. We estimate the effect of resolving the first type of uncertainty through bill arrivals in order to match the reduced-form evidence presented above. 

The model we present expands on the reduced-form evidence in three key ways. First, we estimate household bill effects across the full set of household services, rather than focusing exclusively on shoppable services as index events. The structural estimation of spending signals allows us to generalize beyond the previously narrow definition of index events. Second, the model proposes and estimates mechanisms underlying these effects. These include incorporating differences in household decision-making from noisy signals, but also the potential for households to both incorrectly perceive these signals and learn from them over time. Finally, the model allows for simple counterfactual exercises illustrating how correcting these errors may change decision-making. In particular, we consider how correcting estimated bias in spending signals is predicted to change household decision-making, and compare these results against eliminating all uncertainty in coverage and predicted healthcare costs. Doing so illustrates how one potential mechanism explaining our results---incorrect perceptions of spending relative to a deductible---fits into a much more complex decision-making process given the many uncertainties households face when consuming medical care. 

\subsection{Model Details}\label{subsec:model-details}

In each period $t$, an individual $i$ receives a health shock $\lambda_{it}$, which represents a combination of both acute fluctuations in health status and persistent health needs. Patients then choose an appropriate level of medical spending $m_{it}$---measured in the dollar value of the services---in response to $\lambda_{it}$, spending histories, and individual preferences.\footnote{We assume shocks are measured in dollars to compare health production and OOP spending, consistent with previous versions of this model \citep{einav_selection_2013}. This is useful as it is tractable and incorporates rational responses to nonlinear pricing; individuals close to a deductible will choose to slightly increase their consumption, anticipating the approaching nonlinear change in marginal costs \citep{marone_should_2022}. To be consistent with Section \ref{sec:rf-evidence}, we model spending choices at the per-person week level.} Following \cite{einav_selection_2013}, we calibrate individual patient utility as a quadratic loss function in the distance between selected health spending and the unobserved health shock:
\begin{equation}
    u_{it} = (m_{it}-\lambda_{it})  - \frac{1}{2\omega_i}(m_{it}-\lambda_{it})^2 - c_{ijt}(m_{it}; M_{\mathcal{I}t}). 
\end{equation}

\noindent Here, $\omega_i$ is an individual time-invariant ``moral hazard" parameter capturing individual heterogeneity in demand responsiveness to the price of services.\footnote{The individual-specific moral hazard parameter $\omega_{i}$ has a helpful interpretation as the incremental spending induced by a move from no insurance to full insurance \citep{einav_selection_2013}.} In addition, $c_{ijt}(m_{it};M_{\mathcal{I}t})$ denotes the OOP costs associated with $m_{it}$, which depends on the piecewise-linear cost-sharing contract of individual $i$'s chosen insurance plan, $j$, as well as the OOP spending to date at the household level, $M_{\mathcal{I}t}=\sum_{i \in\mathcal{I}} \sum_{k=1}^{t-1} m_{ik}$. Note that an individual’s OOP costs for services are weakly decreasing in $M_{\mathcal{I},t}$, given the cost-sharing structure. 

Under full information, patients know both the value of $M_{\mathcal{I}_t}$ and how it affects $c_{ij}(\cdot)$. Furthermore, in the case where cost-sharing is linear at all stages of the contract, a patient's marginal OOP cost is given by $c_{ijt} \in [0,1]$, where $c=1$ applies to all services before a deductible has been met and $c=0$ applies for all services after an OOP-max has been met. Between the deductible and the OOP-max, $c$ is typically in the open interval $(0,1)$. The optimal choice of $m_{it}$ in each period is simply the solution to the first order condition: 
\begin{equation} 
1-\frac{1}{\omega_i}(m_{it}-\lambda_{it})-c_{ijt}=0 \Rightarrow m_{it}^* = \max\left[0,\lambda_{it}+\omega_i(1-c_{ijt})\right].
\end{equation} 

\noindent That is, medical expenses in each period are chosen so that the marginal utility of those services is equal to the marginal (known) OOP cost. In particular, as $c$ changes from 1 to $c < 1$, individuals will have a discontinuous increase in their consumption.

We suppose that $M_{\mathcal{I},t}$ is not known with certainty at the time a service is performed. Rather, household spending can be divided into two components: spending for services whose bills have already arrived (where prices are known), and spending for services without accurate pricing information yet available. For ease of notation, suppose that each bill takes $\tau$ weeks to arrive, so that a bill for a service procured in week $t$ would arrive in week $t + \tau$.\footnote{Note that in the empirical estimation of the model, the length of time between a service and bill's arrival is allowed to vary across services; this assumption is only made in this section for ease of exposition.} Based on these components, households respond to a signal of their spending $\theta$: 
\begin{equation}
    \theta_{it} = \underbrace{\sum_{k=0}^{t-\tau} \sum_{i \in \mathcal{I}} c_{ij}(m_{ik})}_{\text{known spending}} + \underbrace{\sum_{k=t-\tau+1}^{t} \sum_{i \in \mathcal{I}} s_i(m_{ik}|x_{ik}),}_{\tilde{\theta}_{it}=\text{unknown spending}} 
\end{equation}

\noindent where $s_i(m_{ik}|x_{ik})$ represents service-specific signals of spending, which may depend on individual, household, and service level characteristics. 

Hence, the timing of the model in each period $t$ is as follows: 
\begin{enumerate}
    \item Individuals form expectations about their spending histories $M_{it}$, based on $\theta_{it}$.
    \item Individual health shocks $\lambda_{it}$ are realized. 
    \item Spending decisions $m^*_{it}$ are made based on realized health shocks and beliefs about spending histories, which govern the perceived marginal cost of additional units of care $\hat{c}_{it}$.
    \item A new signal of spending $s_{it}(m^*_{it})$ is received for the individual and all household members enrolled in the same plan. Household members update their expectations of $M_{it}$, and we proceed to period $t+1$.
\end{enumerate}

\subsection{Parameterizing Price Signals} 
\label{subsec:model-simple}

We model spending signals as noisy, potentially biased signals affecting household beliefs about their future cost-sharing responsibilities. That is, we suppose that signals follow a normal distribution around a constant multiple of the true OOP spending with some noise:
\begin{equation}
    s_i(m_{it}|x_{it}) \sim \mathcal{N}\left(\beta \cdot c_{ijt}(m_{it}), \sigma^2_s\right),
\end{equation}

\noindent where $\beta$ indicates the potential for patients to inflate (or deflate) their true OOP spending by a constant parameter before the bill arrives, and $\sigma^2_s$ quantifies the noise inherent in signals.\footnote{Note that allowing $\beta$ to be a random coefficient varying across individuals is a simple extension of the model; for the present purposes, however, we focus on an average of $\beta$ across the population of interest.} As mentioned in Section \ref{subsec:pricing}, uncertainty about cost-sharing may persist about more than simply OOP spending, as households may not fully internalize negotiated prices or what services are subject to a deductible. Based on these assumptions, household expectations of their OOP spending to date can be written as 
\begin{equation}
\label{eq:beta}
    \mathbbm{E}\left[\theta_{it}\right] = \sum_{i \in \mathcal{I}} \sum_{k=0}^t (1-D_{ik}) \beta c_{ijk}(m_{ik})+D_{ik}c_{ijk}(m_{ik}),
\end{equation}

\noindent where $D_{ik}$ is a binary variable indicating if the bill for services performed in week $k$ has arrived ($D_{ik}=1$) or not ($D_{ik}=0$). Based on the household's expected value of $\theta_{it}$, households have a well-defined expectation over the marginal cost of future consumption given by\footnote{Note that $c < 1$ in general. In practice, we estimate the model on the sample of individuals enrolled in plans with non-zero deductibles, to cleanly capture how OOP misperception may affect discontinuous changes in the marginal cost of spending across deductible thresholds.}  
\begin{equation}
    \mathbbm{E}\left[\hat{c}_{it}\right] = Pr(\theta_{it} \leq d) \cdot 1 + (1-Pr(\theta_{it} \leq d))\cdot c,
\end{equation}

where $d$ is the deductible facing individual $i$. Given the distributional assumptions placed on the signals $s(\cdot)$, this probability can be directly computed as 
\begin{equation}
    Pr(\theta_{it}\leq d) = \Phi\left(\frac{d-\mathbbm{E}[\theta_{it}]}{\sum_{k=t-\tau+1}^t \sigma^2_s}\right),
\end{equation}

\noindent where $\mathbbm{E}[\theta_{it}]$ is as defined in Equation \ref{eq:beta}.

This model allows for households to systematically over- or under-inflate their true OOP spending absent a bill's arrival, measured by the parameter $\beta$; however, lack of pricing information could shape behavior simply due to household risk aversion given an unknown marginal price $\hat{c}$. Intuitively, as signals become noisier ($\sigma^2_s$ increases) or bills take longer to arrive (so that $\sum_{k=t-\tau+1}^t \sigma^2_s$ increases), the probability an individual faces a lower marginal cost increases, potentially resulting in over-spending relative to an individual who immediately receives accurate pricing information. Importantly, this can happen in the model even if signals are unbiased (so that $\beta=1$). 

Estimation of the model requires identification of $\beta$ as well as the size of the noise parameter $\sigma^2_s$. Additional unobservable parameters in the model include heterogeneity in moral hazard $\omega_i$ and individual health shocks $\lambda_{it}$. When estimating the model, we calibrate these hyper-parameters to match moments predicted by both previous research and training data not used in the structural estimation. We use the estimated regression coefficients predicted by \cite{einav_selection_2013} in order to capture variation in moral hazard parameters across households.\footnote{Note that these regression models result in individual-level predictions for $\omega_i$; in estimation, we aggregate these to the household level by taking the mean of $\log(\omega_i)$ across all members $i \in \mathcal{I}.$} We model individual-level health shocks as draws from an individual-specific shifted lognormal distribution; this distribution captures both the skewed nature of the observed spending data and the probability that an individual has zero consumption in a period. That is, each individual in each period draws $\lambda_{it}$ from a distribution $F(\mu_i, \sigma_i, \kappa_i)$ such that
\begin{equation}
    \log(\lambda_{it} - \kappa_i) \sim \mathcal{N}(\mu_i, \sigma_i^2).
\end{equation}

We calibrate the three hyper-parameters $(\mu_i, \sigma_i, \kappa_i)$ to match the empirical distribution of weekly spending using individuals \textit{not} included in the structural estimation. These include individuals enrolled in plans with no deductible, as well as patients enrolled in any type of plan between 2014 to 2018. Individuals in this sample are grouped into cells based on patient demographics---including age, sex, risk score quartile, and relationship to the main employee---and the empirical distribution in each cell is matched to the shifted lognormal moments.\footnote{This is done using three properties of a shifted lognormal distribution: $\overline{\lambda}=\exp(\mu+\frac{1}{2}\sigma^2)+\kappa$, $\lambda^M = \exp(\mu)+\kappa$, and $\frac{\text{sd}(\lambda)}{\overline{\lambda}}=\sqrt{\exp(\sigma^2)-1}$, where $\lambda^M$ denotes the median. The solution to this system of equations given the moments of the empirical distribution of $\lambda$ identifies the three hyperparameters $\mu,\sigma,\kappa$.} Once these parameters are identified, individual-period shocks are drawn for each member of the model sample and then summed to the household-period level.\footnote{In order for shocks to be meaningful, we restrict $\lambda_{\mathcal{I}t} < m_{\mathcal{I}t}$ in each period.}

Given these calibrations, identification of the main parameters $(\beta, \sigma^2_s)$ come from exogenous variation (at the household level) in the length of time required for a bill to arrive after different health services. In contrast to the reduced-form evidence presented in Section \ref{sec:rf-evidence}, the model leverages variation in the waiting periods associated with \textit{all} medical claims in a given household-year. This variation may exist across services as well as across households; importantly, underlying variation in $\tilde{\theta}_{it}$ which artificially moves households above or below their deductible is central to identifying how household expectations of $\hat{c}$ evolve in ways that most closely fit the observed choice data. We therefore can identify the bias parameter $\beta$ using the quasi-randomness of bill timing, where a bill generates a discrete change in marginal costs correcting household expectations (similar to the evidence presented in Section \ref{sec:rf-evidence}). On the other hand, identification of the variance parameter $\sigma^2_s$ leverages variation across households and services to identify the spread of the signal distribution. 

\subsection{Household Learning}
\label{subsec:model-learning}

Given reduced-form results suggesting that households learn from repeated interactions with shoppable services, we model household belief formation about marginal prices over time. To assess this, we incorporate household learning about the potential bias parameter $\beta$, where households may learn from each bill's arrival.\footnote{For now, we model each signal as having equal impact; future extensions of this model could flexibly model heterogeneous signals based on the total cost of a service or by different service types. We also do not model learning about the spread of the distribution of signals over time.} Households are assumed to have prior beliefs about $\beta$ which follow a normal distribution with a mean $\mu_{\beta,0}$ and variance $\sigma^2_{\beta,0}$: 
\begin{equation}
    \hat{\beta}_{i0} \sim \mathcal{N}(\mu_{\beta,0},\sigma^2_{\beta,0}).
\end{equation}

\noindent When a bill arrives for a household conveying information about the prices of medical services, it corrects household perceptions, conveying, in essence, that $\beta = 1$. To model this, we construct learning signals $\ell_{it}$ in each period drawn from a normal distribution centered at 1 and with a variance $\sigma^2_\ell$: 
\begin{equation}
    s_{it} \sim \mathcal{N}(1,\sigma^2_\ell).
\end{equation}

We assume that households update their prior beliefs conditional on their observed signal following Bayes' Rule. Assuming normal distributions for both the prior and posterior allows for closed-form solutions for household beliefs about $\beta$ at each period, and is consistent with previous learning models \citep{crawford_uncertainty_2005}. 

Incorporating learning requires estimating four parameters of interest. First, the average prior mean $\mu_{\beta,0}$ and unchanging spending signal variance $\sigma^2_s$ dictate the spread of OOP spending signals households receive in each period prior to learning (where $\mu_{\beta,0}$ is similar to $\beta$ in the non-learning model). Rather than assume constant household bias, we measure dispersion in potential under-information across households with $\sigma^2_{\beta, 0}$; the speed with which household biases are corrected by bills is governed by the variance of the learning signal, $\sigma^2_\ell$. 

Estimated learning parameters allow analysis of both the spread of beliefs prior to any information and the speed with which beliefs converge. As in the simpler case of the model, identification of the four parameters $(\mu_{\beta,0},\sigma^2_s,\sigma^2_{\beta,0},\sigma^2_s)$ stems from exogenous variation in bill timing. Within-household variation in expenditures relative to pending OOP expenditures serves to identify both the starting point of household beliefs (the prior mean) and the rate of convergence (the signal variance). Similarly, variation in choices across households identifies the spread of beliefs---summarized in the prior variance---and the spread of signals in each subsequent period, $\sigma^2_s$. 

\subsection{Model Results}
\label{subsec:model-results}
We estimate the model presented in Section \ref{sec:model} for 240,111 households in our analytical sample enrolled in plans with nonzero deductibles from 2006 to 2013. For each household-week, we simulate household health shocks and draw household moral hazard parameters; then, for different values of $\beta$, we estimate household signals of underlying OOP spending and associated marginal costs, $\hat{c}_{it}$. Taken together, these estimates produce a prediction of spending $m_{it}(\beta)$, which differs as $\beta$ changes. Our primary measure of model fit is the root mean squared error (RMSE) between observed and predicted levels of weekly spending at the household level. 

\begin{table}[htb]
\centering
\begin{threeparttable}
\begin{tabular}{lcc|cc}
\toprule
& \multicolumn{2}{c}{No Learning} &  \multicolumn{2}{c}{Learning}\\\cmidrule{2-3}\cmidrule{4-5}
& \multicolumn{1}{c}{Estimate} & \multicolumn{1}{c}{95\% C.I.} & \multicolumn{1}{c}{Estimate} & \multicolumn{1}{c}{95\% C.I.}  \\
\midrule
\multicolumn{4}{l}{\textbf{Panel A: Parameter Estimates}} \\
$\beta$ or $\mu_{\beta,0}$ & 1.73 & [1.695,1.765] & 2.58 & [2.532,2.628] \\
$\sigma_s$ & \$15.20 & [\$14.95,\$15.45] & \$14.77 & [\$14.54,\$15.00]\\
$\sigma_\beta$ & --- & --- & 0.12 & [0.109, 0.135] \\
$\sigma_\ell$ & --- & --- & 0.09 & [0.070,0.106] \\
\midrule
\multicolumn{4}{l}{\textbf{Panel B: Implied Spending}} \\ 
Mean & \$2,415 & [\$2,413.86,\$2,416.75] & \$2,350 & [\$2,342.64, \$2,357.37] \\
Median & \$1,302 & [\$1,300.64, \$1,304.91] & \$1,276 & [\$1,268.67, \$1,283.34] \\
\midrule
\multicolumn{4}{l}{\textbf{Panel C: Counterfactual Spending}} \\
Recentered Priors & \\
\hspace{.2cm} \% Households Affected & 98.9\% &  & 85.3\% \\
\hspace{.2cm} $\Delta$ Mean Spending & -\$392 & [-\$394.05,-\$390.63] & -\$364 & [-\$370.64,-\$356.76] \\
& (16.2\%) &  & (15.5\%) &  \\
\hspace{.2cm} $\Delta$ Median Spending & -\$166 & [-\$167.81,-\$164.39] & -\$94 & [-\$95.48, -\$92.60] \\
& (12.7\%) & & (7.4\%) &  \\
Real-time Information \\
\hspace{.2cm} \% Households Affected & 72.0\% &  & 85.0\% \\
\hspace{.2cm} $\Delta$ Mean Spending & -\$1,163 & [-\$1,167.78,-\$1,158.90] & -\$792 & [-\$798.67,-\$784.53] \\
& (48.2\%) &  & (33.7\%) & \\
\hspace{.2cm} $\Delta$ Median Spending & -\$489 & [-\$493.90,-\$485.01] & -\$204 & [-\$205.83, -\$201.97] \\
& (37.6\%) &  & (16.0\%) &  \\
\bottomrule
\end{tabular}
\begin{tablenotes}
    \small
    \item \textit{Notes}: Table presents results from estimating both variations of the model described in Section \ref{sec:model}. Panel A reports the baseline parameter estimates; Panel B reports implied spending at the household-year level given these equilibrium parameters. Panel C reports how these parameters are expected to change under two countrefactual scenarios: recentered priors ($\beta = \mu_{\beta,0}$ = 1) and real-time information ($\beta=1,\sigma^2_s=0)$. All currencies are reported in 2022 USD. 
    \end{tablenotes}
    \caption{\label{tab:cfs} Model Results and Implied Spending}
\end{threeparttable}
\end{table}

Table \ref{tab:cfs} reports the results, based on 50 simulations with different individual health shocks.\footnote{See Appendix Figure \ref{fig:model1-results} for a visual depiction of the parameter space in the non-learning model.} We estimate both significant bias in household perceptions of OOP spending as well as considerable noise in spending signals. The average spending signal in the model without learning is centered at 173\% of true OOP spending, with a standard deviation of \$15.20.\footnote{See Appendix Figure \ref{axfig:implied-probabilities} for a depiction of the implied probabilities a household has met their deductible for a \$1,000 deductible, as a function of expected spending $\mathbbm{E}[\theta_{it}]$.} The equilibrium model parameters have two primary implications. First, there is a great deal of uncertainty at the household level in estimating OOP spending and prices; the large estimated standard deviation of price signals may well lead risk-averse households to have lower expected utility and therefore to change spending decisions. Second, in addition to overall uncertainty, the model implies that households respond to price signals as though they are biased, with the average price signal being inflated by 73\% relative to its true OOP cost. This is consistent with the findings of Section \ref{sec:rf-evidence}, and suggests that households may over-estimate their spending relative to a deductible prior to receiving price information. 

\begin{figure}[htb]
    \caption{Evolution of Beliefs about $\beta$ Across Plan Year}
    \label{fig:learning}
    \centering

    \includegraphics[width=4.5in]{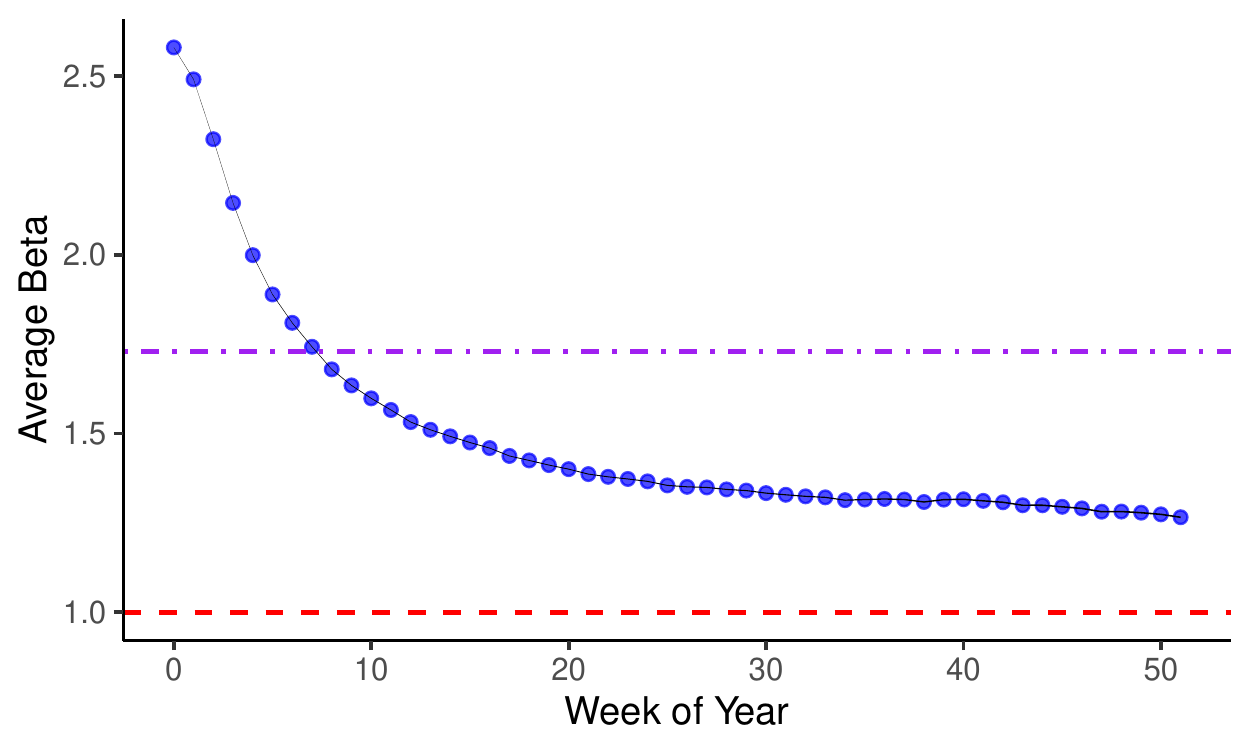}
    \vspace{0.2cm}
    \begin{minipage}{0.95\textwidth} 
	{\footnotesize \textit{Notes}: Figure depicts average value of simulated $\beta$ across the relative week of a plan year for the full sample, with 95\% confidence intervals shaded in black. Simulations are performed based on the median equilibrium parameters of the model discussed in Section \ref{subsec:model-learning}. Dashed purple line represents the average estimated value of $\beta$ in the non-learning model, 1.73.
	\par
	}
	\end{minipage}
\end{figure}

The model results are similar even once household learning is permitted in the model, as reported in the third and fourth columns of Table \ref{tab:cfs}. The initial prior mean for spending signals increases from 1.73 to 2.58, suggesting the greatest mismatch between household perceptions and spending signals at the beginning of the year. This is intuitive, as early in the plan year, the majority of households are attempting to ascertain if a spending event met their deductible, so we would expect noisier spending signals earlier in a plan year. However, household learning takes place relatively quickly: Figure \ref{fig:learning} shows the average value of $\hat{\beta}$ across the sample by week of year.\footnote{There is relatively little variation across households, with the estimated prior standard deviation equal to 0.12. Put into context, we estimate that 95.5\% of households (two standard deviations on either side of the mean) have priors in the interval $(2.34, 2.82)$. Similarly, the model suggests that bills provide precise information about $\beta$, as $\sigma_\ell$ is estimated to be 0.09, meaning 96\% of signals are within the interval $(0.82, 1.18)$. This is well below both the non-learning average of $\beta=1.73$ and the learning model average of $1.47$.} Within the first six weeks of the year, average household beliefs about $\beta$ converge to below the 1.73 estimated in the non-learning model; for the rest of the year, they remain close to the overall average, which is estimated to be about 1.47.\footnote{Appendix Figure \ref{figax:misinformation} presents further results illustrating heterogeneity in beliefs across a year.} Similar to the non-learning model, there remains both potential bias in the center of the spending signal distribution and considerable noise in its spread. 

We therefore assess the extent to which (a) correcting household beliefs and (b) eliminating all uncertainty in price signals affect implied spending. The first simulation is equivalent to aiding consumers to be better informed about healthcare prices prior to consumption; the second simulation completely eliminates all pricing uncertainty---this would require both some form of immediate claims adjudication as well as a hypothetical way to eliminate any other form of consumer uncertainty with regards to spending (as discussed above). Hence, the second simulation is only used for comparison to the first. In both models, the implied levels of household OOP spending are consistent with the baseline averages (Panel B). In the non-learning model, recentering household priors (that is, imposing that $\beta=1$ for all households) reduces predicted spending for 98.9\% of households, with conditional changes in spending of \$392 (16.2\%) on average and \$166 (12.7\%) at the median. Although this constitutes a significant difference in household spending, it represents only a fraction of predicted spending changes from total price uncertainty. Eliminating all such uncertainty is predicted to lower total spending for 85.8\% of households, with only $1/3$ of the change explained by household under-estimation of prices.\footnote{Note that fewer households change spending when uncertainty is eliminated as some households received correct/under-inflated signals randomly, even with a large $\beta$.} We find similar effects even considering a model enriched with household learning; eliminating household bias in $\beta$ results in spending changes for 85.3\% of households with average (median) declines of 15.5\% (7.4\%). 

Taken together, the model results indicate both that households may incorrectly process spending signals prior to a bill's arrival while reinforcing that substantial noise remains in those signals regardless of potential bias. In the interim period while awaiting a bill, households consume medical care as though they are considerably closer to reaching their deductible than they are in reality. What's more, spending signals are quite noisy, leading households with substantial uncertainty in their spending even in the absence of bias. Correcting household bias about OOP prices (relative to a deductible) changes behavior only marginally compared to providing real-time pricing information and eliminating all periods of billing uncertainty.



\section{Discussion \& Conclusion}\label{sec:conclusion}
This paper assesses how households respond to a medical bill as a form of resolving price uncertainty, potentially affecting future household consumption. When a household member consumes a service with a nontrivial---but unknown---amount of OOP spending, other household members increase their spending. However, bill arrivals meaningfully change these increases; we estimate that a bill's arrival leads to a reduction in household health spending of 10.9\%. These results are robust to numerous alternative specifications, and appear consistent with a mechanism where bills reduce or eliminate pricing uncertainty, especially for households close to changes in their insurance contracts such as meeting a deductible. 

We encapsulate our findings in a model of ``imperfect moral hazard" with delayed learning about prices. Our model, just as our reduced-form evidence suggests, indicates that consumers face uncertainty when making consumption decisions with lagged pricing information, and that they may overestimate the probability that they have met a household deductible, particularly early in an insurance contract period. Our model allows us to consider household learning from medical events and place these perception errors into the context of broader pricing uncertainty. 

Our analysis provides several important contributions to models of price uncertainty and household moral hazard in healthcare; however, our results should be viewed in the context of their limitations. First, by limiting our analysis to households enrolled in group ESI plans, we are unable to determine how price uncertainty affects consumption decisions for other populations, such as couples on Medicare or low-income households covered by Medicaid or the ACA Marketplaces. Examining other populations---particularly populations with greater income constraints---would shed additional, important light on the extent to which price uncertainty leads to sub-optimal allocations of care. Second, while our results suggest that households would make different spending decisions without price uncertainty---in particular, consuming less care on average---we are unable to say anything about the welfare effects of these decisions given our current data. Finally, we use an imperfect proxy in our exposure (EOB generation), which may introduce measurement error into our estimates. As discussed in Section \ref{sec:data}, this error is likely to attenuate our estimates, not because the measurement error is classic, but instead because measurement error in true bill arrival introduces contamination bias from the interim period when households do not know OOP spending. If consumers over-estimate OOP prices before the bill arrives, any regressions misclassifying $\mathbbm{1}\{\text{Post\_Bill}\} = 1$ when it should be 0 will attenuate the correction parameter $\beta_\text{Post\_Bill}$ to zero.

The analysis we present could be extended in several meaningful ways. First, future work could incorporate observed payment interactions between patients and physicians, rather than relying on claims data alone. Data on physician practices---including how quickly physicians submit claims to payers for medical claims and send bills to patients---may provide insights into both the source of variation in processing times as well as the potential policy benefits of reducing the length of provider billing cycles. Future work may also consider the spillover effects of bill shock from healthcare consumption on other, non-health household consumption decisions.  

More generally, future research could build on the learning model presented here. This could include a more flexible framework for belief formation, a more thorough treatment of heterogeneity across services, or allowing learning parameters to be covariate-dependent. In particular, exploring the health equity concerns associated with learning about prices could provide valuable insight into the persistence of health disparities in accessing even high-value preventive services \citep{teutsch_health_2020,mcmorrow_determinants_2014}. Finally, future work could explore the impact of real-time claim adjudication on consumer spending responses \citep{orszag_real-time_2020}. This could be especially policy-relevant when exploring how heterogeneity across payers and providers (e.g., integrated care practices) could be used to leverage improved price transparency. 

Increasing understanding of how consumers form expectations about their health needs and utilization is a vital component of designing optimal insurance contracts and health policies. Economic modeling and health policy alike are well-served by incorporating delayed learning as we assess how consumers make health decisions in real time.  

\clearpage

\nocite{naqvi_schemepack_2021} 
\bibliography{IBNR_ImperfectMoralHazard}

\begin{thebibliography}{}

\bibitem[{Agency for Healthcare Research {and} Quality},
  2007]{agency_for_healthcare_research_and_quality_guide_2007}
{Agency for Healthcare Research {and} Quality} (2007).
\newblock Guide to {{Prevention Quality Indicators}}: {{Hospital Admission}}
  for {{Ambulatory Care Sensitive Conditions}}.
\newblock Technical Report AHRQ Pub. No. 02-R0203, {Department of Health and
  Human Services}, {Washington, D.C.}

\bibitem[{Aron-Dine} et~al., 2015]{aron-dine_moral_2015}
{Aron-Dine}, A., Einav, L., Finkelstein, A., and Cullen, M. (2015).
\newblock Moral hazard in health insurance: {{Do}} dynamic incentives matter?
\newblock {\em Review of Economics and Statistics}, 97(4):725--741.

\bibitem[Arrow, 1963]{arrow_uncertainty_1963}
Arrow, K.~J. (1963).
\newblock Uncertainty and the welfare economics of medical care.
\newblock {\em American Economic Review}.

\bibitem[Baicker et~al., 2015]{baicker_behavioral_2015}
Baicker, K., Mullainathan, S., and Schwartzstein, J. (2015).
\newblock Behavioral hazard in health insurance.
\newblock {\em The Quarterly Journal of Economics}, 130(4):1623--1667.

\bibitem[Bloschichak et~al., 2020]{bloschichak_cms-specified_2020}
Bloschichak, A., Milewski, A., and Martin, K. (2020).
\newblock {{CMS-specified}} shoppable services accounted for 12\% of 2017
  health care spending among individuals with employer-sponsored insurance.
\newblock {{HCCI Original Report}}, {Health Care Cost Institute}.

\bibitem[Boyd and Bellemare, 2020]{boyd_microeconomics_2020}
Boyd, C.~M. and Bellemare, M.~F. (2020).
\newblock The microeconomics of agricultural price risk.
\newblock {\em Annual Review of Resource Economics}, 12(1):149--169.

\bibitem[{Brot-Goldberg} et~al., 2023]{brot-goldberg_rationing_2023}
{Brot-Goldberg}, Z.~C., Burn, S., Layton, T., and Vabson, B. (2023).
\newblock Rationing medicine through bureaucracy: Authorization restrictions in
  medicare.
\newblock Technical report, {National Bureau of Economic Research}.

\bibitem[Brown, 2017]{brown_empirical_2017}
Brown, Z.~Y. (2017).
\newblock An empirical model of price transparency and markups in health care.
\newblock {\em Working Paper}.

\bibitem[Cabral, 2017a]{cabral_claim_2017-1}
Cabral, M. (2017a).
\newblock Claim {{Timing}} and {{Ex Post Adverse Selection}}.
\newblock {\em The Review of Economic Studies}, 84(1):1--44.

\bibitem[Cabral, 2017b]{cabral_claim_2017}
Cabral, M. (2017b).
\newblock Claim {{Timing}} and {{Ex Post Adverse Selection}}.
\newblock {\em The Review of Economic Studies}, 84(1):1--44.

\bibitem[Card et~al., 2009]{card_does_2009}
Card, D., Dobkin, C., and Maestas, N. (2009).
\newblock Does {{Medicare}} save lives?
\newblock {\em The quarterly journal of economics}, 124(2):597--636.

\bibitem[Caskey et~al., 2014]{caskey_transition_2014}
Caskey, R., Zaman, J., Nam, H., Chae, S.-R., Williams, L., Mathew, G., Burton,
  M., Li, J.~J., Lussier, Y.~A., and Boyd, A.~D. (2014).
\newblock The {{Transition}} to {{ICD-10-CM}}: {{Challenges}} for {{Pediatric
  Practice}}.
\newblock {\em Pediatrics}, 134(1):31--36.

\bibitem[Cawley and Ruhm, 2011]{cawley2011economics}
Cawley, J. and Ruhm, C.~J. (2011).
\newblock The economics of risky health behaviors.
\newblock In {\em Handbook of health economics}, volume~2, pages 95--199.
  Elsevier.

\bibitem[Chandra et~al., 2010]{chandra_patient_2010}
Chandra, A., Gruber, J., and McKnight, R. (2010).
\newblock Patient {{Cost-Sharing}} and {{Hospitalization Offsets}} in the
  {{Elderly}}.
\newblock {\em The American economic review}, 100(1):193--213.

\bibitem[Chernew et~al., 2007]{chernew_value-based_2007-1}
Chernew, M.~E., Rosen, A.~B., and Fendrick, A.~M. (2007).
\newblock Value-{{Based Insurance Design}}.
\newblock {\em Health Affairs}, 26(Supplement 2):w195--w203.

\bibitem[CMS, 2019]{cms_medicare_2019}
CMS (2019).
\newblock Medicare and {{Medicaid Programs}}: {{CY}} 2020 {{Hospital Outpatient
  PPS Policy Changes}} and {{Payment Rates}} and {{Ambulatory Surgical Center
  Payment System Policy Changes}} and {{Payment Rates}}. {{Price Transparency
  Requirements}} for {{Hospitals}} to {{Make Standard Charges Public}}.

\bibitem[Colla and Mainor, 2017]{colla_choosing_2017}
Colla, C.~H. and Mainor, A.~J. (2017).
\newblock Choosing {{Wisely Campaign}}: {{Valuable For Providers Who Knew About
  It}}, {{But Awareness Remained Constant}}, 2014\textendash 17.
\newblock {\em Health Affairs}, 36(11):2005--2011.

\bibitem[Colla et~al., 2015]{colla_choosing_2015}
Colla, C.~H., Morden, N.~E., Sequist, T.~D., Schpero, W.~L., and Rosenthal,
  M.~B. (2015).
\newblock Choosing {{Wisely}}: {{Prevalence}} and correlates of low-value
  health care services in the {{United States}}.
\newblock {\em Journal of General Internal Medicine}, 30(2):221--228.

\bibitem[Cooper et~al., 2019]{cooper_price_2019}
Cooper, Z., Craig, S.~V., Gaynor, M., and Van~Reenen, J. (2019).
\newblock The price ain't right? {{Hospital}} prices and health spending on the
  privately insured.
\newblock {\em The Quarterly Journal of Economics}, 134(1):51--107.

\bibitem[Correia et~al., 2020]{correia_fast_2020}
Correia, S., Guimar{\~a}es, P., and Zylkin, T. (2020).
\newblock Fast {{Poisson}} estimation with high-dimensional fixed effects.
\newblock {\em The Stata Journal}, 20(1):95--115.

\bibitem[Crawford and Shum, 2005]{crawford_uncertainty_2005}
Crawford, G.~S. and Shum, M. (2005).
\newblock Uncertainty and {{Learning}} in {{Pharmaceutical Demand}}.
\newblock {\em Econometrica}, 73(4):1137--1173.

\bibitem[Cutler and Zeckhauser, 2000]{cutler_anatomy_2000}
Cutler, D.~M. and Zeckhauser, R.~J. (2000).
\newblock The anatomy of health insurance.
\newblock In {\em Handbook of Health Economics}, volume~1, pages 563--643.
  {Elsevier}.

\bibitem[Dave and Kaestner, 2009]{dave_health_2009}
Dave, D. and Kaestner, R. (2009).
\newblock Health insurance and ex ante moral hazard: Evidence from
  {{Medicare}}.
\newblock {\em International journal of health care finance and economics},
  9(4):367--390.

\bibitem[{Davis-Jacobsen} and Plemons, 2008]{davis-jacobsen_real-time_2008}
{Davis-Jacobsen}, D. and Plemons, C. (2008).
\newblock Real-{{Time Claim Adjudication}}: {{Deal}} or {{No Deal}}?
\newblock {\em The Journal of Medical Practice Management}, 23(4):222--224.

\bibitem[Denning, 2014]{denning_kansas_2014}
Denning, K. S.~J. (2014).
\newblock Kansas {{Legislature Passes Health Care Cost Transparency Bill}}:
  {{Real Time Explanation}} of {{Benefits}}.
\newblock {\em Missouri Medicine}, 111(3):162.

\bibitem[{Diaz-Campo}, 2022]{diaz-campo_dynamic_2022}
{Diaz-Campo}, . (2022).
\newblock Dynamic moral hazard in nonlinear health insurance contracts.
\newblock {\em Working Paper}.

\bibitem[Duarte, 2012]{duarte_price_2012}
Duarte, F. (2012).
\newblock Price elasticity of expenditure across health care services.
\newblock {\em Journal of Health Economics}, 31(6):824--841.

\bibitem[Dunn, 2016]{dunn_health_2016}
Dunn, A. (2016).
\newblock Health insurance and the demand for medical care: {{Instrumental}}
  variable estimates using health insurer claims data.
\newblock {\em Journal of Health Economics}, 48:74--88.

\bibitem[Dunn et~al., 2020]{dunn_costs_2020}
Dunn, A., Gottlieb, J.~D., Shapiro, A., and Tebaldi, P. (2020).
\newblock The costs of payment uncertainty in healthcare markets.
\newblock {\em Federal Reserve Bank of San Francisco}.

\bibitem[Einav and Finkelstein, 2018]{einav_moral_2018}
Einav, L. and Finkelstein, A. (2018).
\newblock Moral hazard in health insurance: What we know and how we know it.
\newblock {\em Journal of the European Economic Association}, 16(4):957--982.

\bibitem[Einav et~al., 2013]{einav_selection_2013}
Einav, L., Finkelstein, A., Ryan, S.~P., Schrimpf, P., and Cullen, M.~R.
  (2013).
\newblock Selection on moral hazard in health insurance.
\newblock {\em American Economic Review}, 103(1):178--219.

\bibitem[Einav et~al., 2015]{einav_response_2015}
Einav, L., Finkelstein, A., and Schrimpf, P. (2015).
\newblock The {{Response}} of {{Drug Expenditure}} to {{Nonlinear Contract
  Design}}: {{Evidence}} from {{Medicare Part D}} *.
\newblock {\em The Quarterly Journal of Economics}, 130(2):841--899.

\bibitem[Fadlon and Nielsen, 2019]{fadlon_family_2019}
Fadlon, I. and Nielsen, T.~H. (2019).
\newblock Family {{Health Behaviors}}.
\newblock {\em American Economic Review}, 109(9):3162--3191.

\bibitem[Fadlon and Nielsen, 2021]{fadlon_family_2021}
Fadlon, I. and Nielsen, T.~H. (2021).
\newblock Family labor supply responses to severe health shocks: {{Evidence}}
  from {{Danish}} administrative records.
\newblock {\em American Economic Journal: Applied Economics}, 13(3):1--30.

\bibitem[Fronsdal et~al., 2020]{fronsdal_variation_2020}
Fronsdal, T.~L., Bhattacharya, J., and Tamang, S. (2020).
\newblock Variation in health care prices across public and private payers.
\newblock Technical report, {National Bureau of Economic Research}.

\bibitem[Fu et~al., 2021]{fu_out--pocket_2021}
Fu, S.~J., Rose, L., Dawes, A.~J., Knowlton, L.~M., Ruddy, K.~J., and Morris,
  A.~M. (2021).
\newblock Out-of-pocket costs among patients with a new cancer diagnosis
  enrolled in high-deductible health plans vs traditional insurance.
\newblock {\em JAMA network open}, 4(12):e2134282--e2134282.

\bibitem[Gaffney et~al., 2020]{gaffney_high-deductible_2020}
Gaffney, A., White, A., Hawks, L., Himmelstein, D., Woolhandler, S.,
  Christiani, D.~C., and McCormick, D. (2020).
\newblock High-deductible health plans and healthcare access, use, and
  financial strain in those with chronic obstructive pulmonary disease.
\newblock {\em Annals of the American Thoracic Society}, 17(1):49--56.

\bibitem[Ganguli et~al., 2020]{ganguli_assessment_2020}
Ganguli, I., Lupo, C., Mainor, A.~J., Wang, Q., Orav, E.~J., Rosenthal, M.~B.,
  Sequist, T.~D., and Colla, C.~H. (2020).
\newblock Assessment of prevalence and cost of care cascades after routine
  testing during the {{Medicare}} annual wellness visit.
\newblock {\em JAMA network open}, 3(12):e2029891--e2029891.

\bibitem[Geyman, 2012]{geyman_cost-sharing_2012}
Geyman, J.~P. (2012).
\newblock Cost-{{Sharing}} under {{Consumer-Driven Health Care Will Not Reform
  U}}.{{S}}. {{Health Care}}.
\newblock {\em The Journal of Law, Medicine \& Ethics}, 40(3):574--581.

\bibitem[Goldman and Philipson, 2007]{goldman_integrated_2007}
Goldman, D. and Philipson, T. (2007).
\newblock Integrated {{Insurance Design}} in the {{Presence}} of {{Multiple
  Medical Technologies}}.
\newblock {\em American Economic Review}, 97(2):427--432.

\bibitem[Gondi et~al., 2021]{gondi_early_2021-1}
Gondi, S., Beckman, A.~L., Ofoje, A.~A., Hinkes, P., and McWilliams, J.~M.
  (2021).
\newblock Early {{Hospital Compliance With Federal Requirements}} for {{Price
  Transparency}}.
\newblock {\em JAMA Internal Medicine}, 181(10):1396--1397.

\bibitem[Goodman-Bacon et~al., 2022]{goodman2022bacondecomp}
Goodman-Bacon, A., Goldring, T., and Nichols, A. (2022).
\newblock Bacondecomp: Stata module to perform a bacon decomposition of
  difference-in-differences estimation.

\bibitem[Gross et~al., 2022]{gross_liquidity_2022}
Gross, T., Layton, T.~J., and Prinz, D. (2022).
\newblock The liquidity sensitivity of healthcare consumption: {{Evidence}}
  from social security payments.
\newblock {\em American Economic Review: Insights}, 4(2):175--190.

\bibitem[Grubb, 2015]{grubb_consumer_2015}
Grubb, M.~D. (2015).
\newblock Consumer {{Inattention}} and {{Bill-Shock Regulation}}.
\newblock {\em The Review of Economic Studies}, 82(1):219--257.

\bibitem[Grubb and Osborne, 2015]{grubb_cellular_2015}
Grubb, M.~D. and Osborne, M. (2015).
\newblock Cellular {{Service Demand}}: {{Biased Beliefs}}, {{Learning}}, and
  {{Bill Shock}}.
\newblock {\em American Economic Review}, 105(1):234--271.

\bibitem[Gruber, 2022]{gruber_financing_2022}
Gruber, J. (2022).
\newblock Financing {{Health Care Delivery}}.
\newblock {\em Annual Review of Financial Economics}, 14(1).

\bibitem[Hansen, 2017]{hansen_truven_2017}
Hansen, L. (2017).
\newblock The {{Truven Health Marketscan Databases}} for {{Life Sciences
  Researchers}}.
\newblock White {{Paper}}, {Truven Health Analytics -- IBM Watson Health}.

\bibitem[Hartzema et~al., 2011]{hartzema_utilizing_2011}
Hartzema, A.~G., Racoosin, J.~A., MaCurdy, T.~E., Gibbs, J.~M., and Kelman,
  J.~A. (2011).
\newblock Utilizing {{Medicare}} claims data for real-time drug safety
  evaluations: Is it feasible?
\newblock {\em Pharmacoepidemiology and Drug Safety}, 20(7):684--688.

\bibitem[Hettinger, 2022]{hettinger_intertemporal_2022}
Hettinger, K. (2022).
\newblock Intertemporal substitution in response to non-linear health insurance
  contracts.
\newblock {\em Working Paper}.

\bibitem[Hoagland, 2022]{hoagland_ounce_2022}
Hoagland, A. (2022).
\newblock An ounce of prevention or a pound of cure? {{The}} value of health
  risk information.
\newblock {\em Working Paper}.

\bibitem[Hoagland and Shafer, 2021]{hoagland_out--pocket_2021}
Hoagland, A. and Shafer, P. (2021).
\newblock Out-of-pocket costs for preventive care persist almost a decade after
  the {{Affordable Care Act}}.
\newblock {\em Preventive Medicine}, 150:106690.

\bibitem[Ippolito and Vabson, 2023]{ippolito_how_2023}
Ippolito, B. and Vabson, B. (2023).
\newblock How {{Should Policymakers Respond To Growth In Cost Sharing That
  Often Goes Unpaid}}?
\newblock {\em Health Affairs Forefront}.

\bibitem[Johansson et~al., 2023]{johansson_reductions_2023}
Johansson, N., Sonja, C., Kunz, J.~S., Petrie, D., and Svensson, M. (2023).
\newblock Reductions in out-of-pocket prices and forward-looking moral hazard
  in health care demand.
\newblock {\em Journal of health economics}, 87:102710.

\bibitem[Karlsson et~al., 2009]{karlsson_ostrich_2009}
Karlsson, N., Loewenstein, G., and Seppi, D. (2009).
\newblock The ostrich effect: {{Selective}} attention to information.
\newblock {\em Journal of Risk and uncertainty}, 38(2):95--115.

\bibitem[Klein et~al., 2022]{klein_response_2022}
Klein, T.~J., Salm, M., and Upadhyay, S. (2022).
\newblock The response to dynamic incentives in insurance contracts with a
  deductible: {{Evidence}} from a differences-in-regression-discontinuities
  design.
\newblock {\em Journal of Public Economics}, 210:104660.

\bibitem[Kowalski, 2016a]{kowalski_censored_2016-1}
Kowalski, A. (2016a).
\newblock Censored quantile instrumental variable estimates of the price
  elasticity of expenditure on medical care.
\newblock {\em Journal of Business \& Economic Statistics}, 34(1):107--117.

\bibitem[Kowalski, 2016b]{kowalski_censored_2016}
Kowalski, A. (2016b).
\newblock Censored {{Quantile Instrumental Variable Estimates}} of the {{Price
  Elasticity}} of {{Expenditure}} on {{Medical Care}}.
\newblock {\em Journal of Business \& Economic Statistics}, 34(1):107--117.

\bibitem[Kullgren et~al., 2010]{kullgren_health_2010}
Kullgren, J.~T., Galbraith, A.~A., Hinrichsen, V.~L., Miroshnik, I., Penfold,
  R.~B., Rosenthal, M.~B., Landon, B.~E., and Lieu, T.~A. (2010).
\newblock Health care use and decision making among lower-income families in
  high-deductible health plans.
\newblock {\em Archives of internal medicine}, 170(21):1918--1925.

\bibitem[League, 2022]{league_administrative_2022}
League, R. (2022).
\newblock Administrative burden and consolidation in health care: {{Evidence}}
  from medicare contractor transitions.
\newblock Technical report, {Technical report}.

\bibitem[Lieber, 2017]{lieber_does_2017}
Lieber, E.~M. (2017).
\newblock Does it pay to know prices in health care?
\newblock {\em American Economic Journal: Economic Policy}, 9(1):154--79.

\bibitem[Marone and Sabety, 2022]{marone_should_2022}
Marone, V. and Sabety, A. (2022).
\newblock Should there be vertical choice in health insurance markets?
\newblock {\em American Economic Review}, 112(1):304--342.

\bibitem[McMorrow et~al., 2014]{mcmorrow_determinants_2014}
McMorrow, S., Kenney, G.~M., and Goin, D. (2014).
\newblock Determinants of receipt of recommended preventive services:
  Implications for the {{Affordable Care Act}}.
\newblock {\em American Journal of Public Health}, 104(12):2392--2399.

\bibitem[Mehrotra et~al., 2017]{mehrotra_americans_2017}
Mehrotra, A., Dean, K.~M., Sinaiko, A.~D., and Sood, N. (2017).
\newblock Americans {{Support Price Shopping For Health Care}}, {{But Few
  Actually Seek Out Price Information}}.
\newblock {\em Health Affairs}, 36(8):1392--1400.

\bibitem[Meyler et~al., 2007]{meyler2007health}
Meyler, D., Stimpson, J.~P., and Peek, M.~K. (2007).
\newblock Health concordance within couples: a systematic review.
\newblock {\em Social science \& medicine}, 64(11):2297--2310.

\bibitem[Mohama, 2021]{mohama_more_2021}
Mohama, A. (2021).
\newblock More hospitals want patients to pay in advance. {{Is}} that radical
  transparency\textemdash or unfair to patients?
\newblock {\em Advisory Board}.

\bibitem[Muir et~al., 2012]{muir_clarifying_2012}
Muir, M.~A., Alessi, S.~A., and King, J.~S. (2012).
\newblock Clarifying costs: {{Can}} increased price transparency reduce
  healthcare spending.
\newblock {\em William \& Mary Policy Review}, 4:319.

\bibitem[Mullahy and Norton, 2022]{mullahy_why_2022}
Mullahy, J. and Norton, E.~C. (2022).
\newblock Why {{Transform Y}}? {{A Critical Assessment}} of
  {{Dependent-Variable Transformations}} in {{Regression Models}} for
  {{Skewed}} and {{Sometimes-Zero Outcomes}}.
\newblock {\em NBER Working Paper}.

\bibitem[Naqvi, 2021]{naqvi_schemepack_2021}
Naqvi, S. A.~A. (2021).
\newblock {{SCHEMEPACK}}: {{Stata}} module providing ready-to-use graph
  schemes.
\newblock Boston College Department of Economics.

\bibitem[Ngangoue, 2021]{ngangoue_learning_2021}
Ngangoue, M.~K. (2021).
\newblock Learning {{From Unrealized}} versus {{Realized Prices}}.
\newblock {\em American Economic Journal: Microeconomics}, 13(2):174--201.

\bibitem[Orszag and Rekhi, 2020]{orszag_real-time_2020}
Orszag, P. and Rekhi, R. (2020).
\newblock Real-{{Time Adjudication}} for {{Health Insurance Claims}}.
\newblock Technical report, {1\% Steps for Health Care Reform}.

\bibitem[Patel et~al., 2023]{patel_financial_2023}
Patel, N., Moniz, M., Dalton, V., Fendrick, A.~M., and Horn{\'y}, M. (2023).
\newblock The financial risk for privately insured patients in the era of
  health care price transparency.
\newblock {\em Working Paper}.

\bibitem[Pauly and Blavin, 2008]{pauly_moral_2008}
Pauly, M.~V. and Blavin, F.~E. (2008).
\newblock Moral hazard in insurance, value-based cost sharing, and the benefits
  of blissful ignorance.
\newblock {\em Journal of Health Economics}, 27(6):1407--1417.

\bibitem[Peng, 2005]{peng_learning_2005}
Peng, L. (2005).
\newblock Learning with information capacity constraints.
\newblock {\em Journal of Financial and Quantitative Analysis}, 40(2):307--329.

\bibitem[Reed et~al., 2005]{reed_care-seeking_2005}
Reed, M., Fung, V., Brand, R., Fireman, B., Newhouse, J.~P., Selby, J.~V., and
  Hsu, J. (2005).
\newblock Care-{{Seeking Behavior}} in {{Response}} to {{Emergency Department
  Copayments}}.
\newblock {\em Medical Care}, 43(8):810--816.

\bibitem[Santos~Silva and Tenreyro, 2006]{santos_silva_log_2006}
Santos~Silva, J. M.~C. and Tenreyro, S. (2006).
\newblock The log of gravity.
\newblock {\em The Review of Economics and Statistics}, 88(4):641--658.

\bibitem[Shafer et~al., 2021]{shafer_trends_2021}
Shafer, P.~R., Hoagland, A., and Hsu, H.~E. (2021).
\newblock Trends in {{Well-Child Visits With Out-of-Pocket Costs}} in the {{US
  Before}} and {{After}} the {{Affordable Care Act}}.
\newblock {\em JAMA Network Open}, 4(3):e211248.

\bibitem[Shi, 2022]{shi_monitoring_2022}
Shi, M. (2022).
\newblock Monitoring for waste: {{Evidence}} from medicare audits.
\newblock {\em URL https://mshi311. github.
  io/website2/Shi\_MedicareAudits\_2022\_09\_15. pdf}.

\bibitem[Shukla et~al., 2021]{shukla_retrospective_2021}
Shukla, D., Patel, S., Clack, L., Smith, T.~B., and Shuler, M.~S. (2021).
\newblock Retrospective analysis of trends in surgery volumes between 2016 and
  2019 and impact of the insurance deductible: {{Cross-sectional}} study.
\newblock {\em Annals of Medicine and Surgery}, 63:102176.

\bibitem[Siegfried et~al., 2019]{siegfried_adult_2019}
Siegfried, I., Jacobs, J., and Olympia, R.~P. (2019).
\newblock Adult emergency department referrals from urgent care centers.
\newblock {\em The American Journal of Emergency Medicine}, 37(10):1949--1954.

\bibitem[Snowbeck, 2022]{snowbeck_risk_2022}
Snowbeck, C. (2022).
\newblock Risk adjustment, claims processing woes still troubling {{Bright
  Health Group}}.
\newblock {\em Star Tribune}.

\bibitem[Teutsch et~al., 2020]{teutsch_health_2020}
Teutsch, S., Carey, T.~S., and Pignone, M.~P. (2020).
\newblock Health equity in preventive services: The role of primary care.
\newblock {\em American Family Physician}, 102(5):264--265.

\bibitem[Webb~Hooper et~al., 2019]{webb_hooper_understanding_2019}
Webb~Hooper, M., Mitchell, C., Marshall, V.~J., Cheatham, C., Austin, K.,
  Sanders, K., Krishnamurthi, S., and Grafton, L.~L. (2019).
\newblock Understanding multilevel factors related to urban community trust in
  healthcare and research.
\newblock {\em International Journal of Environmental Research and Public
  Health}, 16(18):3280.

\bibitem[White and Eguchi, 2014]{white_reference_2014}
White, C. and Eguchi, M. (2014).
\newblock Reference pricing: {{A}} small piece of the health care price and
  quality puzzle.
\newblock {\em National Institute for Health Care Reform}.

\bibitem[Wooldridge, 1999]{wooldridge_distribution-free_1999}
Wooldridge, J.~M. (1999).
\newblock Distribution-free estimation of some nonlinear panel data models.
\newblock {\em Journal of Econometrics}, 90(1):77--97.

\bibitem[Wooldridge, 2021]{wooldridge2021two}
Wooldridge, J.~M. (2021).
\newblock Two-way fixed effects, the two-way mundlak regression, and
  difference-in-differences estimators.
\newblock {\em Available at SSRN 3906345}.

\bibitem[Zhang et~al., 2020]{zhang_impact_2020}
Zhang, A., Prang, K.-H., Devlin, N., Scott, A., and Kelaher, M. (2020).
\newblock The impact of price transparency on consumers and providers: A
  scoping review.
\newblock {\em Health Policy}, 124(8):819--825.

\bibitem[Zhang et~al., 2018]{zhang_does_2018}
Zhang, X., Haviland, A., Mehrotra, A., Huckfeldt, P., Wagner, Z., and Sood, N.
  (2018).
\newblock Does enrollment in high-deductible health plans encourage price
  shopping?
\newblock {\em Health Services Research}, 53(S1):2718--2734.

\end{thebibliography}

\clearpage

\appendix

\section{Appendix}\label{apdx:results}
\counterwithin{figure}{section}
\counterwithin{table}{section}
\setcounter{figure}{0}
\setcounter{table}{0}

\begin{figure}[htbp]
    \caption{Variation in Prices for CPT 59400: Routine Vaginal Delivery}
    \label{figax:price-hist}
    \centering

	\subfloat[Total Costs]{
	    \includegraphics[width=3.0in]{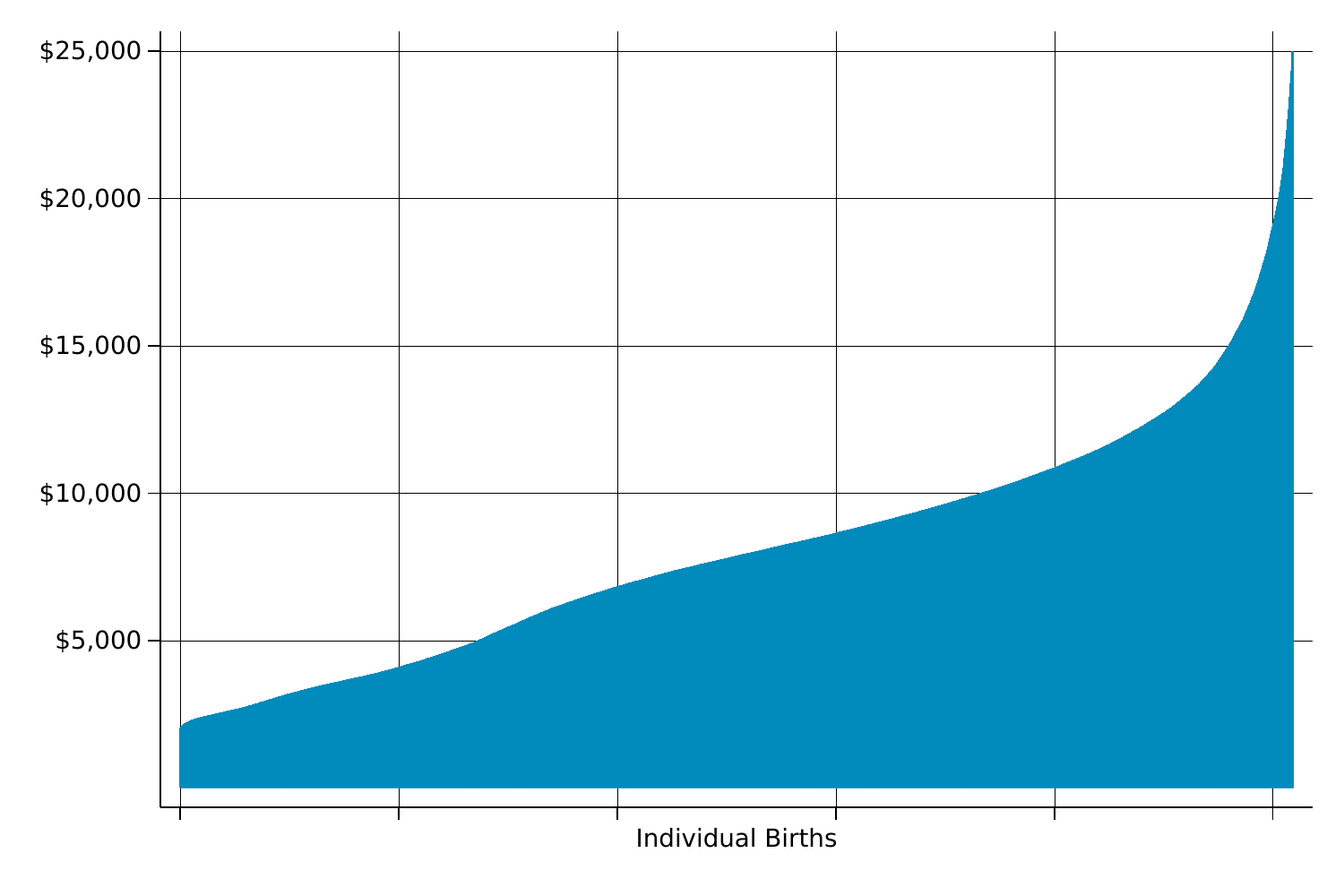}
	 }
	\subfloat[OOP Costs]{
	    \includegraphics[width=3.0in]{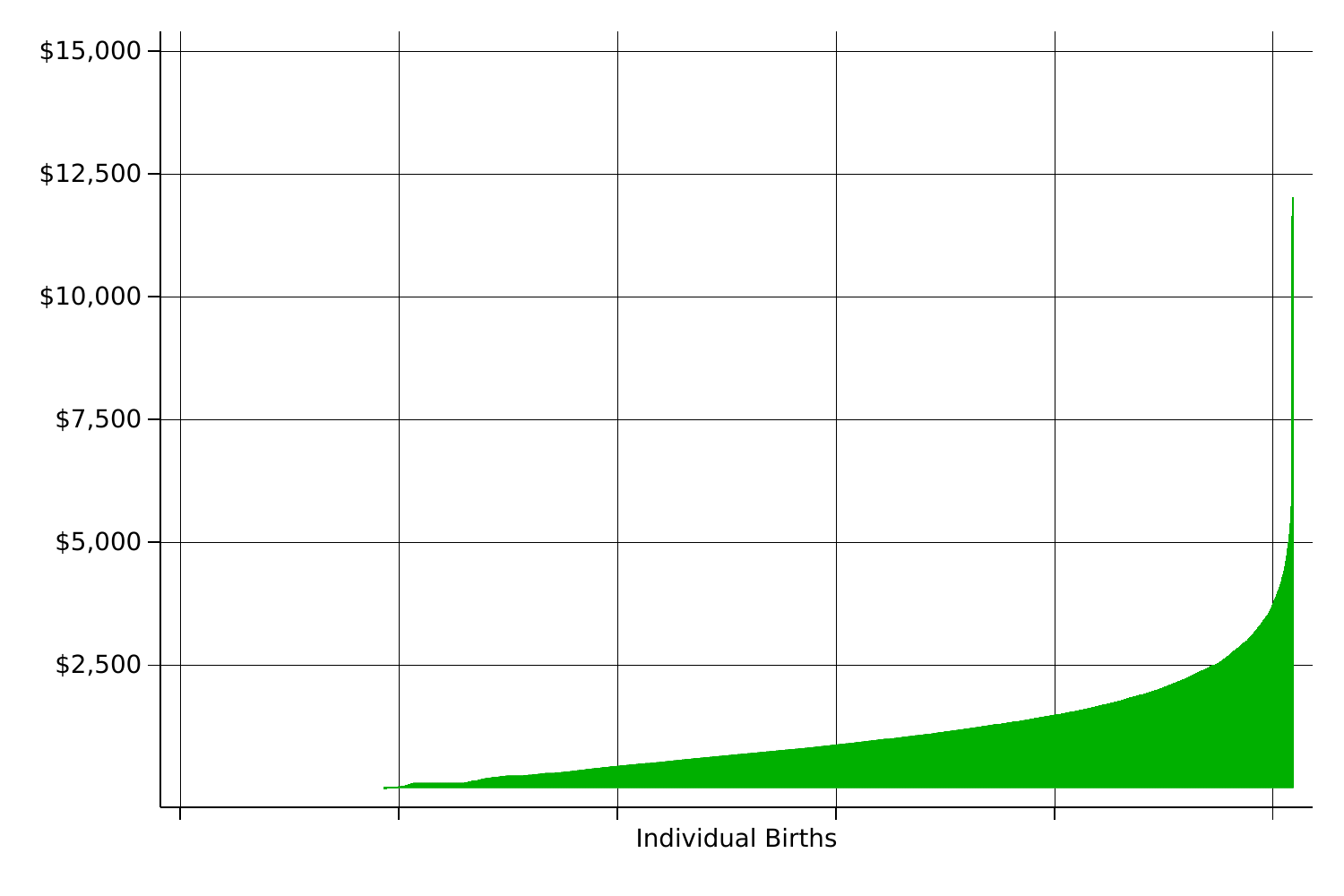}
	 }
    \vspace{0.2cm}
    
    \begin{minipage}{0.95\textwidth} 
	{\footnotesize \textit{Notes}: Figures show variation in total and OOP costs associated with CPT code 59400: ``Routine obstetric care including antepartum care, vaginal delivery (with or without episiotomy, and/or forceps) and postpartum care." Each vertical bar represents a unique encounter in our analytical data set, with the height of the bar corresponding to the price (all measured in 2022 USD). 
	\par
	}
	\end{minipage}
\end{figure}
\clearpage

\begin{figure}[htbp]
    \caption{Variation in Wait Times for Bills} 
    \label{figax:waittimes-months}
    \centering

	\subfloat[All Services]{
	    \includegraphics[width=3.0in]{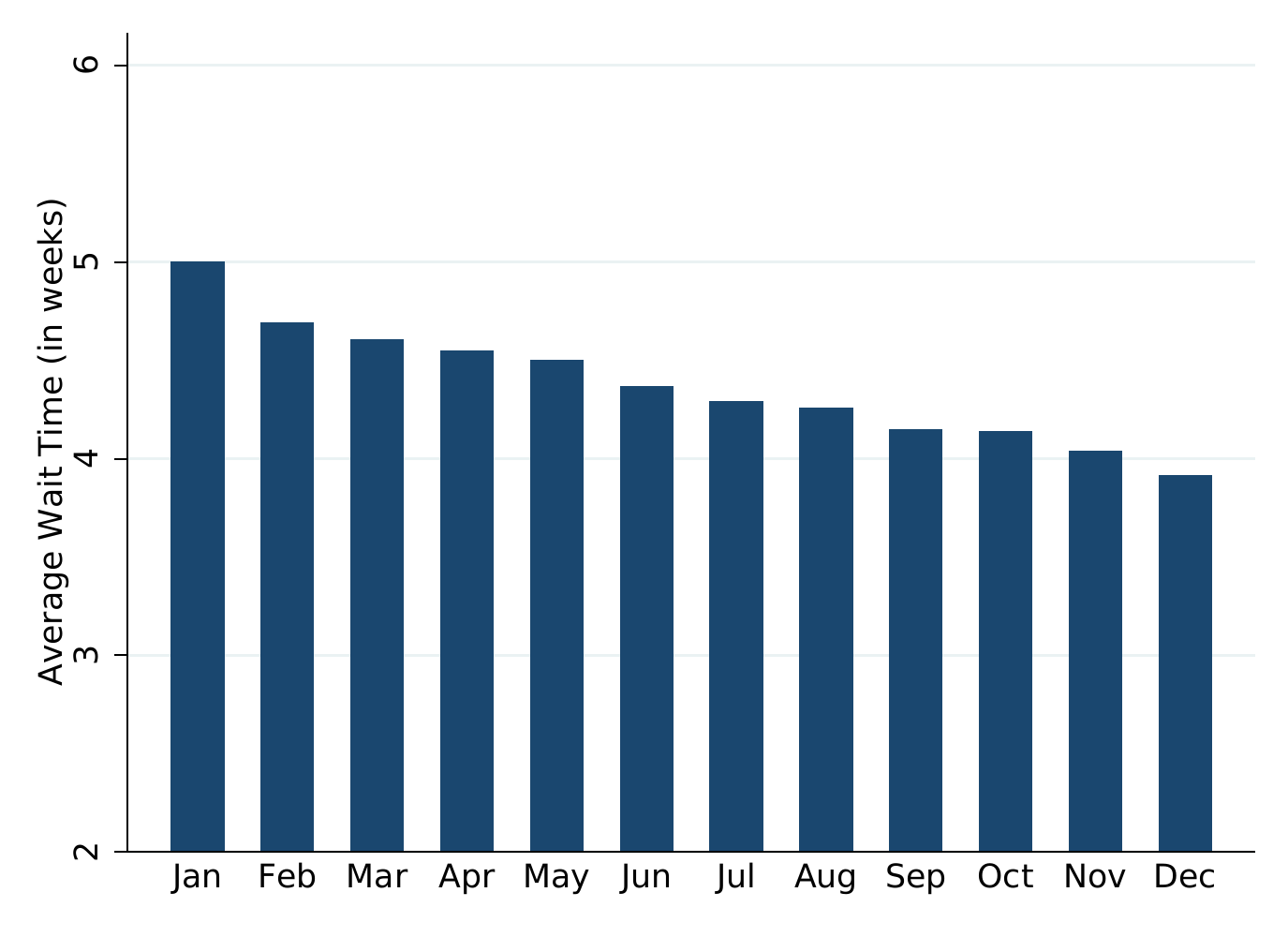}
	 }
	\subfloat[CMS Shoppable Services]{
	    \includegraphics[width=3.0in]{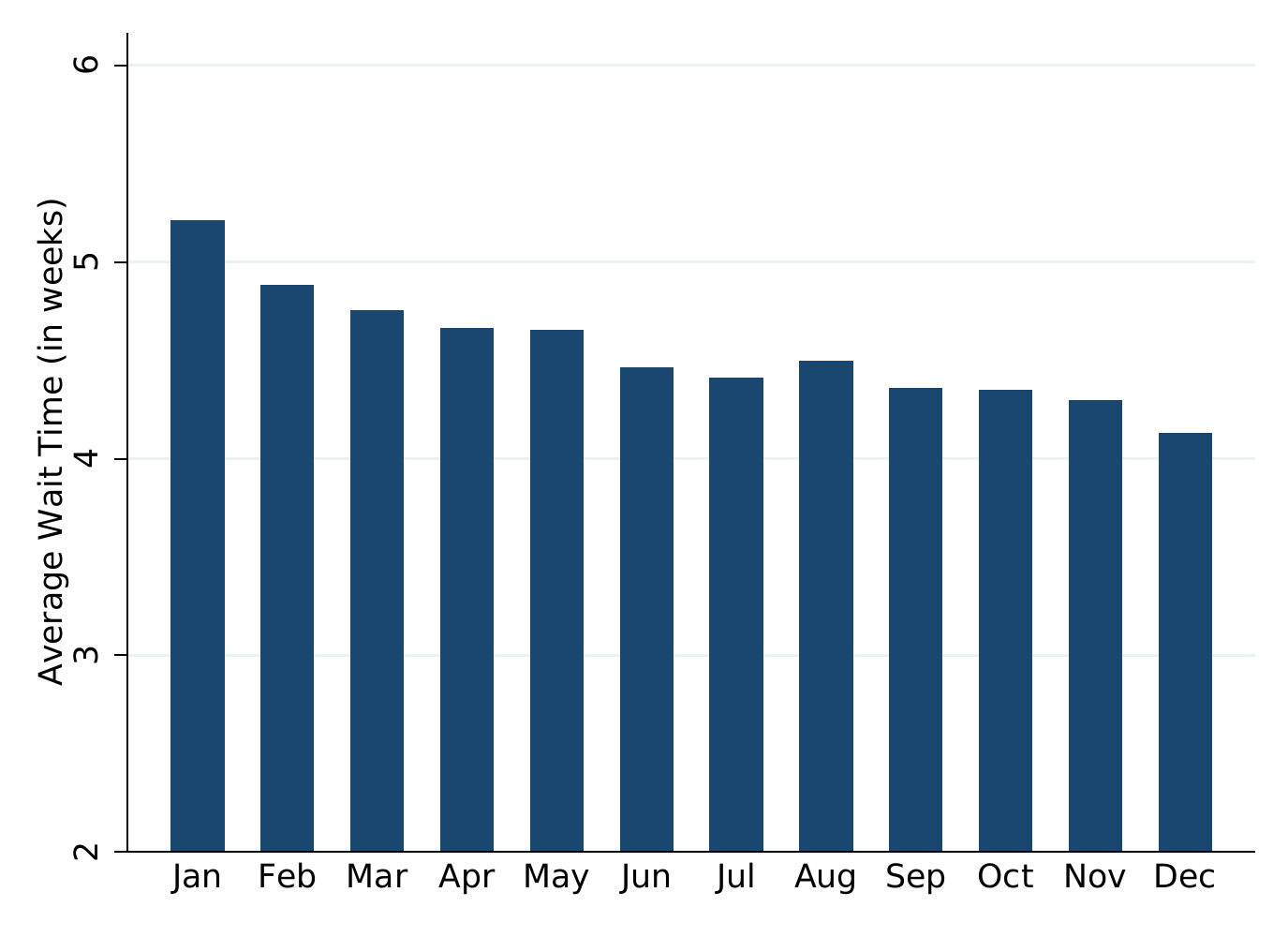}
	 }
    \vspace{0.2cm}
    
    \begin{minipage}{0.95\textwidth} 
	{\footnotesize \textit{Notes}: Indicates average wait time (in weeks) between date of service and date the insurer paid their portion of the claim (the earliest date at which definitive OOP information is known). Panel (a) illustrates variation in average wait times across months of the year (pooled across all years) for all claims in the analytical data; panel (b) limits the sample to only the shoppable services used as major health events in the text. 
	\par
	}
	\end{minipage}
\end{figure}
\clearpage

\begin{figure}[htbp]
    \caption{Difference between Realized and Predicted Wait Times for Bills}
    \label{axfig:billshock-desc}
    \centering

	\subfloat[Histogram]{
	    \includegraphics[width=3.0in]{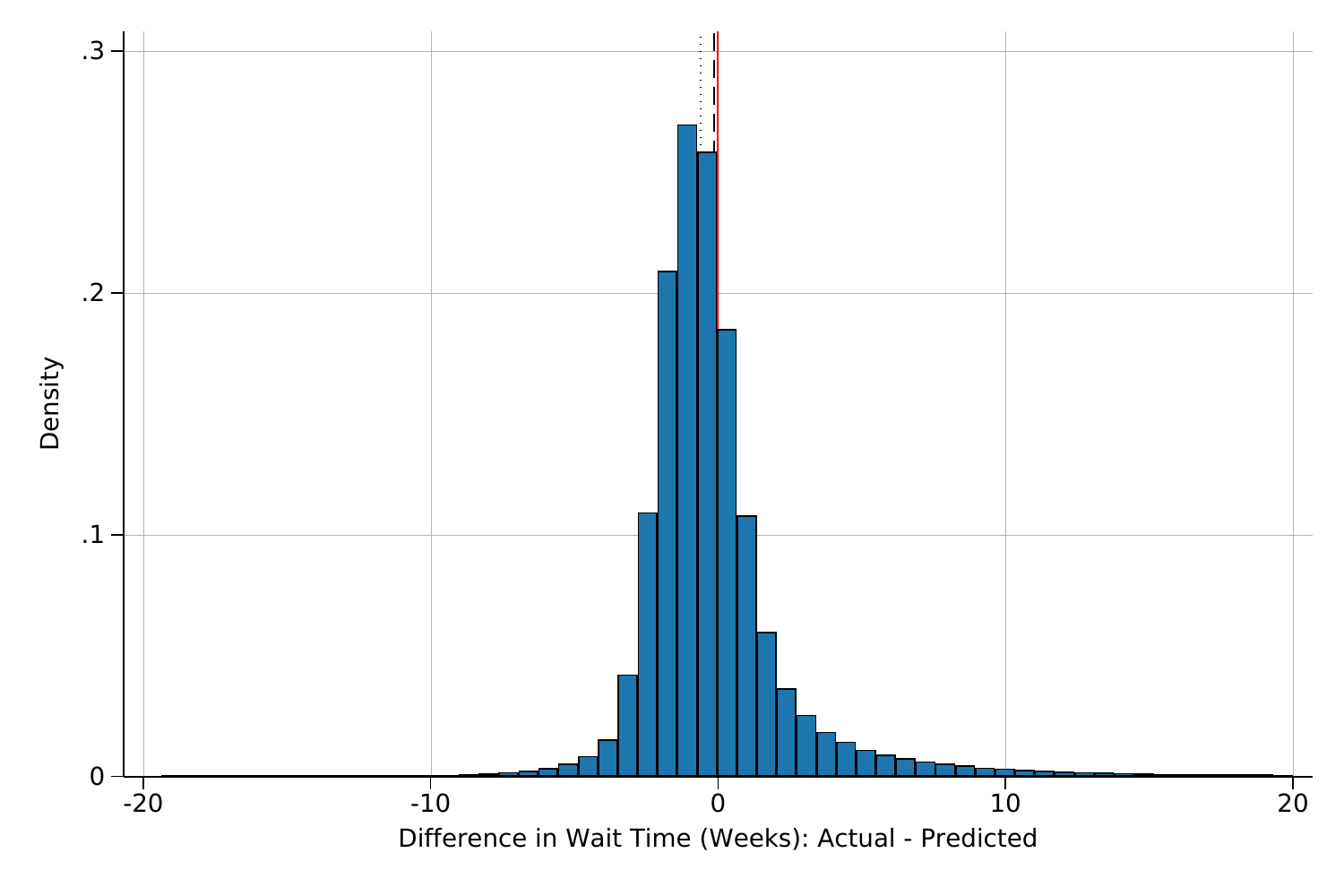}
	 }
	\subfloat[Binscatter]{
	    \includegraphics[width=3.0in]{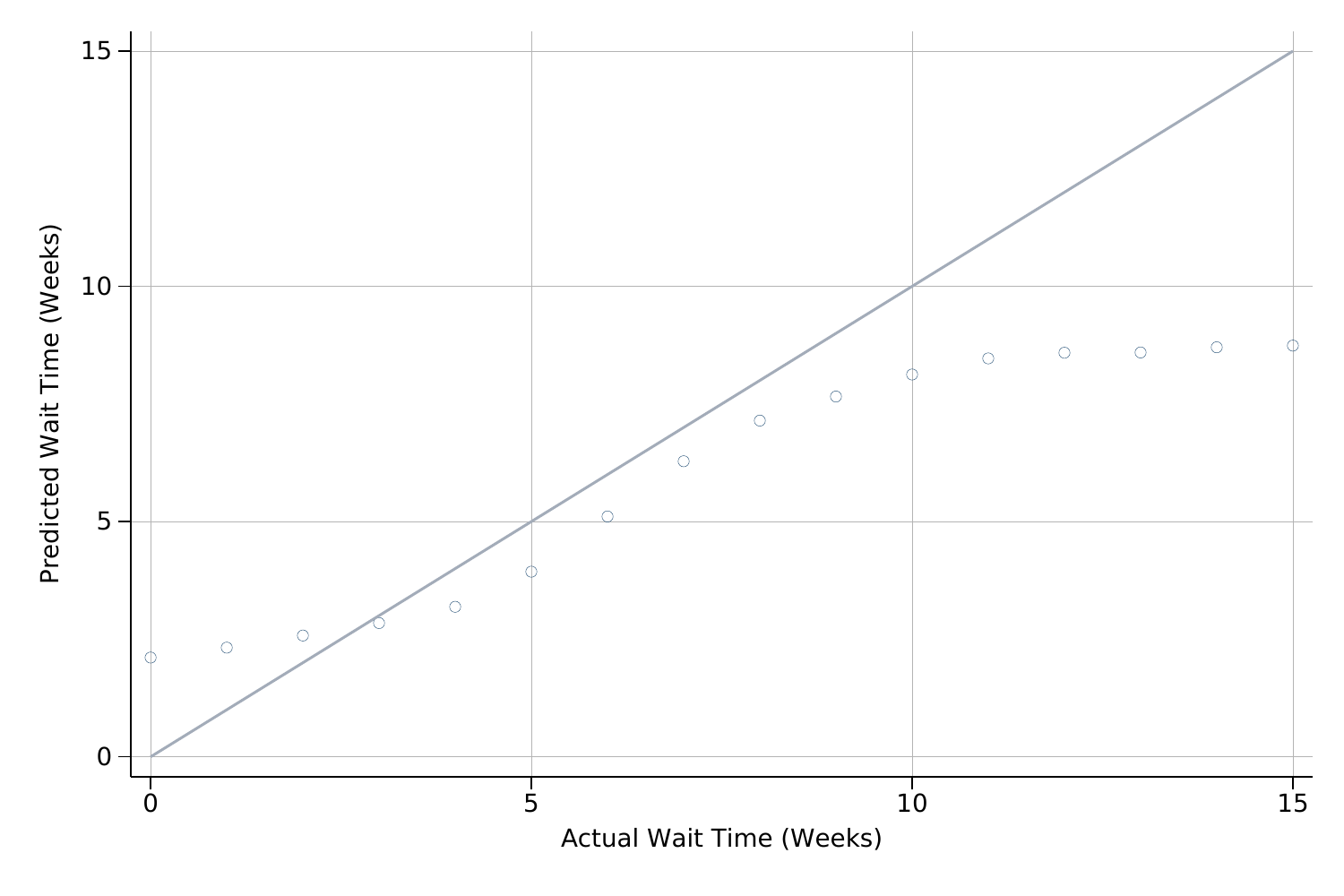}
	 }
    \vspace{0.2cm}
    
    \begin{minipage}{0.95\textwidth} 
	{\footnotesize \textit{Notes}: Figures each show differences between realized wait times (measured as the weeks waited between a service date and the claim paid date) and predicted wait times across all shoppable services in analytic data. Predicted wait times use regression on year, week of year, service type, and provider fixed effects. Panel (a) shows differences as a histogram, while panel (b) shows them as a binscatter. 
	\par
	}
	\end{minipage}
\end{figure}
\clearpage

\begin{figure}[htpb]
    \caption{Relationship between Bill Delay and Total/OOP Cost for Shoppable Services}
    \label{axfig:billbalance}
    \centering
    
    \subfloat[Bill Total Cost]{
	    \includegraphics[width=3.0in]{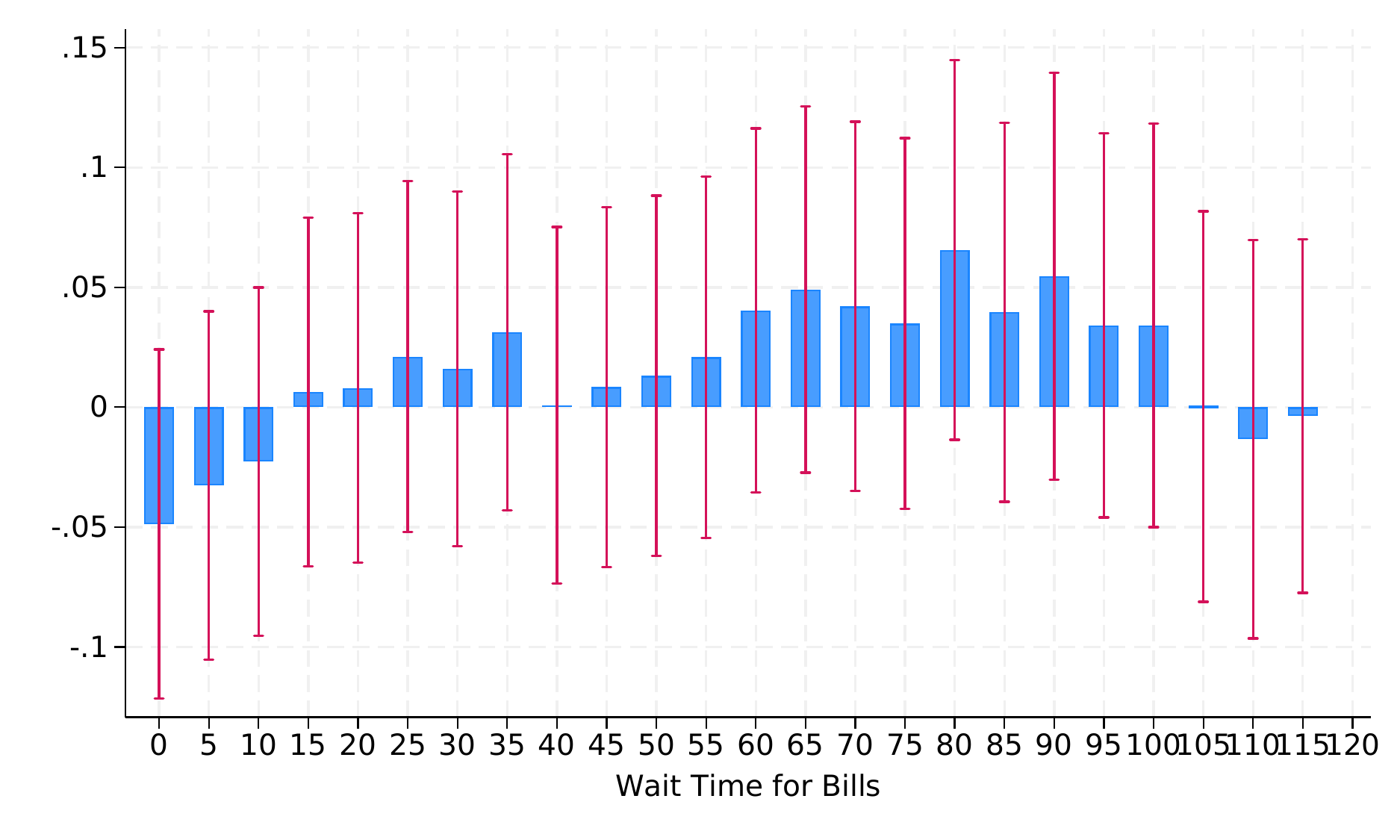}
	 }
	\subfloat[Bill OOP Cost]{
	    \includegraphics[width=3.0in]{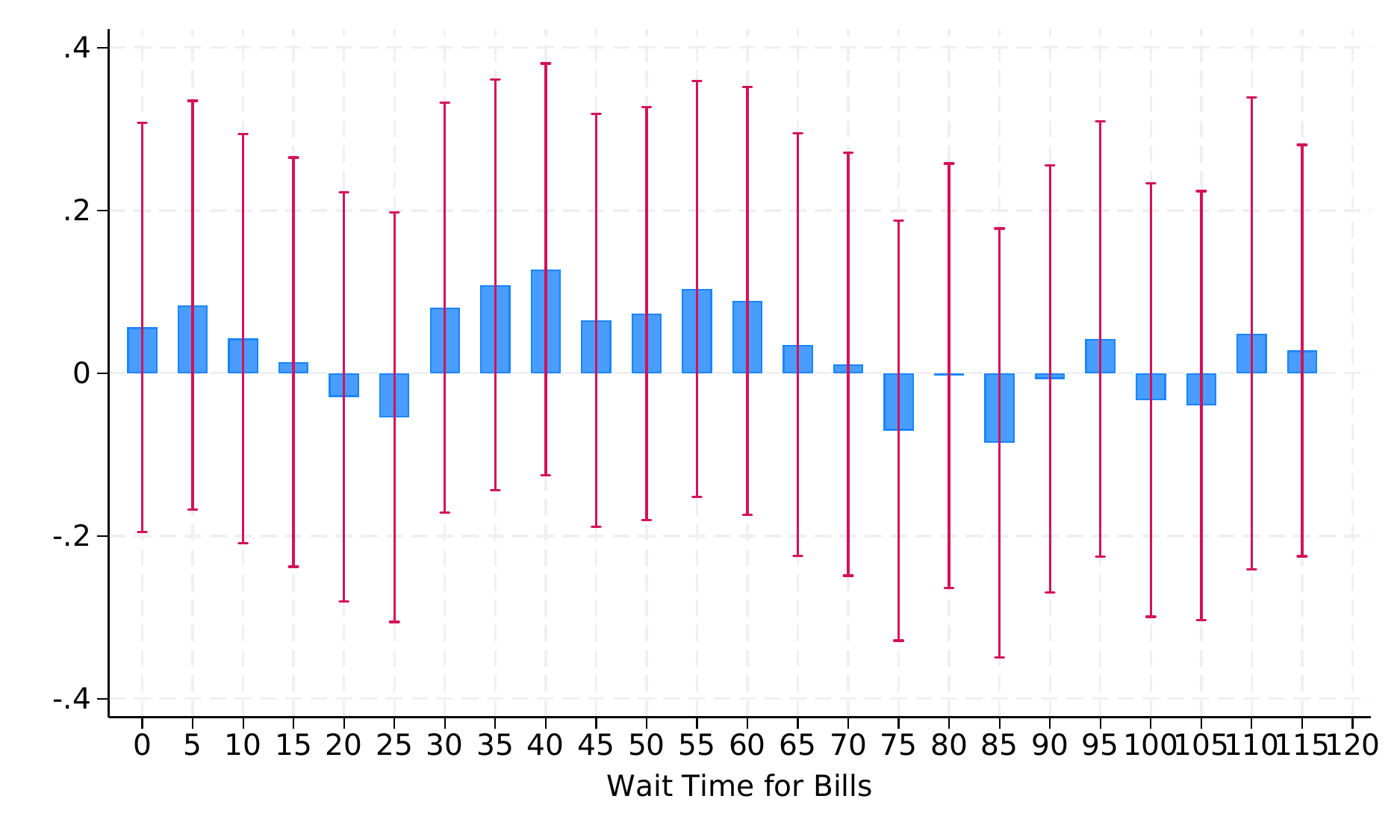}
	 }
    \vspace{0.2cm}

    \begin{minipage}{0.95\textwidth} 
	{\footnotesize \textit{Notes}: Figure shows estimated regression coefficients from Poisson regressions relating the overall spending associated with a shoppable service to bins for wait times (in days) for the bill to arrive. Regressions include fixed effects for year and week-of-year of shoppable service as well as service types and physicians. Panel (a) shows total billed cost, while panel (b) uses OOP cost only as the outcome variable. Wait times longer than 120 days ($<$2.5\% of sample) are included in the final bin. 95\% confidence intervals are shown (with unclustered standard errors). Results presented here are robust to varying bin width or to scaling outcome variables by pre-event levels (e.g., OOP/(pre-event OOP)).
	\par
	}
	\end{minipage}
\end{figure}

\clearpage 

\begin{figure}[htbp]
    \caption{Examining Correlations Between Waiting Times and Service Type/Cost}
    \label{axfig:corr-wait-cost}
    \centering
    
    \subfloat[All Services: Total Payments]{
	    \includegraphics[width=3.0in]{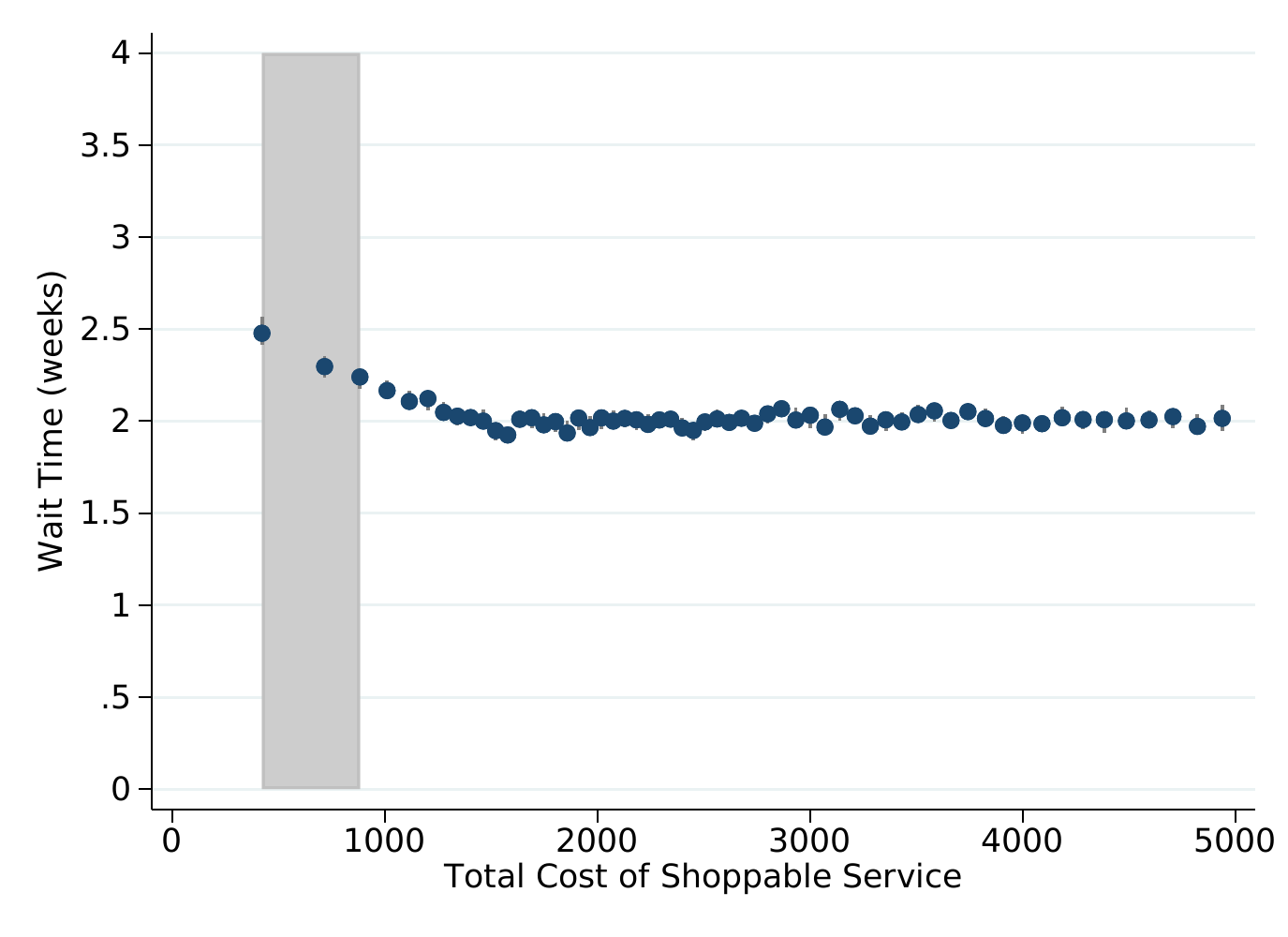}
	 }
	\subfloat[All Services: OOP]{
	    \includegraphics[width=3.0in]{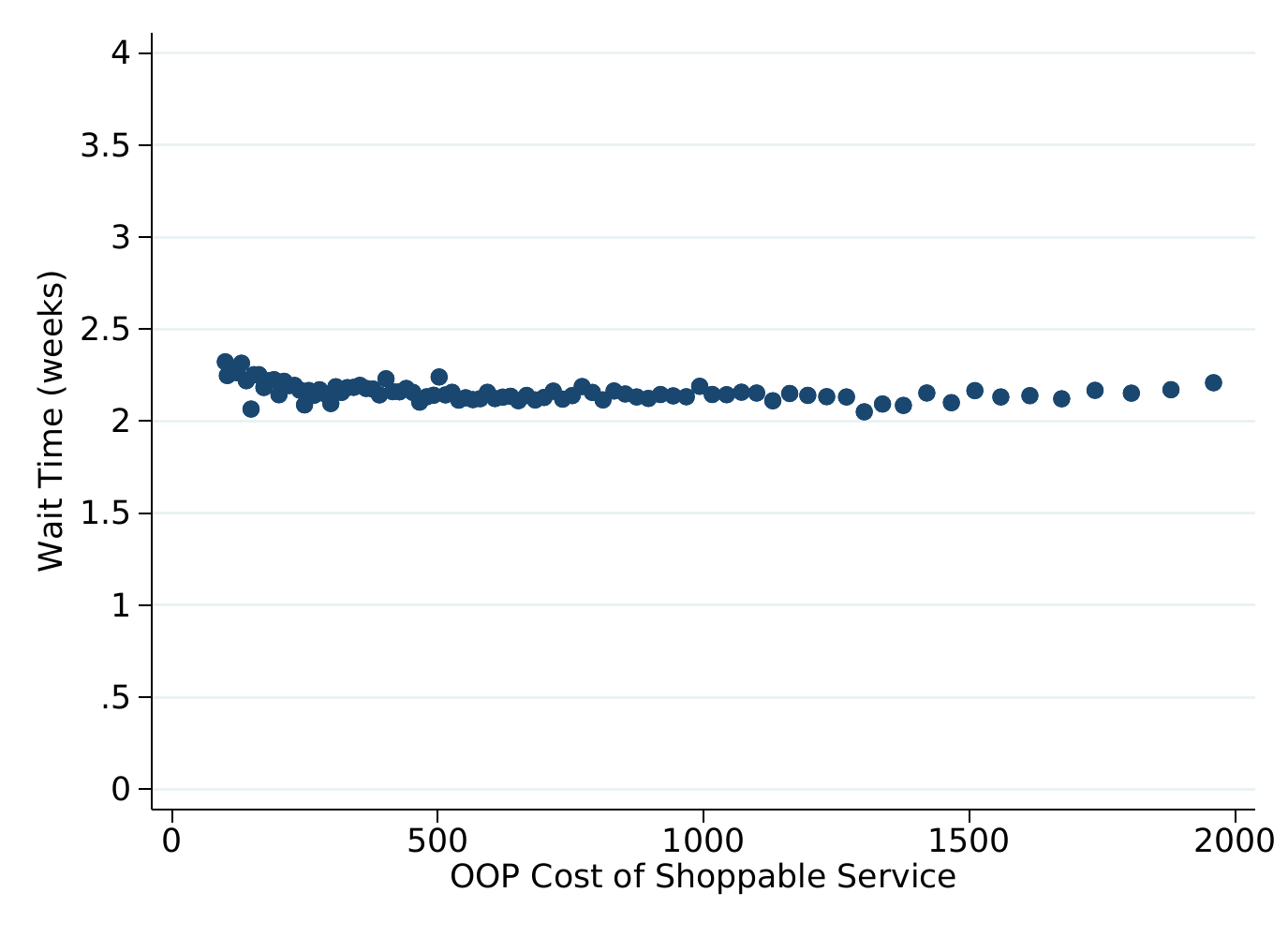}
	 }
    \vspace{0.2cm}
        \subfloat[Selected Services: Total Payments]{
	    \includegraphics[width=3.0in]{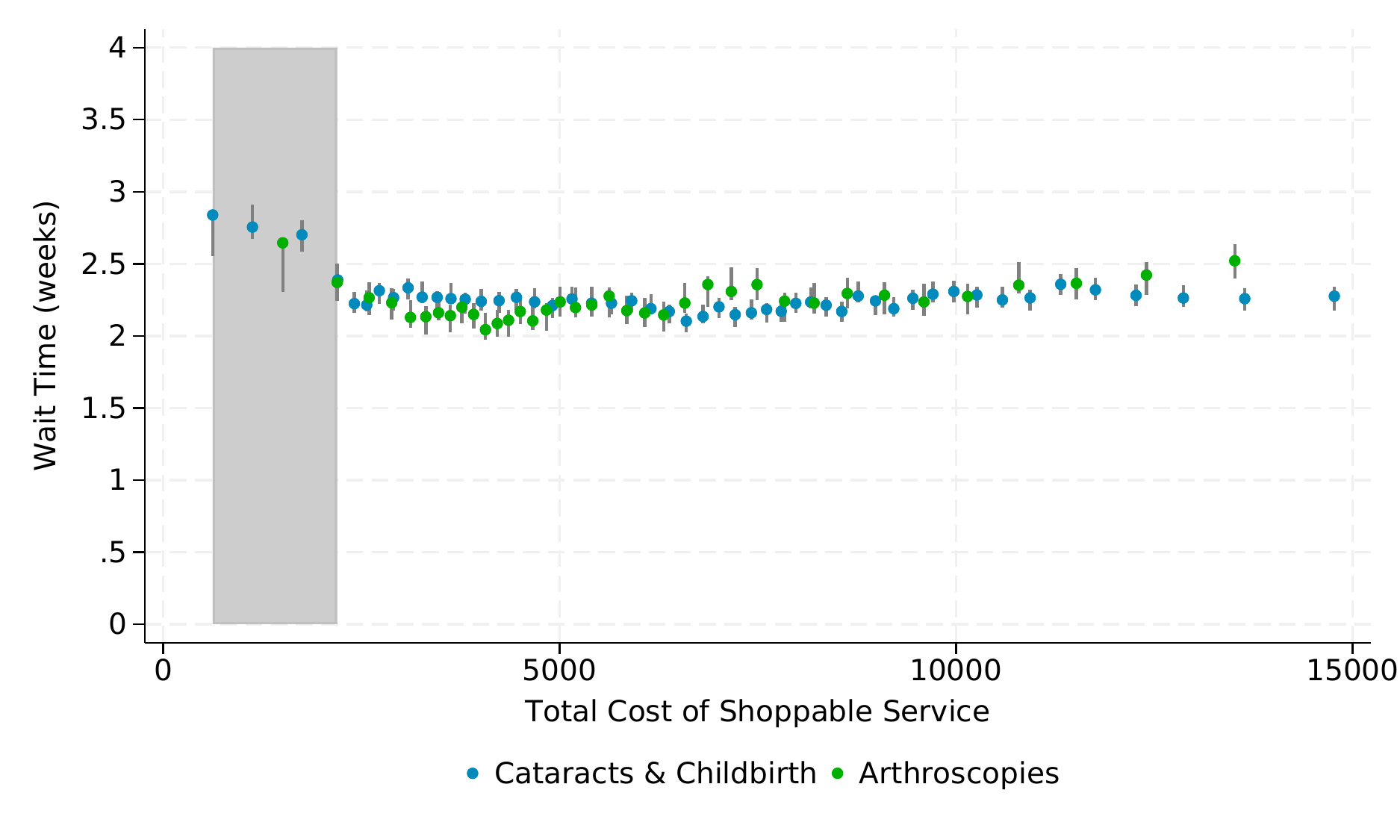}
	 }
	\subfloat[Selected Services: OOP]{
	    \includegraphics[width=3.0in]{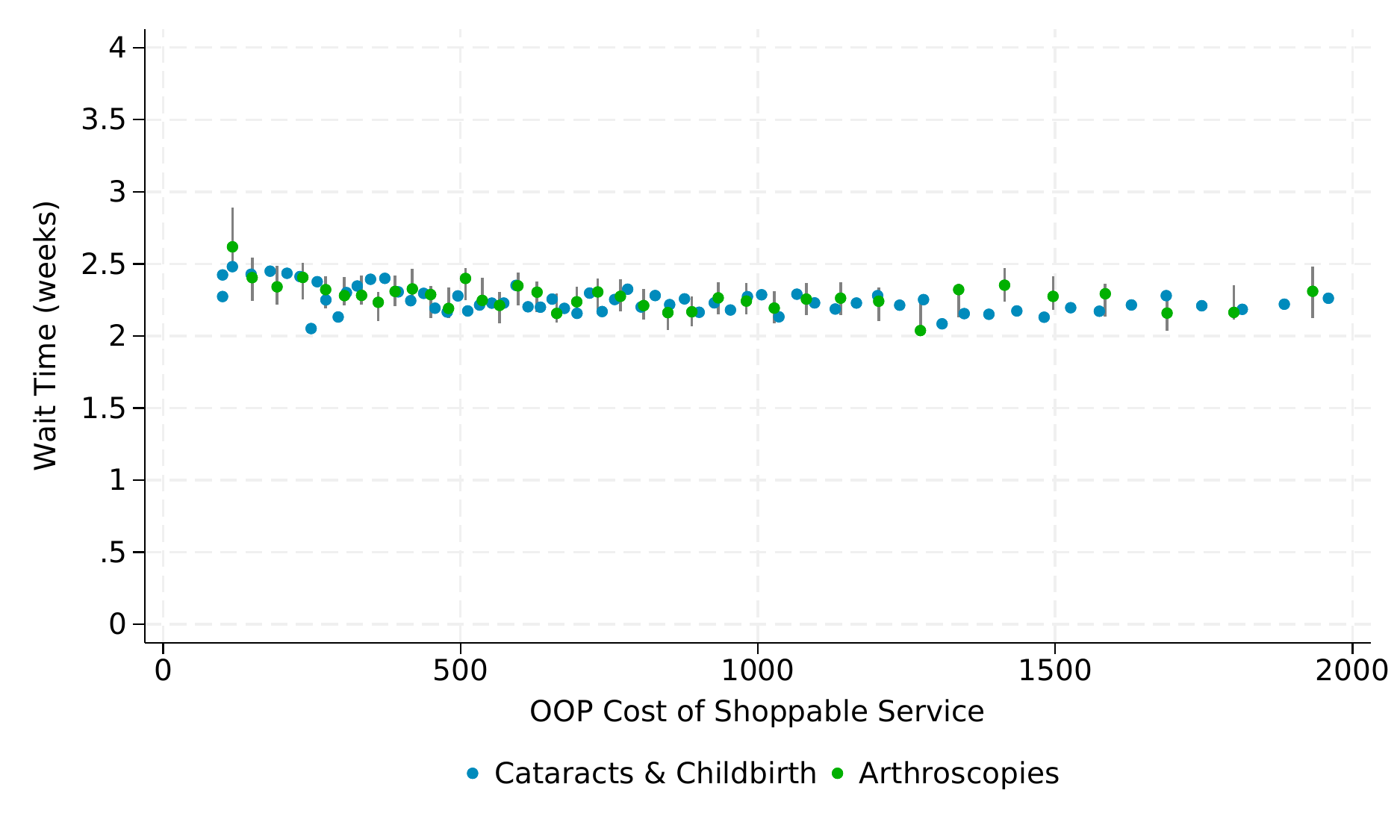}
	 }
    \vspace{0.2cm}

    \begin{minipage}{0.95\textwidth} 
	{\footnotesize \textit{Notes}: Figures show binscatter regressions (using the ``binsreg" package in Stata) between the size of a bill and the time elapsed between service and plan paid date. Selected services shown in the second row are those which we estimated to have the largest cost differences (in absolute value) between bills arriving under 30 days and those exceeding a month. These include arthroscopies (which had the largest positive differences) and cataract removals and vaginal deliveries (which had the largest negative differences). Panels (a) and (b) adjust for provider, procedure type, and year fixed effects; panels (c) and (d) adjust for provider and year fixed effects only. 
	\par
	}
	\end{minipage}
\end{figure}

\clearpage


\begin{figure}[htbp]
    \caption{Distribution of Placebo Regression Coefficients for $\beta_{\text{post\_bill}}$}
    \label{fig:placebo}
    \centering

    \includegraphics[scale=0.9]{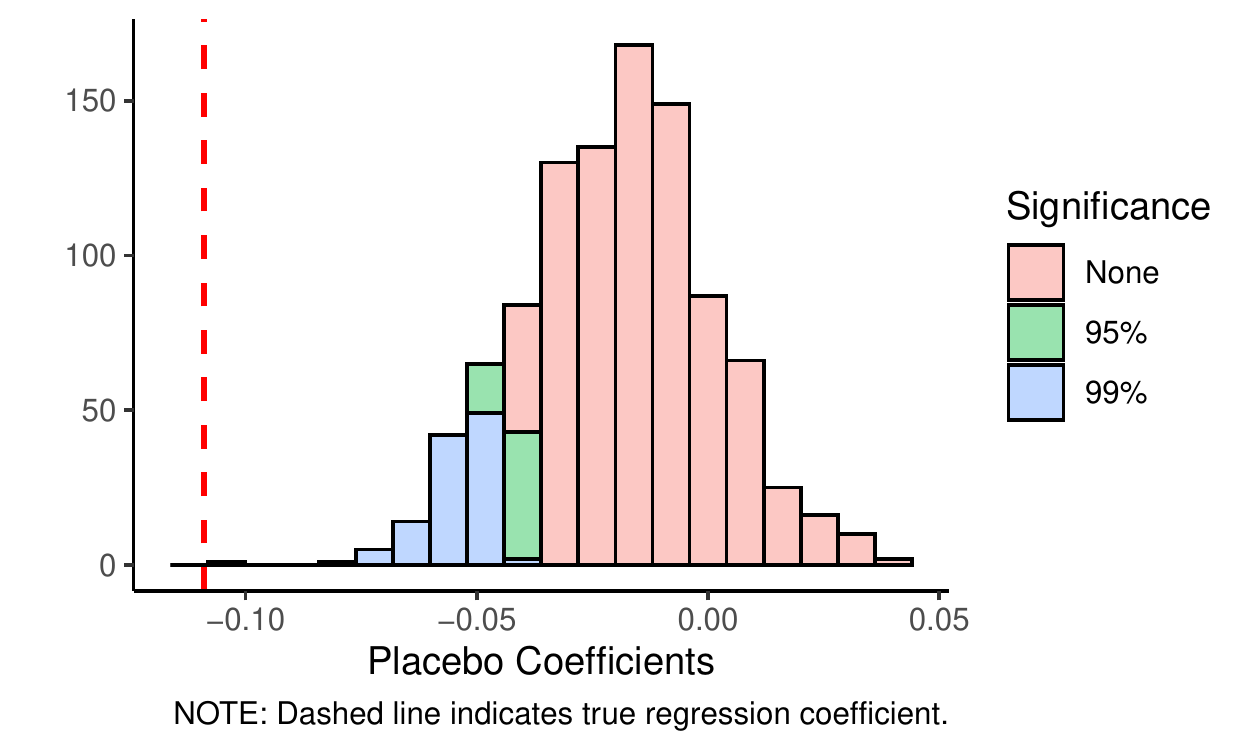}
    \vspace{0.2cm}
    \begin{minipage}{0.95\textwidth} 
	{\footnotesize \textit{Notes}: Figure shows the distribution of placebo regression coefficients for the dummy variable $\text{Post\_Bill}_{it}$ in Equation \ref{eq:reg} ($n=1,000$). Each placebo is constructed by artificially varying wait times for bills based on the empirical distribution of wait times in the analytical sample. Standard errors are clustered at the household level. Coefficients are color-coded based on statistical significance. The vertical dashed red line indicates the estimated coefficient of the preferred specification (Table \ref{tab:ddd-table}). 
	\par
	}
	\end{minipage}
\end{figure}

\clearpage

\begin{figure}[htb]
    \caption{Robustness of Dynamic Treatment Effects to New DID Estimation}
    \label{axfig:event-study-wooldridge}
    \centering

	    \includegraphics[width=\textwidth]{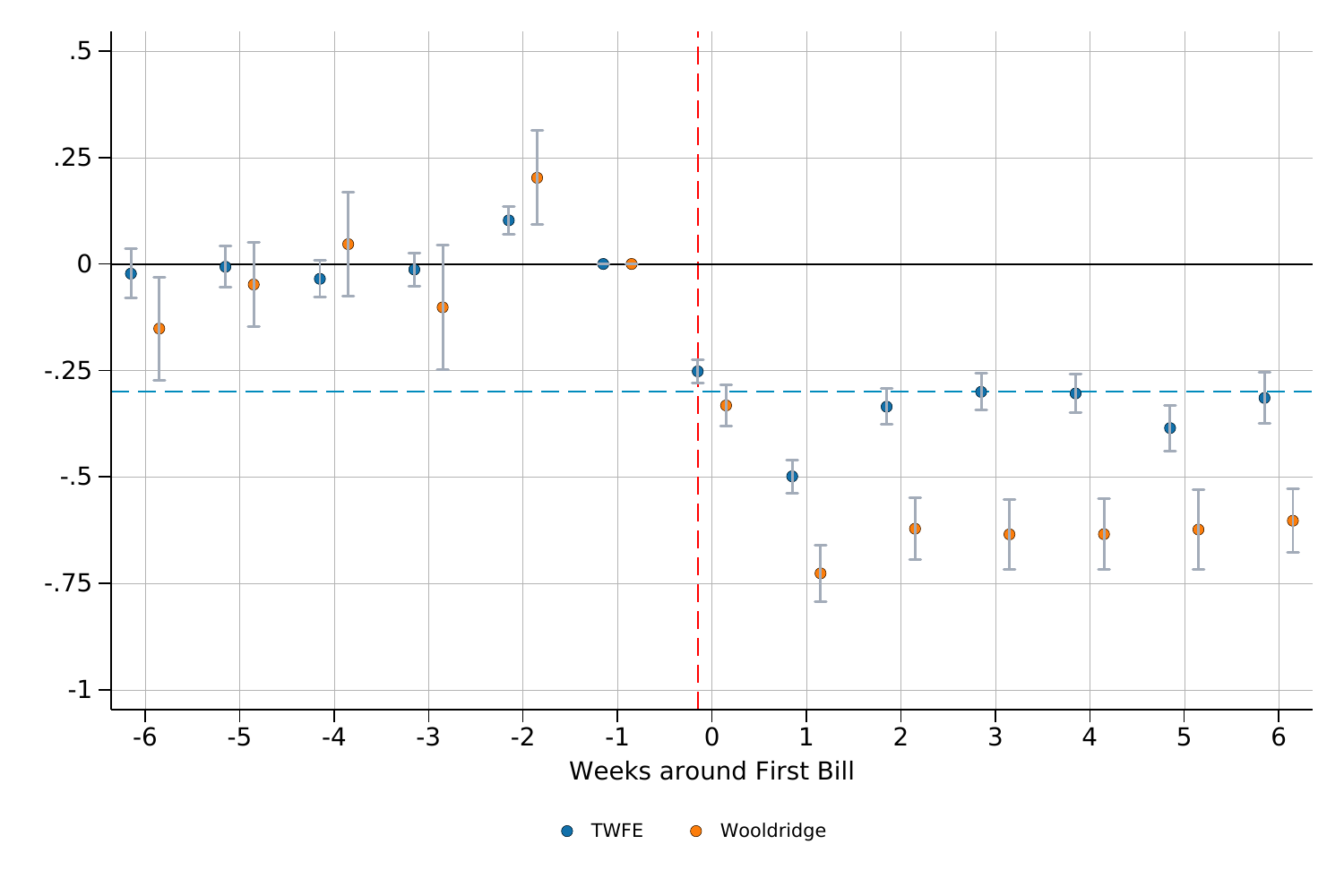}
	 
    \vspace{0.2cm}
    
    \begin{minipage}{0.95\textwidth} 
	{\footnotesize \textit{Notes}: Figure shows estimated coefficients and 95\% confidence intervals for the TWFE regression estimated in Equation \ref{eq:twfe} on the matched sample, as discussed in Section \ref{subsec:dynamic}. Coefficients are compared to those obtained from using the Mundlak approach to TWFE estimation described by \cite{wooldridge2021two}, incorporating a full set of household/year/week-of-year interactions in the main regression. Note that due to computational limits, the Mundlak approach uses a random 50\% sample of the analytic dataset. Dynamic treatment effects are calculated as weighted linear combinations of these coefficients, with standard errors are clustered at the household level. 
	\par
	}
	\end{minipage}
\end{figure}
\clearpage

\begin{figure}[htbp]
    \caption{Deductible-Stratified Results Measured in Dollars}
    \label{axfig:ded-dollars}
    \centering

	\subfloat[One-Way Stratification (Figure \ref{fig:dedhet})]{
	    \includegraphics[width=5.5in]{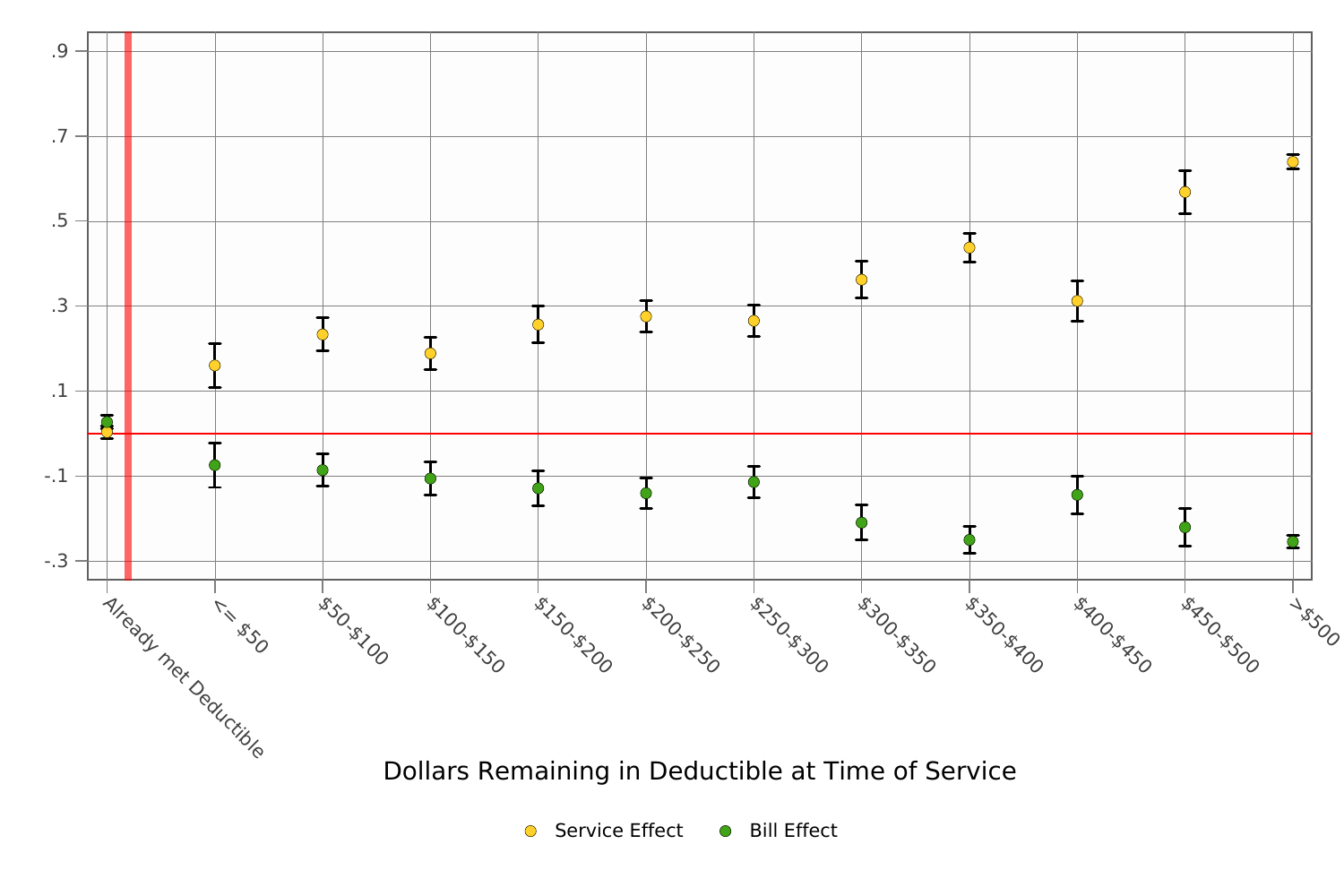}
	 }
  \clearpage
  
	\subfloat[Two-Way Stratification (Figure \ref{fig:dedhet-twoway})]{
	    \includegraphics[width=6.0in]{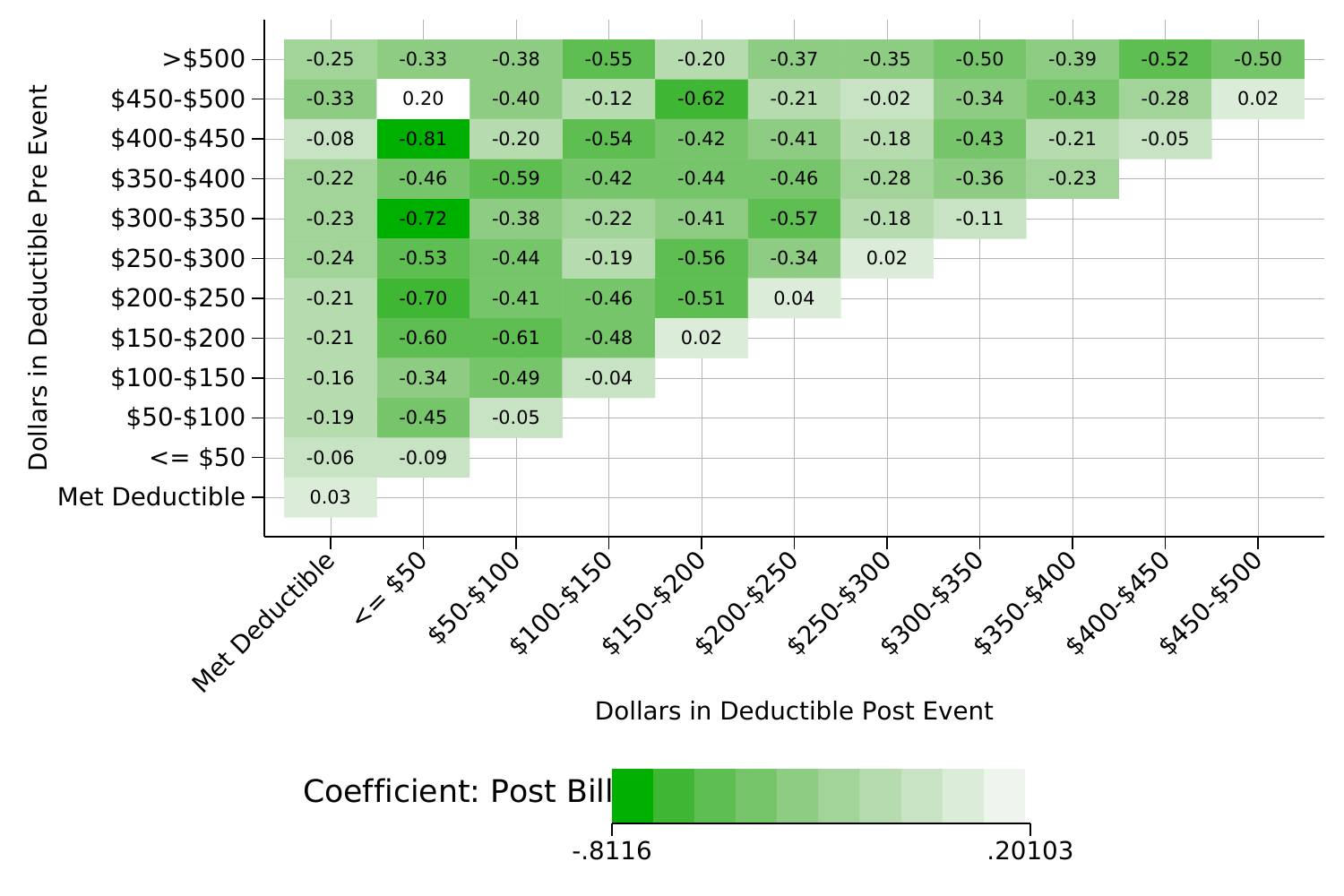}
	 }
    \vspace{0.2cm}
    
    \begin{minipage}{0.95\textwidth} 
	{\footnotesize \textit{Notes}: Figures show equivalent versions of Figures \ref{fig:dedhet} and \ref{fig:dedhet-twoway} with bins measured in levels (dollars) of deductible. Both panels follow methodology of main figures discussed in text. Panel (a) stratifies across deductible spending prior to an event, while panel (b) stratifies across spending prior to \textit{and} following events. Here, sample is restricted to those with non-zero, unmet deductibles. 
	\par
	}
	\end{minipage}
\end{figure}
\clearpage

\begin{figure}[htb]
    \caption{Heterogeneous Service Effects By Household Deductibles and Service Cost}
    \label{axfig:dedhet-twoway-svc}
    \centering

    \includegraphics[width=4.5in]{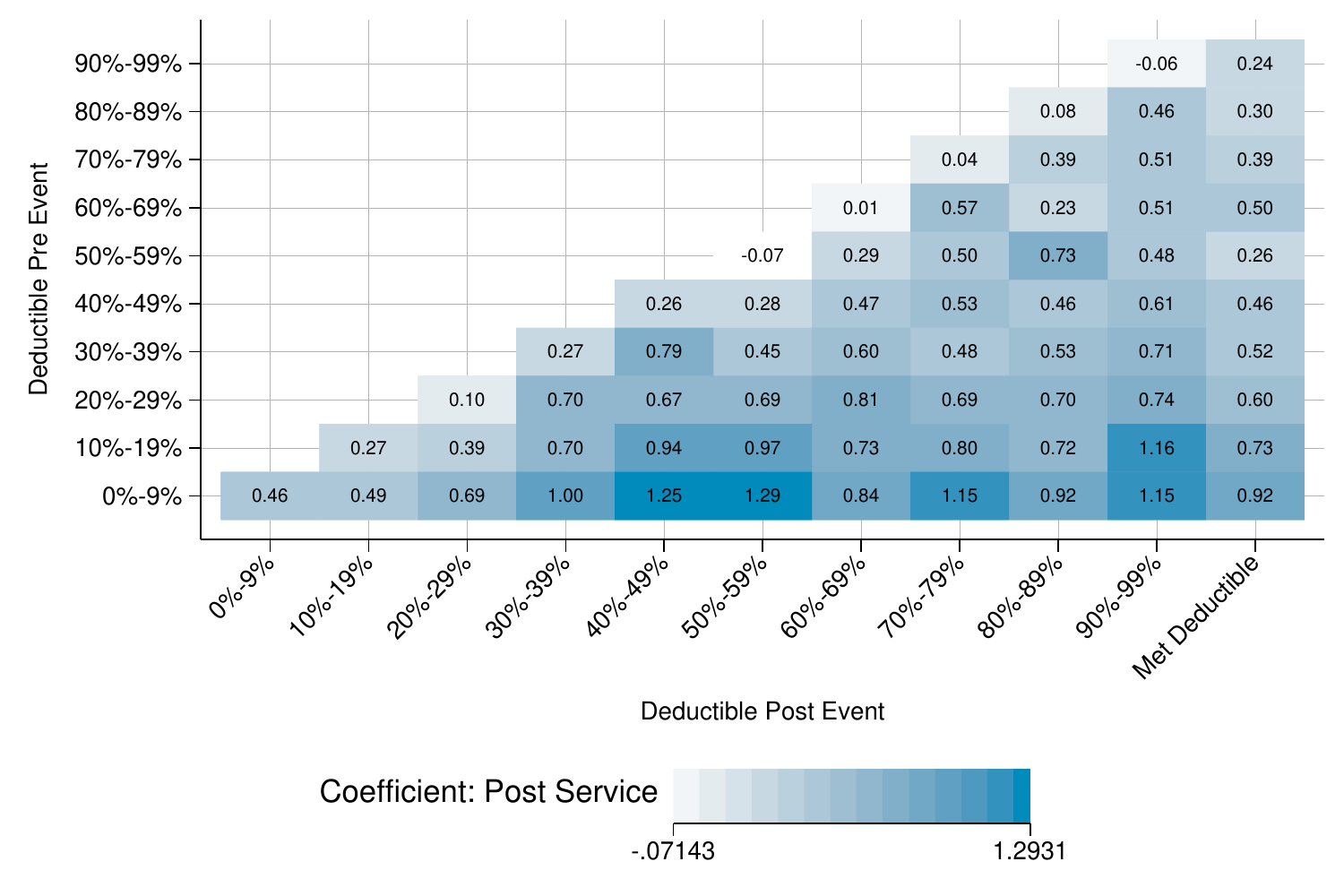}
    \vspace{0.2cm}
    \begin{minipage}{0.95\textwidth} 
	{\footnotesize \textit{Notes}: Figure depicts estimated coefficients for $\mathbbm{1}\{\text{Post\_Service}_{it}\}$ in Equation \ref{eq:reg} across deciles of household deductible spending prior to \textit{and} following an event. Here, sample is restricted to individuals in a non-zero deductible plan who have not yet met their deductible at the time of service. Each row indicates a different decile of deductible spending prior to the event, while each column indicates deciles following the event. Regressions include linear time trends by group as additional controls. Standard errors are clustered at the household level.
	\par
	}
	\end{minipage}
\end{figure}

\clearpage

\begin{figure}[htbp]
    \caption{Variation in Estimated Bill Effects Across Calendar Year}
    \label{fig:calendar-year}
    \centering
    
    \subfloat[Service Consumption Coefficients]{
	    \includegraphics[width=3.0in]{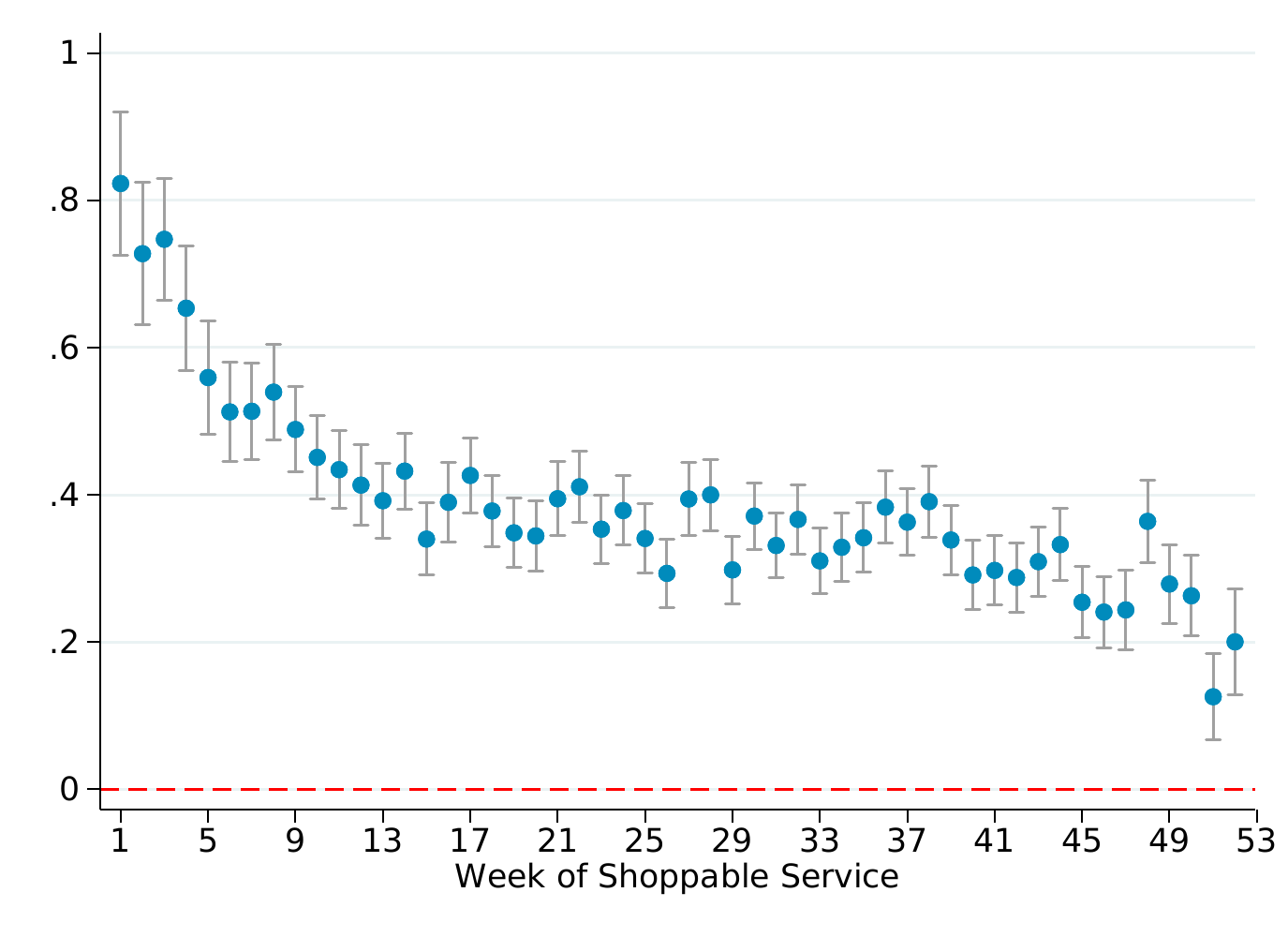}
	 }
	\subfloat[Bill Arrival Coefficients]{
	    \includegraphics[width=3.0in]{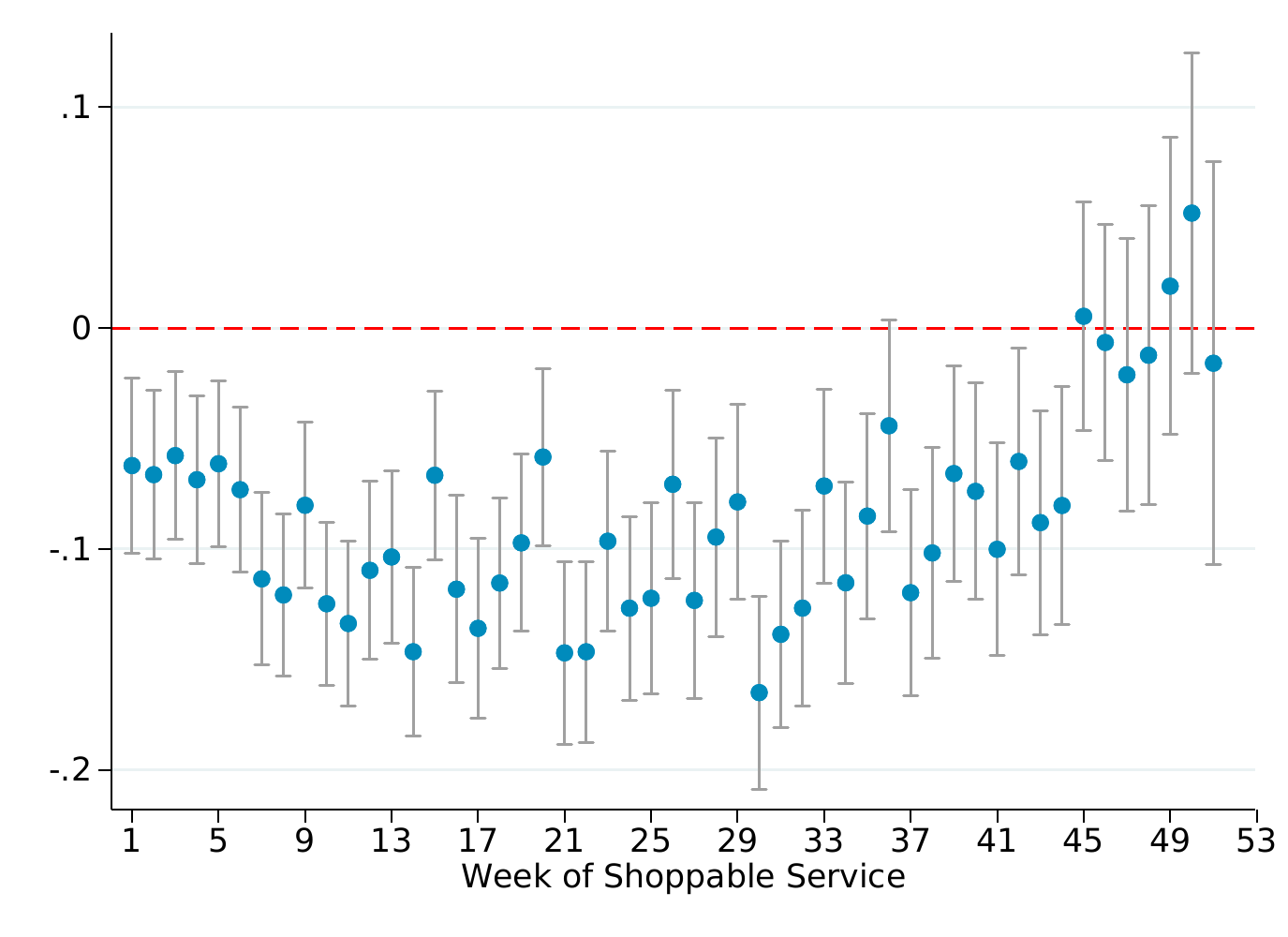}
	 }
  \vspace{0.2cm}

    \begin{minipage}{0.95\textwidth} 
	{\footnotesize \textit{Notes}: Separate estimation of service and bill effects stratified by relative week of year that shoppable service was consumed. Regressions adjust for household, provider, procedure type, and year fixed effects (results are robust to both inclusion and exclusion of relative week of year fixed effects). 
	\par
	}
	\end{minipage}
\end{figure}

\clearpage

\begin{figure}[htbp]
    \caption{Model Results: Parameter Space Heatmap}
    \label{fig:model1-results}
    \centering

    \includegraphics[width=4.5in]{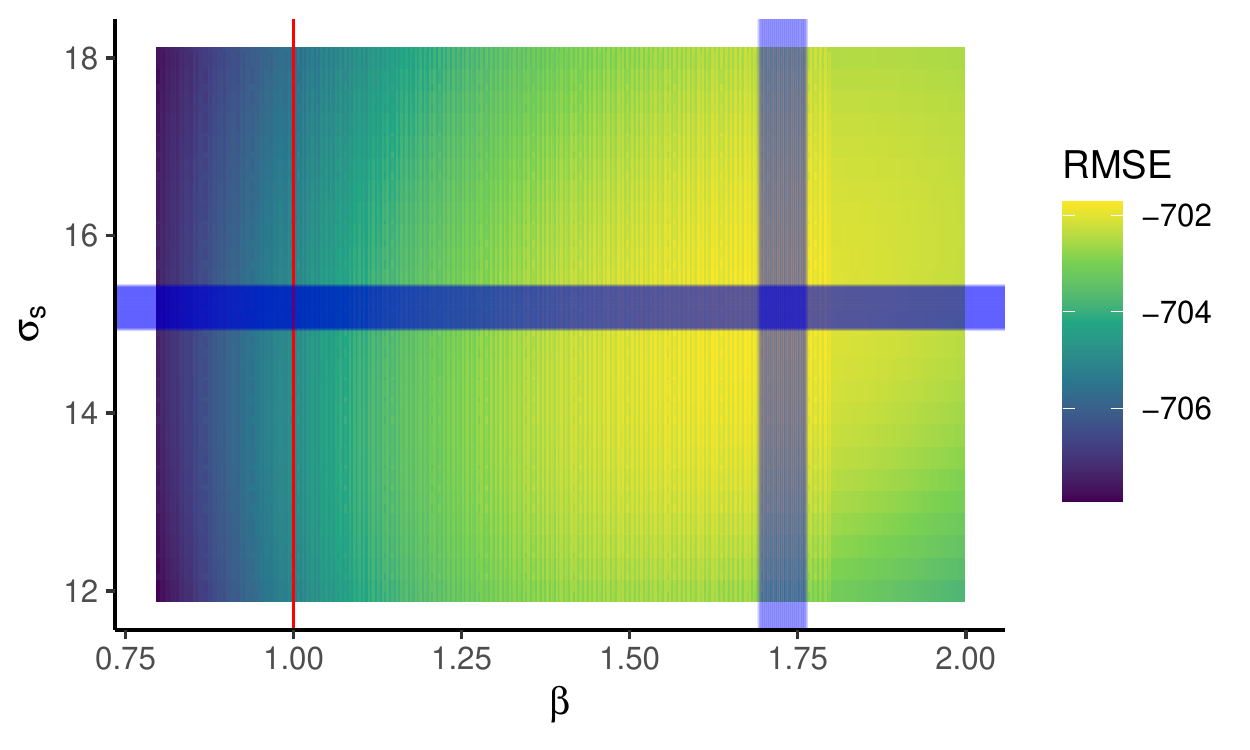}
    \vspace{0.2cm}
    \begin{minipage}{0.95\textwidth} 
	{\footnotesize \textit{Notes}: Figure depicts the relationship between chosen level of household pre-bill discounting parameter $\beta$ and spending signal variance $\sigma^2_s$, and the root mean squared error (RMSE) of the model presented in Section \ref{sec:model}. RMSE is measured as the square root of the mean squared error between observed and predicted household spending at the weekly level. For each point in the parameter space, fill color denotes the median result of 50 simulations with independently drawn health shocks. The 95\% bootstrapped confidence interval for both parameters is shaded in blue. 
	\par
	}
	\end{minipage}
\end{figure}

\clearpage

\begin{figure}[htbp]
    \caption{Implied Probability of Being Under Deductible, Equilibrium Parameters}
    \label{axfig:implied-probabilities}
    \centering
    
    \includegraphics[width=6in]{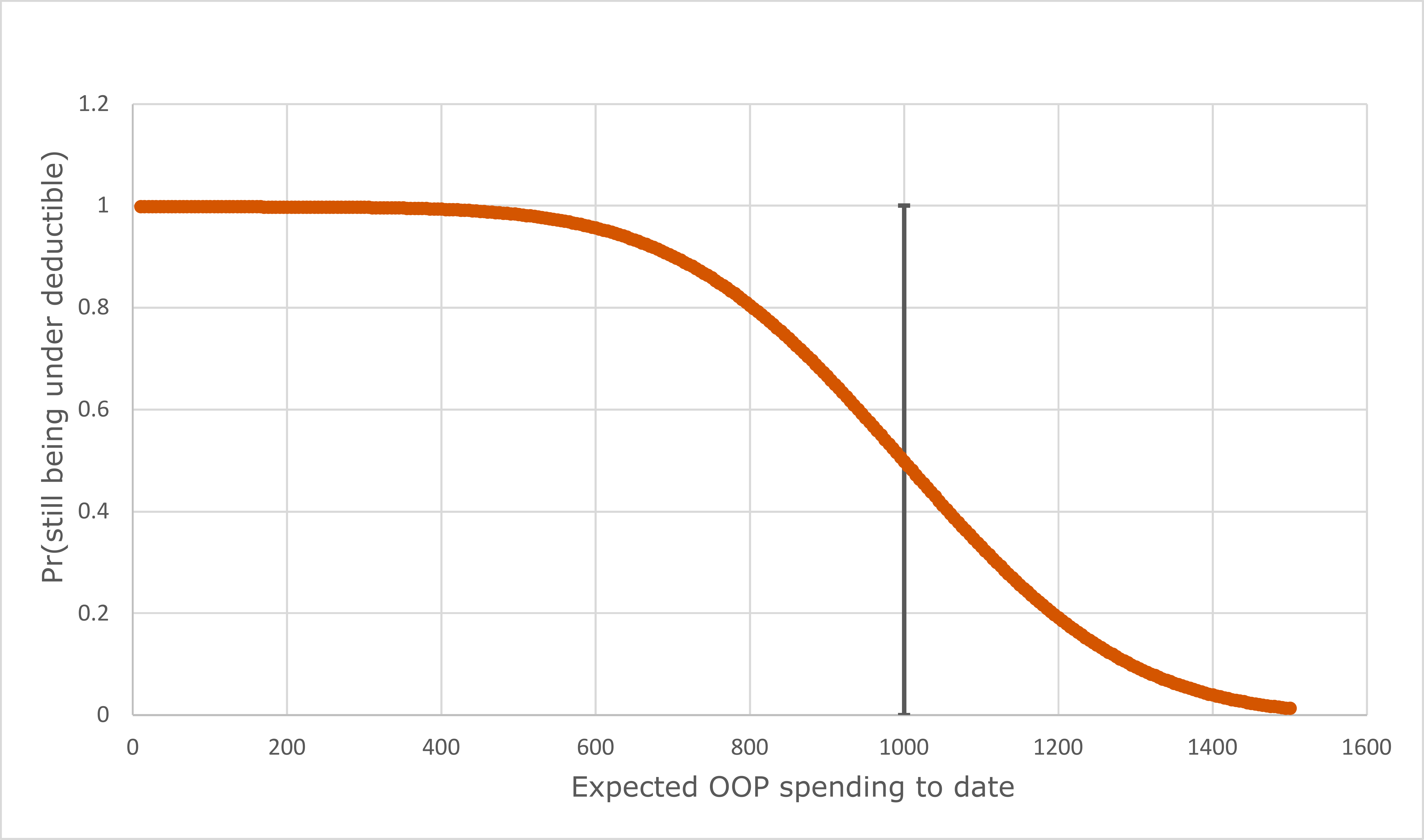}
    \vspace{0.2cm}

    \begin{minipage}{0.95\textwidth} 
	{\footnotesize \textit{Notes}: Figure shows implied probability of being still under the deductible for a household with the equilibrium model parameters $(\beta=1.73,\sigma_s=15.2)$ and a \$1,000 deductible, as described in Section \ref{sec:model}.
	\par
	}
	\end{minipage}
\end{figure}

\clearpage


\begin{figure}[htb]
    \caption{Evolution of Beliefs about $\beta$ Across Plan Year}
    \label{figax:misinformation}
    \centering

    \includegraphics[width=5in]{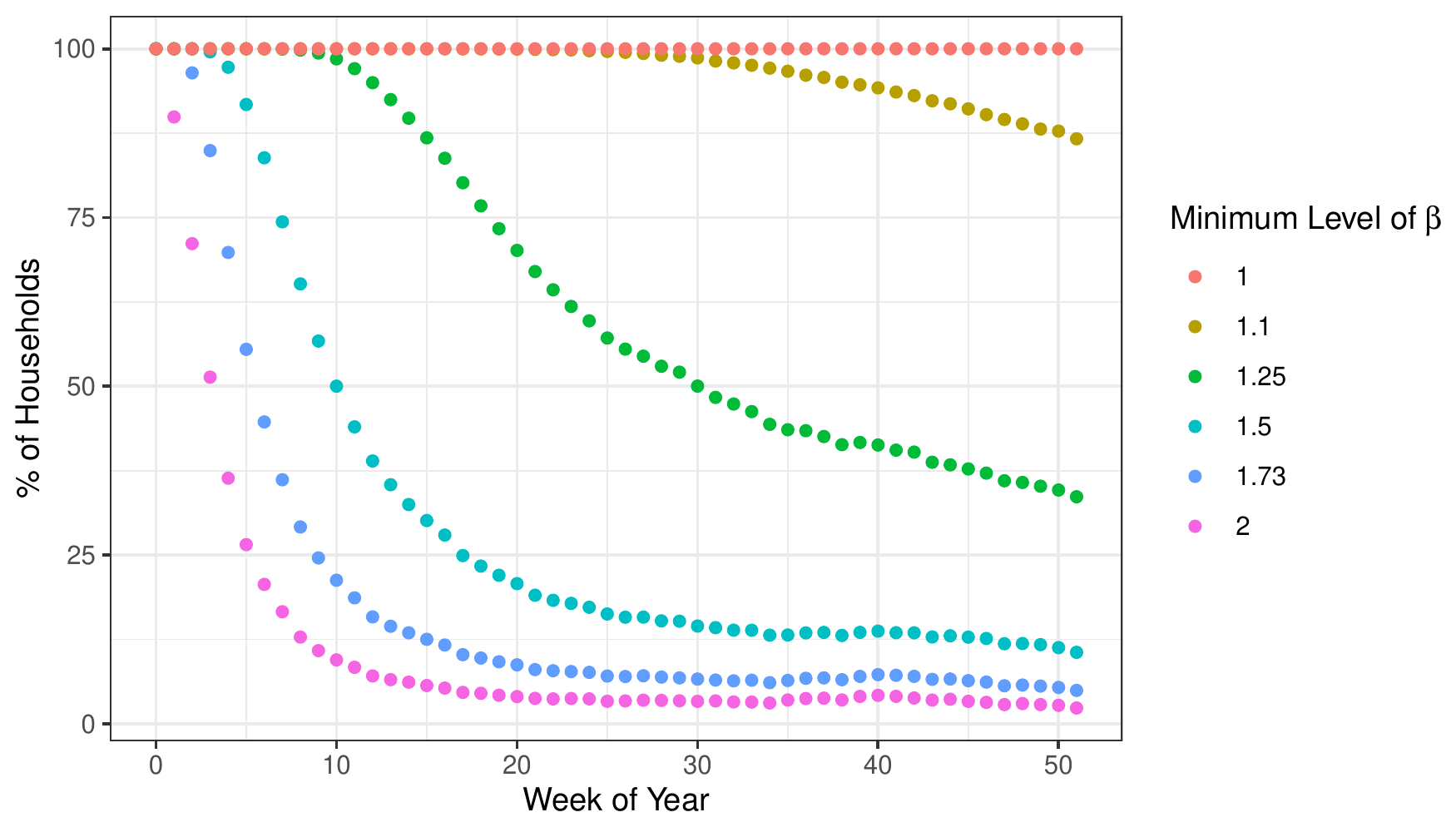}
    \vspace{0.2cm}
    \begin{minipage}{0.95\textwidth} 
	{\footnotesize \textit{Notes}: Figure depicts the fraction of households in the sample with simulated $\beta$ greater than or equal to some threshold $\beta_\text{min}$ for various thresholds. Simulations are performed based on the median equilibrium parameters of the model discussed in Section \ref{subsec:model-learning}. 
	\par
	}
	\end{minipage}
\end{figure}




\begin{table}[H]
\centering
\begin{threeparttable}
\begin{tabular}{ll|l}
\toprule
\multicolumn{1}{l}{\textbf{Type}} & \multicolumn{1}{l}{\textbf{Code}} & \multicolumn{1}{l}{\textbf{Service Description}   } \\
\midrule
DRG                & 216                  & Cardiac valve and other major cardiothoracic procedures w/ cardiac catheterization            \\
DRG                & 460                  & Spinal fusion, except cervical                                                                            \\
DRG                & 470                  & Major joint replacement or   reattachment of lower extremity                                           \\
DRG                & 473                  & Cervical spinal fusion                                                                                 \\
DRG                & 743                  & Uterine and adnexa procedures for   non-malignancy                                                       \\
\midrule
CPT                & 19120                & Removal of 1 or more breast growth,   open procedure                                                   \\
CPT                & 29826                & Shaving of shoulder bone using an   endoscope                                                         \\
CPT                & 29881                & Removal of one knee cartilage using   an endoscope                                                     \\
CPT                & 42820                & Removal of tonsils and adenoid   glands (patient younger than age 12)                           \\
CPT                & 43235                & Diagnostic examination of   esophagus, stomach, and/or upper small bowel \\
CPT                & 43239                & Biopsy of the esophagus, stomach,   and/or upper small bowel using an endoscope                      \\
CPT                & 45378                & Diagnostic examination of large   bowel using an endoscope                                    \\
CPT                & 45380                & Biopsy of large bowel using an   endoscope                                                             \\
CPT                & 45385                & Removal of polyps or growths of   large bowel using an endoscope                                       \\
CPT                & 45391                & Ultrasound examination of lower   large bowel using an endoscope                                       \\
CPT                & 47562                & Removal of gallbladder using an   endoscope                                                            \\
CPT                & 49505                & Repair of groin hernia (patient age   5 years or older)                                                \\
CPT                & 55700                & Biopsy of prostate gland                                                                               \\
CPT                & 55866                & Surgical removal of prostate and   surrounding lymph nodes using an endoscope                          \\
CPT                & 59400                & Routine obstetric care for vaginal   delivery                                                          \\
CPT                & 59510                & Routine obstetric care for cesarean   delivery                                                         \\
CPT                & 59610                & Routine obstetric care for vaginal   delivery after prior cesarean delivery                            \\
CPT                & 64483                & Injections of anesthetic and/or   steroid drug into lower or sacral spine nerve root  \\
CPT                & 66821                & Removal of recurring cataract in   lens capsule using laser                                            \\
CPT                & 66984                & Removal of cataract with insertion   of lens                                                           \\
CPT                & 93000                & Electrocardiogram, routine, with   interpretation and report                                           \\
CPT                & 93452                & Insertion of catheter into left   heart for diagnosis                                                  \\
CPT                & 62322         & Injection of substance into spinal   canal of lower back or sacrum \\   
CPT                & 62323          & Injection of substance into spinal   canal of lower back or sacrum  \\   
\bottomrule
\end{tabular}
\begin{tablenotes}
    \small
    \item \textit{Notes:} Table shows list of procedures used to identify non-urgent ``shoppable services," which are the exposure of interest in the primary reduced-form specifications. Services are identified based on lists provided by the Center for Medicare and Medicaid Services (CMS), using the relevant Diagnostic Related Groups (DRGs) or Current Procedural Terminology (CPT) codes to identify procedures. 
\end{tablenotes}
\caption{\label{tab:procs} Shoppable Services Used in Analytical Sample}
\end{threeparttable}
\end{table}
\clearpage


\begin{table}[htb]
\centering
\begin{threeparttable}
\begin{tabular}{l|c|cccc}
\toprule
& \multicolumn{1}{c}{Main} & \multicolumn{4}{c}{Alternative Specifications} \\
\midrule
Post Service & 0.137***  & 0.365***  & 0.367***  & 0.366***  & 0.300***  \\
& (0.0103)  & (0.0102)  & (0.0102)  & (0.0102)  & (0.105)   \\
Post Bill & -0.074*** & -0.066*** & -0.069*** & -0.067*** & -0.062*** \\
& (0.0099)  & (0.0101)  & (0.0100)  & (0.0100)  & (0.0010) \\
\midrule
$\overline{\text{spend}_{it}}$ & \$93.82 & \$93.82 & \$93.82 & \$93.82 & \$93.82 \\
Household FEs &  X & X & X & X & X \\
Year FEs &  X &  & X & X & X \\
Week of Year FEs &  X &  & & X & X \\
Provider FEs &  X & & & & X\\
Event Type FEs & X &  & & & \\
Observations & 126,747,805 &  126,747,805 & 126,747,805 & 126,747,805 & 126,747,805 \\
\bottomrule
\end{tabular}
\begin{tablenotes}
    \small
    \item \textit{Notes}: Table presents results from triple-difference Poisson regressions highlighting the role of a bill's arrival on health spending of affected household members (Equation \ref{eq:reg}). Here, we report results for households affected by unplanned hospitalizations for injuries and appendectomies. We identify index events as any household member admitted to a hospital with a principal diagnosis contained in the ICD chapter heading ``Injury, poisoning and certain other consequences of external causes" (in ICD-10-CM, S00-T88), or whose primary diagnosis is in the heading ``Diseases of appendix" (K35--K38). Event type fixed-effects are defined by diagnosis group using the first three characters of the ICD-10-CM diagnosis code. Throughout, standard errors were clustered at the household level. 
    
    $^{*} p < 0.05, ^{**} p < 0.01, ^{***} p < 0.001$ 
    \end{tablenotes}
    \caption{\label{axtab:expanded-services} Robustness of Effects to Expanded Services: Injuries and Appendectomies} 
\end{threeparttable}
\end{table}

\clearpage

\begin{table}[htb]
\centering
\begin{threeparttable}
\begin{tabular}{l|c|c|c}
\toprule
& \multicolumn{1}{c}{Full Specification} & \multicolumn{1}{c}{Event-Specific Time Trends} & \multicolumn{1}{c}{No Household FEs} \\
\midrule
Post Service &  0.218***& 0.253***  & 0.726***   \\
& (0.0032)  & (0.0051)& (0.0036)  \\
Post Bill & -0.109***  & -0.071*** & -0.016***\\
& (0.0030) & (0.0043)  & (0.0038) \\

\midrule
$\overline{\text{spend}_{it}}$ & \$120.49 & \$120.49 & \$120.49  \\
Observations & 61,860,735 &  61,860,735 & 61,860,735 \\
\bottomrule
\end{tabular}
\begin{tablenotes}
    \small
    \item \textit{Notes}: Robustness models include (a) a model including weekly time trends for each type of shoppable service in the sample, and (b) a model omitting household fixed effects. Throughout, standard errors were clustered at the household level, and regressions included family, year, relative week of year, event type, and provider (of shoppable service) fixed effects. 
    
    $^{*} p < 0.05, ^{**} p < 0.01, ^{***} p < 0.001$ 
    \end{tablenotes}
    \caption{\label{axtab:ddd-table-ext} Robustness of Main Specification}
\end{threeparttable}
\end{table}

\clearpage

\begin{table}[htb]
\centering
\begin{threeparttable}
\begin{tabular}{l|cc|cccc}
\toprule
& \multicolumn{2}{c}{Main Models} & \multicolumn{4}{c}{Alternative Specifications} \\
\midrule
Post Service &  0.402*** & 0.224*** &  0.753*** &  0.626*** &  0.625*** &  0.599*** \\
& (0.0022) & (0.0033) & (0.0033) & (0.0034) & (0.0034) & (0.0033) \\
Post Bill & & -0.067*** & -0.075*** & -0.082*** & -0.084*** & -0.085*** \\
& & (0.0031) & (0.0033) & (0.0032) & (0.0033) & (0.0032) \\
\midrule
$\overline{\text{spend}_{it}}$ & \$120.49 & \$120.49 & \$120.49 & \$120.49 & \$120.49 & \$120.49 \\
Household FEs & X & X & X & X & X & X \\
Year FEs & X & X &  & X & X & X \\
Week of Year FEs & X & X &  & & X & X \\
Provider FEs & X & X & & & & X\\
Event Type FEs & X & X &  & & & \\
Observations & 61,860,735 &   61,860,735 &  61,860,735 &   61,860,735 & 61,860,735 & 61,860,735 \\
\bottomrule
\end{tabular}
\begin{tablenotes}
    \small
    \item \textit{Notes}: Table presents results from triple-difference Poisson regressions highlighting the role of a bill's arrival on health spending of affected household members. Includes linear controls for time trends before and after service. Regression coefficients displayed illustrate the expected change in log household spending (measured per person-week) associated with the service date and bill arrival (both measured as dummy variables). Throughout, standard errors were clustered at the household level. 
    
    $^{*} p < 0.05, ^{**} p < 0.01, ^{***} p < 0.001$ 
    \end{tablenotes}
    \caption{\label{axtab:timetrends} Robustness of DDD Regressions to Linear Time Controls} 
\end{threeparttable}
\end{table}

\clearpage

\begin{table}[htb]
\centering
\begin{threeparttable}
\begin{tabular}{l|c|c}
\toprule
& \multicolumn{1}{c}{Original Specification (Table \ref{tab:ddd-table})} & \multicolumn{1}{c}{Alternative Specification} \\
\midrule
Post Service &  0.218*** & 0.283*** \\
& (0.0032) & (0.0043) \\
Post Bill & -0.109*** & -0.077*** \\
& (0.0030) & (0.0042) \\
Post Bill $\times$ Service Met Deductible & & 0.122*** \\
& & (0.0212) \\
\midrule
Household FEs & X & X \\
Year FEs & X & X \\
Week of Year FEs  & X & X \\
Provider FEs & X & X \\
Event Type FEs & X & X \\
Observations & 61,860,735 & 44,403,451 \\
\bottomrule
\end{tabular}
\begin{tablenotes}
    \small
    \item \textit{Notes}: Table presents results from triple-difference Poisson regressions highlighting the role of a bill's arrival on health spending of affected household members. Results are split across households whose event did and did not move them across the threshold of the household deductible. Sample is restricted to households enrolled in plans with a nonzero deductible, which was unmet at the time of the shoppable service. Standard errors are clustered at the household level.
    
    $^{*} p < 0.05, ^{**} p < 0.01, ^{***} p < 0.001$ 
    \end{tablenotes}
    \caption{\label{axtab:metded} Bill Arrival Effects Across Deductible Threshold}
\end{threeparttable}
\end{table}

\clearpage

\begin{table}[htb]
\centering
\begin{threeparttable}
\begin{tabular}{l|c|cccc}
\toprule
& \multicolumn{1}{c}{Main} & \multicolumn{4}{c}{Alternative Specifications} \\
\midrule
Post Service & -0.216 & 0.698*** & 0.465**  & 0.427**  & 0.450**    \\
&(0.1114) & (0.1455) & (0.1603) & (0.1580) & (0.1366)  \\
Post Bill & 0.122 & 0.363*   & 0.415**  & 0.415**  & 0.205    \\
& (0.1227) & (0.1635) & (0.1587) & (0.1571) & (0.1529) \\
\midrule
$\overline{\text{spend}_{it}}$ & \$93.82 & \$93.82 & \$93.82 & \$93.82 & \$93.82 \\
Household FEs &  X & X & X & X & X \\
Year FEs &  X &  & X & X & X \\
Week of Year FEs &  X &  & & X & X \\
Provider FEs &  X & & & & X\\
Event Type FEs & X &  & & & \\
Observations & 31,905,563 &  31,905,563 & 31,905,563 & 31,905,563 & 31,905,563 \\
\bottomrule
\end{tabular}
\begin{tablenotes}
    \small
    \item \textit{Notes}: Table presents results from triple-difference Poisson regressions highlighting the role of a bill's arrival on health spending of affected household members. In this exercise, sample is limited to households who have met their OOP max prior to the consumption of the index service, plus the control group of not-yet-treated household-years. Includes linear controls for time trends before and after service. Regression coefficients displayed illustrate the expected change in log household spending (measured per person-week) associated with the service date and bill arrival (both measured as dummy variables). Throughout, standard errors were clustered at the household level. 
    
    $^{*} p < 0.05, ^{**} p < 0.01, ^{***} p < 0.001$ 
    \end{tablenotes}
    \caption{\label{axtab:oop-max} Falsification Test: Bill Effects for OOP-Capped Households} 
\end{threeparttable}
\end{table}

\clearpage

\begin{table}[htb]
\centering
\begin{threeparttable}
\begin{tabular}{l|cccc}
\toprule
& \multicolumn{4}{c}{Week of Month} \\
& \multicolumn{1}{c}{First} & \multicolumn{1}{c}{Second} & \multicolumn{1}{c}{Third} & \multicolumn{1}{c}{Last} \\
\midrule
Post Service &  0.230*** & 0.221*** & 0.211*** & 0.232***  \\
& (0.0066) & (0.0061) & (0.0066) & (0.0066)  \\
Post Bill & -0.074*** & -0.065*** & -0.064*** &  -0.080*** \\
& (0.0062) & (0.0058) & (0.0064) & (0.0064) \\
\midrule
$\overline{\text{spend}_{it}}$ & \$101.04 & \$153.30 & \$153.25 & \$102.26 \\ 
Observations & 38,633,474 & 39,994,383 & 39,069,522 & 38,935,377 \\ 
\bottomrule
\end{tabular}
\begin{tablenotes}
    \small
    \item \textit{Notes}: Table presents results from triple-difference Poisson regressions highlighting the role of a bill's arrival on health spending of affected household members (Equation \ref{eq:reg}). Columns stratify bill arrivals by relative week of month to test the hypothesis that households may respond differently to bills arriving at the end of the month versus the beginning due to liquidity constraints. Standard errors are clustered at the household level.
    
    $^{*} p < 0.05, ^{**} p < 0.01, ^{***} p < 0.001$ 
    \end{tablenotes}
    \caption{\label{axtab:liquidity} Heterogeneity by Week of Bill Arrival}
\end{threeparttable}
\end{table}

\clearpage

\begin{table}[htb]
\centering
\begin{threeparttable}
\begin{tabular}{l|c|cccc}
\toprule
& \multicolumn{1}{c}{Main} & \multicolumn{4}{c}{Alternative Specifications} \\
\midrule
Post Service &  0.262***  & 0.252*** &  0.254*** &  0.250*** &  0.251***  \\
& (0.0102) & (0.0042) & (0.0042) & (0.0042) & (0.0041)  \\
Post Bill & -0.090*** & -0.084*** & -0.086*** & -0.081*** & -0.083*** \\
& (0.0096) & (0.0040) & (0.0039) & (0.0040) & (0.0039)  \\
\midrule
$\overline{\text{spend}_{it}}$ &  \$167.01 & \$167.01 & \$167.01 & \$167.01 & \$167.01 \\
Household FEs &  X & X & X & X & X \\
Year FEs &  X &  & X & X & X \\
Week of Year FEs &  X &  & & X & X \\
Provider FEs &  X & & & & X\\
Event Type FEs & X &  & & & \\
Observations & 61,860,735 &  61,860,735 &   61,860,735 & 61,860,735 & 61,860,735 \\
\bottomrule
\end{tabular}
\begin{tablenotes}
    \small
    \item \textit{Notes}: Table presents results from triple-difference Poisson regressions highlighting the role of a bill's arrival on health spending of affected household members. Regressions omit all claims consumed at the provider of the focal index event (note that this means control group years without a focal index event are excluded). Includes linear controls for time trends before and after service. Regression coefficients displayed illustrate the expected change in log household spending (measured per person-week) associated with the service date and bill arrival (both measured as dummy variables). Throughout, standard errors were clustered at the household level. 
    
    $^{*} p < 0.05, ^{**} p < 0.01, ^{***} p < 0.001$ 
    \end{tablenotes}
    \caption{\label{axtab:focal-provider} Robustness to Omitting Claims from the Index Service Provider} 
\end{threeparttable}
\end{table}

\clearpage


\begin{table}[p]
\centering
\begin{threeparttable}
\begin{tabular}{l|cc|cc}
\toprule
& \multicolumn{2}{c}{\textbf{Regression Coefficients}} & \multicolumn{2}{c}{\textbf{Pre-Treatment Averages}} \\
& \multicolumn{1}{c}{Post Service} & \multicolumn{1}{c}{Post Bill} &\multicolumn{1}{c}{\% $\geq 0 $} & \multicolumn{1}{c}{Conditional Mean} \\
\midrule
\textbf{Hospital Care} \\
Emergency Department &  0.123*** & 0.017 & 0.67\% &      \$929.98 \\
& (0.0141) & (0.0147) \\
Preventable Hospitalizations &  0.412*** & -0.170 & 0.04\% &  \$19,979.89 \\
& (0.0880) & (0.0879) \\
\midrule
\textbf{Outpatient Care} \\
Behavioral Health & -0.032* &  0.029* & 1.19\% &     \$119.47  \\
& (0.0144) & (0.0142) \\
Chiropractic Care & -0.015 &  0.017 & 1.86\% &    \$133.39 \\
& (0.0160) & (0.0161) \\
Evaluation \& Management &  1.469*** & -0.272*** &1.05\%  &     \$121.45\\
& (0.0076) & (0.0064) \\
Imaging &  0.102*** & 0.005 & 2.55\%  &     \$265.52 \\
& (0.0119) & (0.0123) \\
Lab Services &  0.196*** & -0.147*** & 3.96\% &      \$62.14  \\
& (0.0123) & (0.0130) \\
Low-Value Services &  0.066*** &  0.028** & 6.58\% &     \$148.61 \\
& (0.0094) & (0.0097) \\
Preventive Care &  0.349*** & -0.211*** & 11.89\% &     \$120.79  \\
& (0.0039) & (0.0040) \\
Specialist Care &  0.546*** & -0.108*** & 0.57\% &     \$114.70 \\
& (0.0221) & (0.0226) \\
\midrule
\textbf{Prescriptions} &  0.013*** & 0.010 & 18.30\% &     \$147.14  \\
& (0.0049) & (0.0050) \\
\bottomrule
\end{tabular}
\begin{tablenotes}
\small
\item \textit{Notes}: Table shows coefficients from triple-difference regressions capturing service-specific effects of pricing information ($N=59,177,995$, see Equation \ref{eq:reg}). Columns (1) and (2) present regression coefficients; column (3) indicates the fraction of pre-treatment weeks when spending was positive; and column (4) presents pre-treatment weekly averages, conditional on positive spending. See Appendix Table \ref{tab:outpatient-services} for a complete list of the CPT codes for each of the outpatient categories. All models include fixed effects for households, years, relative week of year, and providers, as well as linear time trends before and after the event; standard errors were clustered at the household level. $^{*} p < 0.05, ^{**} p < 0.01, ^{***} p < 0.001$
\end{tablenotes}
\caption{\label{tab:services} Estimated Impact of Bill Arrival on Service-Specific Spending}
\end{threeparttable}
\end{table}

\clearpage

\begin{table}[htb]
\centering
\begin{threeparttable}
\begin{tabular}{lll}
\toprule
\multicolumn{1}{l}{\textbf{Service Description}} & \multicolumn{2}{c}{\textbf{Code}} \\
\midrule
\textbf{Panel A:} Infection Diagnoses & (ICD-9-CM) & (ICD-10-CM) \\
Acute Respiratory Infections & 460-466 & J00-J06 \\
Pneumonia and Influenza	& 480-488 & J09-J18\\
Otitis media	&381--382 & H65--H66\\
Streptococcal sore throat and scarlet fever	& 034 & A38, J02\\
Whooping cough & 033 & A37 \\
Infectious mononucleosis & 075 & B27 \\
Chickenpox & 052 & B01 \\
Urinary Tract Infections &	590, 595, 599 & N10--N12,N15,N16,N28,N30,N36,N39,R31\\
Food poisoning (bacterial) & 005 &	A05 \\
Other Intestinal Infections	& 008--009 & 	A04,A08,A09 \\
Nausea and Vomiting	& 536.2, 787	& R11,R12,R13,R14,R15,R19 \\ 
Dyspepsia & 	536.8,536.9	& K30--K31 \\
\midrule 
\textbf{Panel B:} Injuries & (ICD-9-CM) & (ICD-10-CM) \\
Fracture of Upper Limb	& 810-819	& S12,S22,S32,S42,S52,S62 \\
Fracture of Lower Limb	& 820-829	& S72,S82,S92 \\ 
Dislocation & 	830-839	& S13,S23,S33,S43,S53,S63,S73,S83,S93 \\
Sprains and Strains & 	840-848 &	Combined with above \\ 
\midrule
\textbf{Panel C:} Place of Service Codes & (POS) & -- \\
Physician Office & 11, 72, 95 \\ 
Urgent Care Center & 17,20 \\ 
Emergency Department & 23 \\ 
Hospital (including on-campus outpatient) & 21, 22, 28 \\
\bottomrule
\end{tabular}
\begin{tablenotes}
    \small
    \item \textit{Notes:} Table shows list of diagnoses used to identify care for non-delayable injuries and infections. Services are identified based on lists provided by the Center for Medicare and Medicaid Services (CMS), using the ICD-9-CM and ICD-10-CM diagnostic codes and place of service codes.  
\end{tablenotes}
\caption{\label{tab:infections} Identifying Injuries, Infections, and Places of Service}
\end{threeparttable}
\end{table}
\clearpage

\centering
\begin{ThreePartTable}

\begin{longtable}{p{2.5cm}|p{2.5cm}|p{11cm}}
\toprule
Outpatient & \multicolumn{2}{c}{\textbf{CPT Codes}}  \\
Category & Code Range & Code Values \\ 
\midrule

\hline Behavioral Health & 90000-99999 &  90791-90792, 90801-90802, 90805-90807, 90832-90834, 90836-90840, 90845-90847, 90849, 90853, 96105, 96112-96113, 96116, 96121, 96125, 96130-96133, 96136-96139, 96156, 96158-96159, 96164-96165, 96167-96168, 96170-96171, 99483-99494 \\ 
\hline Chiropractic Care & 90000-99999 & 97001, 97010-97014, 97018, 97022, 97026, 97032-97035, 97039, 97110-97113, 97116, 97124, 97140, 97161-97162, 97530, 97535, 97750, 98940-98943, 99211 \\ 
\hline Evaluation \& &  10000-19999 & 11976, 11981-11983 \\
Management  & 30000-39999 & 36415-36416 \\
 & 40000-49999 & 44388-44389, 44392-44394, 45300, 45303-45309, 45315-45317, 45320, 45330-45335, 45338-45340, 45378-45386 \\
& 50000-59999 & 57170, 58300-58301, 58340, 58565, 58600, 58605, 58611, 58615, 58670-58671 \\
& 70000-79999 & 71250, 74263, 74740, 76070-76071, 76075-76078, 76497, 76977, 77078-77083, 78350 \\
& 80000-89999 & 80061, 82270, 82274, 82465, 82947-82952, 83036, 83718-83721, 84478, 86580, 86592-86593, 86631-86632, 86689, 86701-86703, 86803-86804, 87110, 87270, 87320, 87340-87341, 87390-87391, 87490-87492, 87590-87592, 87620-87622, 87801, 87810, 87850, 88141-88143, 88147-88155, 88164-88167, 88174-88175, 88304-88305 \\
& 90000-99999 & 92015, 92507, 92551-92553, 92558, 92567, 92585-92588, 96040, 96110, 96127, 96160-96161, 96372, 97802-97804, 99173-99174, 99201-99205, 99211-99215, 99381, 99385-99387, 99395-99397, 99401-99404, 99411-99412, 99420 \\ 
\hline Imaging & 10000-19999 & 10005-10006, 19081-19084 \\
& 20000-29999 & 29881 \\
& 70000-79999 &  70030, 70110, 70130, 70150, 70160, 70200, 70210, 70220, 70260, 70330, 70336, 70360, 70450, 70460, 70470, 70480-70482, 70486-70491, 70496-70498, 70540, 70543-70553, 71010, 71020, 71045-71048, 71100-71101, 71110, 71120, 71130, 71250, \\
& & 71260, 71275, 71550-71552, 71555, 72040, 72050-72052, 72070, 72082, 72100, 72110, 72114, 72125-72132, 72141-72142, 72146-72149, 72156-72159, 72170, 72191-72202, 72220, 73000, 73010, 73030, 73050, 73060, 73070, 73090, 73100, 73110, 73120, 73130, 73140, 73200-73202, 73206, 73218-73225, 73501-73503, 73521-73523, 73552, 73560-73564, 73590, 73600, 73610, 73620, 73630, 73650, 73660, 73700-73702, 73706, 73718-73725, 74000, 74018-74021, 74150, 74160, 74170, 74174-74178, 74181-74185, 74210, 74220, 74241, 74245-74250, 74261-74263, 74270, 74280, 74400, 75635, 76010, 76390-76391, 76536, 76641-76642, 76645, 76700, 76705-76706,76770, 76775-76776, 76801, 76812, 76817, 76830, 76856-76857, 76870, 76881-76882, 76981, 77021, 77046-77049, 77052, 77057, 77063-77067, 77072-77077, 77080, 77085, 78012-78014, 78070-78071, 78206, 78215, 78226-78227, 78290, 78306, 78315, 78452, 78472, 78607-78608, 78707-78708, 78800, 78804, 78814-78816 \\ 
& 90000-99999 & 91200, 93000, 93005, 93010-93018, 93024-93025, 93040-93042, 93050, 93201-93205, 93208-93210, 93220-93237, 93241-93248, 93260-93261, 93264, 93268-93272, 93278-93299, 93303-93308, 93312-93321, 93325, 93350-93352, 93355-93356, 93451-93464, 93501-93505, 93508-93511, 93514, 93524-93533, 93536, 93539-93545, 93555-93556, 93561-93568, 93571-93572, 93580-93583, 93590-93603, 93607-93624, 93631, 93640-93644, 93650-93657, 93660-93662, 93668, 93701-93702, 93720-93724, 93727, 93731-93745, 93750, 93760-93762, 93770, 93784-93793, 93797-93799, 93880, 93926, 93970-93971, 93975 \\ 
\hline Lab Services &  20000-29999 & 20610 \\
& 30000-39999 & 36415-36416 \\
& 80000-89999 &  80048, 80050, 80053, 80061, 80076, 81000-81003, 81025, 82000, 82003, 82009-82010, 82013, 82016-82017, 82024, 82030, 82040, 82042-82045, 82055, 82075, 82077, 82085, 82088, 82101, 82103-82108, 82120, 82127-82128, 82130-82131, 82135-82136, 82139-82140, 82143, 82145, 82150, 82154, 82157, 82160, 82163-82164, 82172, 82175, 82180, 82190, 82205, \\
& & 82232, 82239-82240, 82247-82248, 82250-82252, 82261, 82270-82274, 82286, 82300, 82306-82308, 82310, 82330-82331, 82340, 82355, 82360, 82365, 82370, 82373-82376, 82378-82380, 82382-82384, 82387, 82390, 82397, 82415, 82435-82436, 82438, 82441, 82465, 82480, 82482, 82485-82489, 82491-82492, 82495, 82507, 82520, 82523, 82525, 82528, 82530,  82533, 82540-82544, 82550, 82552-82554, 82565, 82570, 82575, 82585, 82595, 82600, 82607-82608, 82610, 82615, 82626-82627, 82633-82634, 82638, 82642, 82646, 82649, 82651-82654, 82656-82658, 82664, 82666, 82668, 82670-82672, 82677, 82679, 82681, 82690, 82693, 82696, 82705, 82710, 82715, 82725-82726, 82728, 82731, 82735, 82742, 82746-82747, 82757, 82759-82760, 82775-82777, 82784-82785, 82787, 82800, 82803, 82805, 82810, 82820, 82926, 82928, 82930, 82938, 82941, 82943, 82945-82948, 82950-82953, 82955, 82960, 82962-82963, 82965, 82975, 82977-82980, 82985, 83001-83003, 83006, 83008-83010, 83012-83015, 83018-83021, 83026, 83030, 83033, 83036-83037, 83045, 83050-83051, 83055, 83060, 83065, 83068-83071, 83080, 83088, 83090, 83150, 83491, 83497-83500, 83505, 83516, 83518-83521, 83525, 83527-83529, 83540, 83550, 83570, 83582, 83586, 83593, 83605, 83615, 83625, 83630-83634, 83655, 83661-83664, 83670, 83690, 83695, 83698, 83700-83701, 83704, 83715-83719, 83721-83722, 83727, 83735, 83775, 83785, 83788-83789, 83805, 83825, 83835, 83840, 83857-83858, 83861, 83864, 83866, 83872-83874, 83876, 83880, 83883, 83885, 83887, 83890-83894, 83896-83898, 83900-83909, 83912-83916, 83918-83919, 83921, 83925, 83930, 83935, 83937, 83945, 83950-83951, 83970, 83986-83987, 83992-83993, 84022, 84030, 84035, 84060-84061, 84066, 84075, 84078, 84080-84081, 84085, 84087, 84100, 84105-84106, 84110, 84112, 84119-84120, 84126-84127, 84132-84135, 84138, 84140, 84143-84146, 84150, 84152-84157, 84160, 84163, 84165-84166, 84181-84182, 84202-84203, 84206-84207, 84210, 84220, 84228, 84233-84235, 84238, 84244, 84252, 84255, 84260, 84270, 84275, 84285, 84295, 84300, 84302, \\
& & 84305, 84307, 84311, 84315, 84375-84379, 84392, 84402-84403, 84410, 84425, 84430-84432, 84436-84437, 84439, 84442-84443, 84445-84446, 84449-84450, 84460, 84466, 84478-84482, 84484-84485, 84488, 84490, 84510, 84512, 84520, 84525, 84540, 84545, 84550, 84560, 84577-84578, 84580, 84583, 84585-84586, 84588, 84590-84591, 84597, 84600, 84620, 84630, 84681, 84702-84704, 84830, 84999, 85007, 85014, 85018, 85025, 85027, 85610, 85651-85652, 85730, 86003, 86038, 86140, 86580, 86592, 86880, 86900-86901, 87040, 87070, 87077, 87081, 87086, 87088, 87186, 87491, 87591, 87621, 87804, 87880, 88142, 88175, 88304-88305, 88312-88313, 88342, 88720 \\
& 90000-99999 & 94760, 99000-99001 \\ 
Low-Value & 20000-29999 & 29877-29879 \\
Services & 30000-39999 & 36222-36224 \\
& 70000-79999 & 70450, 70460, 70470, 70498, 70547-70553, 71010, 71015, 71020-71023, 71030, 71034-71035, 72010, 72020, 72052, 72100, 72110, 72114, 72120, 72131-72133, 72141-72142, 72146-72149, 72156-72158, 72200-72202, 72220, 78451-78454, 78460-78461, 78464-78465, 78472-78473, 78481-78483, 78491-78492 \\
& 80000-89999 & 82306, 82652, 87620-87625, 88141-88143, 88147-88155, 88164-88167, 88174-88175 \\
& 90000-99999 & 93000, 93005, 93010, 93015-93018, 93303-93308, 93312, 93315, 93318, 93350-93351, 93880-93882, 94010 \\ 
Preventive Care & 10000-19999 & 11976, 11981-11983 \\
& 30000-39999 & 36415-36416  \\
& 40000-49999 & 44388-44389, 44392-44394, 45300, 45303-45309, 45315-45317, 45320, 45330-45335, 45338-45340, 45378-45386  \\
& 50000-59999 & 57170, 58300-58301, 58340, 58565, 58600, 58605, 58611, 58615, 58670-58671  \\
& 70000-79999 & 71250, 74263, 74740, 76070-76071, 76075-76078, 76497, 76977, 77078-77083, 78350 \\
& 80000-89999 & 80061, 82270, 82274, 82465, 82947-82952, 83036, 83718-83721, 84478, 86580, 86592-86593, 86631-86632, 86689, 86701-86703, 86803-86804, 87110, 87270, 87320, 87340-87341, 87390-87391, 87490-87492, 87590-87592, 87620-87622, 87801, 87810, 87850, 88141-88143, 88147-88155, 88164-88167, 88174-88175, 88304-88305  \\
& 90000-99999 & 92015, 92507, 92551-92553, 92558, 92567, 92585-92588, 96040, 96110, 96127, 96160-96161, 96372, 97802-97804, 99173-99174, 99201-99205, 99211-99215, 99381, 99385-99387, 99395-99397, 99401-99404, 99411-99412, 99420 \\ 
Specialist Care & 10000-19999 & 11100, 17000, 17003-17004, 17110-17111, 17250 \\
& 40000-49999 & 43239, 47562 \\
& 80000-89999 & 82962  \\
& 90000-99999 & 92012-92014, 92587, 93010, 94010 \\ 
\midrule
\bottomrule
\caption{\label{tab:outpatient-services} Identifying Types of Outpatient Services}

\end{longtable}
\end{ThreePartTable}
\clearpage


\newgeometry{scale=1}
\thispagestyle{empty}
{%
  \centering
  \includegraphics[page=1,scale=.95]{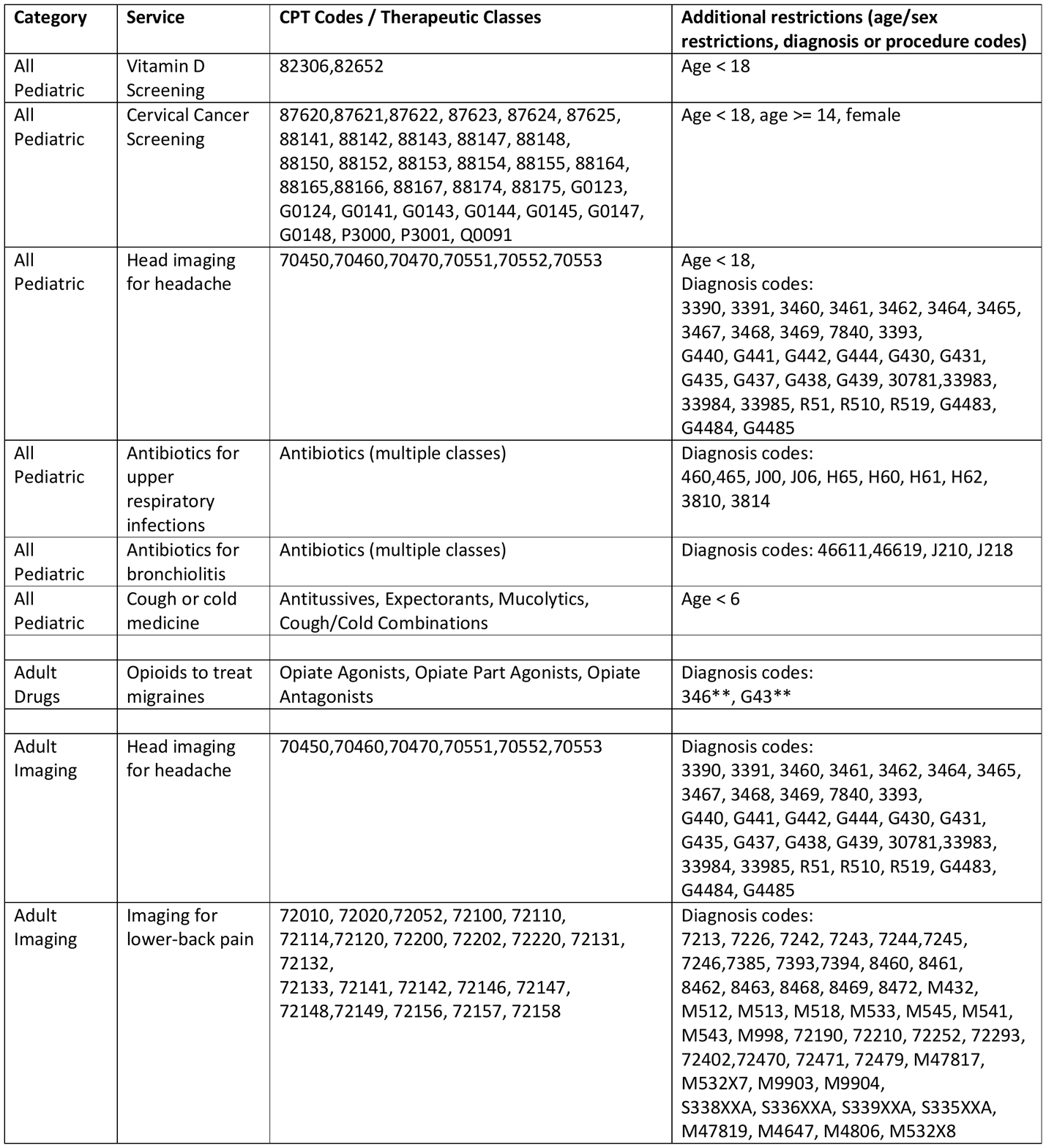}
  \captionof{table}{Identifying Low-Value Health Services}\label{tab:lv}
  \par
}
\includepdf[pages=2-, scale=1]{Figures/Appendix_LowValueServices.pdf}
\restoregeometry


\end{document}